\documentclass[3p]{elsarticle}

\usepackage{hyperref}

\journal{Digital Signal Processing, Elsevier}
\date{November 6, 2020}

\usepackage{amsmath,amssymb,amsfonts,amsbsy}
\usepackage{graphicx}
\usepackage{textcomp}
\usepackage{xcolor}
\usepackage{algorithm}
\usepackage[noend]{algorithmic}
\usepackage{booktabs}
\usepackage{multirow}
\usepackage{subfig}
\usepackage{siunitx}
\usepackage{mathtools}

\DeclareMathOperator*{\diag}{diag}
\DeclareMathOperator*{\cond}{cond}
\DeclareMathOperator*{\argmin}{argmin}

\DeclareMathOperator*{\card}{card}
\DeclareMathOperator*{\expect}{E}
\DeclarePairedDelimiter{\abs}{\lvert}{\rvert}
\newcommand{\norm}[1]{\left\lVert#1\right\rVert}
\newcommand{\ca}{C\textsubscript{A}}

\newcommand{\inlinecomment}[1]{\hfill \texttt{\textbackslash \textbackslash #1}}

\newcommand{\bX}{\mathbf{X}}
\newcommand{\bA}{\mathbf{A}}
\newcommand{\bS}{\mathbf{S}}
\newcommand{\bN}{\mathbf{N}}
\newcommand{\bH}{\mathbf{H}}

\begin{document}

\begin{frontmatter}

\title{Joint deconvolution and unsupervised source separation for data on the sphere}

\author{R. Carloni Gertosio\corref{cor}}
\ead{remi.carlonigertosio@cea.fr}
\author{J. Bobin}
\ead{jerome.bobin@cea.fr}
\cortext[cor]{Corresponding author.}
\address{IRFU, CEA, Université Paris-Saclay, F-91191, Gif-sur-Yvette, France}

\begin{abstract}
Tackling unsupervised source separation jointly with an additional inverse problem such as deconvolution is central for the analysis of multi-wavelength data. This becomes highly challenging when applied to large data sampled on the sphere such as those provided by wide-field observations in astrophysics, whose analysis requires the design of dedicated robust and yet effective algorithms. We therefore investigate a new joint deconvolution/sparse blind source separation method dedicated for data sampled on the sphere, coined SDecGMCA. It is based on a projected alternate least-squares minimization scheme, whose accuracy is proved to strongly rely on some regularization scheme in the present joint deconvolution/blind source separation setting. To this end, a regularization strategy is introduced that allows designing a new robust and effective algorithm, which is key to analyze large spherical data. Numerical experiments are carried out on toy examples and realistic astronomical data.

\end{abstract}


\end{frontmatter}

\paragraph{Notations} Vectors are written in bold lowercase letters, such as $\mathbf{x}$. $\mathbf{x}_i$ denotes the $i$\textsuperscript{th} entry of $\mathbf{x}$. Matrices are written in bold uppercase letters, such as $\mathbf{X}$. $\mathbf{X}_i$ and $\mathbf{X}^j$ denote respectively the $i$\textsuperscript{th} row and $j$\textsuperscript{th} column of $\mathbf{X}$, and $\mathbf{X}_i^j$ denotes the ($i$, $j$)\textsuperscript{th} entry of $\mathbf{X}$. $\norm{\cdot}_p$ denotes the $p$-norm for vectors and the induced $p$-norm for matrices. $\norm{\cdot}_{\ell_p}$ denotes the "entrywise" $p$-norm for matrices. The transpose operator is written with a superscript $\top$, such as $\mathbf{X}^\top$. The transpose-conjugate operator is noted with a superscript $\dagger$, such as $\mathbf{X}^\dagger$. $\bX \geq 0$ means that the coefficients of $\bX$ are non-negative. $\diag(\mathbf{x})$ returns the diagonal matrix constituted of the coefficients of $\mathbf{x}$. $\odot$ denotes the element-wise product, also known as the Hadamard product. $*$ denotes the convolution operator on the sphere. The spherical harmonic projection of a vector $\mathbf{x}$ is written $\mathbf{\hat{x}}$. $\mathbf{\hat{X}}$ is defined as the stack of the spherical harmonic projections of the rows of $\mathbf{X}$. The estimate of a signal at iteration $i$ is noted with a superscript in brackets, such as $\mathbf{X}^{(i)}$. The ground truth signal is written with an asterisk, such as $\mathbf{X}^*$. 

\section{Introduction}

In a large number of applications, ranging from biomedical imaging to astrophysics, retrieving the relevant information from multichannel observations requires tackling an unsupervised matrix factorization problem dubbed Blind Source Separation (BSS). More precisely, we assume that such data are composed of $N_c$ channels. For channel $\nu$, a single observation, whether it represents an image or a 1D signal, can be described by the following linear mixture model:
\begin{equation}
    \label{eq:LMM}
    \bX_\nu = \sum_{n=1}^{N_s} \bA_\nu^n \bS_n + \bN_\nu,
\end{equation}
where the source $\bS_n$ is composed of $N_p$ entries or samples, $\bA_\nu^n$ is a scalar that quantifies the mixture weight of source $n$ in the observation $\nu$, and $\bN_\nu$ stands for some additive noise. The number of channels $N_c$ is assumed greater or equal to the number of sources $N_s$; the resulting problem is complete or over-complete. Unsupervised source separation seeks estimates of $\bA$ and $\bS$ from the knowledge of $\bX$ only. As this problem boils down to an unsupervised matrix factorisation problem, BSS is an ill-posed inverse problem, which requires additional prior information about the sources $\bS$ and/or the so-called mixing matrix $\bA$. So far, common assumptions are the statistical independence of the sources (Independent Component Analysis \cite{HBSS}), the non-negativity of the sources and the mixing matrix (Non-negative Matrix Factorisation \cite{Cichocki_07_HierarchicalALSAlgorithms, Gillis_12}). In the present paper, we will more specifically focus on imposing the sparsity of the sources in some representation domain, which has been widely showed to provide efficient separation procedures in a large number of applications \cite{ica:zibu_relnewton,Li_03_Sparserepresentationand,AMCA15,KervazoPALM19}.\\

While this problem has now become standard and widely studied during the last three decades, it is much less commonplace when the observations are further distorted by channel-dependent measurement operators $\bH_\nu$. For instance, large-band multi-wavelength observations have the particularity to have resolutions which can be significantly different between the observation channels. In this case, coping with now heterogeneous data requires tackling an extra deconvolution step, thus leading to a joint deconvolution and blind source separation (DBSS) problem. A mathematically similar problem arises when the observations are composed of incomplete measurements such as in interferometric measurements \cite{Chapman15,Jiang_2017} or compressive hyperspectral imaging \cite{Golbabaee13,Kobarg14}. The above mixture model is then substituted with the following:
\begin{equation}
\label{eq:imagmodeldirspace}
\bX_{\nu} = \mathbf{H}_{\nu} (\mathbf{A}_{\nu} \mathbf{S}) + \mathbf{N}_{\nu}.
\end{equation}
Being an ill-posed matrix factorization problem, BSS alone is already a challenging inverse problem. This is all-the-more complex when a channel-dependent operator further needs to be inverted as it can be ill-conditioned or even not invertible. Furthermore, standard BSS methods cannot be applied directly to the problem in Eq.~\eqref{eq:imagmodeldirspace} unless the data $\bX$ are pre-processed so as to obtain new data with a common resolution. However, jointly solving both deconvolution and separation is expected to yield much better results, allowing to more precisely account for the forward observation model and noise in a single pass.\\
To the best of our knowledge, joint DBSS has been seldom investigated. The closest work known so far has been introduced by Kleinsteuber \textit{et al.}, who proposed a BSS algorithm to analyze incomplete data in the framework of compressed sensing \cite{Kleinsteuber12}. This can be regarded as a special case of DBSS where the measurement operator is defined as a projection onto a low-dimensional measurement subspace.  However, the proposed method is not compatible in our case since it only applies to compressively sensed measurements.\\
More recently, we introduced the first joint DBSS method \cite{Jiang_2017}. The proposed DecGMCA algorithm enforces the sparsity of the sources in some domain, that shall be represented by its transfer matrix $\bf \Phi$, by seeking a stationary point of the following cost function:
\begin{equation}
    \min_{\bA,\bS} \; \sum_{n=1}^{N_s} \lambda_n \left\|\bS_n {\bf \Phi}^\top \right\|_{1} + \frac{1}{2} \sum_{\nu=1}^{N_c} \left\| \bX_\nu - {\bH}_\nu\left({\bA_\nu \bS} \right) \right \|_2^2.
\end{equation} 
For that purpose, DecGMCA builds upon a projected Alternate Least-Squares (projected ALS) minimization procedure, where the sources and the mixing matrix are estimated sequentially. Projected ALS has been showed to provide computationally efficient and fast BSS algorithms. Furthermore, it allows the use of heuristics to automatically tune the regularization parameters $\{\lambda_n,~n\in[1, N_s]\}$ \cite{KervazoPALM19}. However, in the setting of DBSS, resorting to a projected ALS optimization scheme also raises a major difficulty: the least-squares problem with respect to $\bS$ is generally ill-conditioned -- if not ill-posed -- and needs to be regularized, which has a significant impact on the quality of the separation results.\\

\paragraph{\bf Contributions} 
In this article, we investigate a new joint DBSS algorithm to analyze data that are sampled on the sphere. This is essential to cope with the kind of wide-field spherical data, which are now common in scientific fields such as astronomy. This includes the analysis of the forthcoming Square Kilometer Array radiotelescope (SKA\footnote{\url{https://www.skatelescope.org/}}) or X-ray observatories such as the future European mission Athena\footnote{\url{https://sci.esa.int/web/athena}} to cite only two examples. In contrast to the standard case, analyzing spherical data raises extra difficulties due to the high computational cost of their manipulation, which makes essential the design of a computationally efficient and reliable algorithm. We therefore first aim at extending the algorithm DecGMCA \cite{Jiang_2017} to tackle joint deconvolution and separation problems from spherical data. As described in Section~\ref{sec:sdecgmca}, the method is based on a projected alternate least-squares minimization in order to combine rapidity and precision. Compared to the BSS problem, the procedure calls for an extra regularization to deal with a naturally ill-conditioned if not ill-posed problem. Yet, the regularization strategy and its impact on the solution has not been examined. Fot that purpose, we introduce in Section~\ref{sec:regstrat} several regularization strategies, which significantly improve the separation quality. Based on these results, we introduce in Section~\ref{sec:algo} a new algorithm, coined SDecGMCA, to tackle efficiently joint deconvolution and blind source separation problems. Finally, in Section~\ref{sec:numerics}, numerical experiments, which involve both toy examples and realistic astrophysical simulations, are presented.\\

\section{The spherical DecGMCA algorithm}
\label{sec:sdecgmca}

In this section, we first adapt the DBSS method proposed in \cite{Jiang_2017} to tackle spherical data. Furthermore, we investigate in depth the \textit{ad hoc}, yet necessary, regularization procedure in the update of the sources used in the DecGMCA algorithm. We show that it has a significant impact on the reconstruction quality of the sources. We then introduce two new regularization strategies, which noticeably outperform the one of DecGMCA.

\subsection{Toward a DBSS method for spherical data}
\label{sec:spherebss}
\begin{figure}
	\centering
	\includegraphics[width=0.33\linewidth]{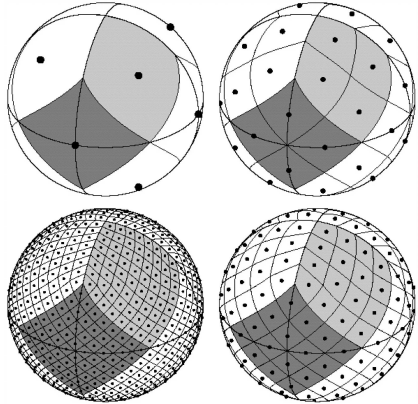}
	\caption{HEALPix discretization of the sphere (source: \cite{pixel:healpix}). Moving clockwise from the top left, $N_{side} = 1$, 2, 4, 8}
	\label{fig:healpix_discretization}
\end{figure}

In the following, the multichannel data $\bX$ are assumed to be sampled on the sphere with HEALPix \cite{pixel:healpix}. As showed in Fig.~\ref{fig:healpix_discretization}, the sphere is divided in 12 quadrilateral sections of equal area, which are hierarchically subdivided in $N_{side}^2$ pixels. The resolution parameter $N_{side}$ is the number of pixels along the sides of the 12 initial sections. HEALPix is the most commonly used pixelization of the sphere in astrophysics and geophysics, since it is well adapted for hierarchical analysis and spherical harmonic projections.\\
The measurement operators $\bH_\nu$ are assumed linear and isotropic. Therefore, the mixture model \eqref{eq:imagmodeldirspace} becomes:
\begin{equation}
\label{eq:imagmodeldirect}
\bX_{\nu} = (\mathbf{A}_{\nu} \mathbf{S})*\mathbf{H}_{\nu} + \mathbf{N}_{\nu},
\end{equation}
where $*$ denotes the convolution operator on the sphere between a spherical function and an isotropic kernel \cite{Driscoll94FourierConvolutionSphere}.  The number of channels is recalled to be greater or equal to the number of sources. \\
Quite similarly to the Euclidean case, the spherical convolution operator is diagonal in the spherical harmonics basis.  Spherical harmonics are ordered by degree $l \geq 0$ and mode $m \in [-l, l]$. The harmonic coefficients $\mathcal{D} = \{(l,m)\in\mathbb{N}\times\mathbb{Z}\text{, }  |m| \leq l \}$, also called multipoles, are comparable to spatial frequencies. Equivalently to a Fourier decomposition, any spherical function is uniquely decomposed in a linear combination of spherical harmonics. Let $\mathbf{y_{l,m}}$ denote the $(l,m)^\text{th}$ normalized spherical harmonic sampled with HEALPix. The projection in the spherical harmonic space of a vector $\mathbf{x}$ is given by: 
\begin{equation}
\label{eq:SHproj}
\mathbf{\hat{x}}_{l,m} = \frac{4\pi}{N_p} \mathbf{y_{l,m}}^\dagger \mathbf{x},
\end{equation}
where $N_p = 12 N_{side}^2$ is the number of pixels. The spherical harmonics sampled with HEALPix form a linearly independent set up to $l_{max} = 3 N_{side} - 1$. Therefore, the projections will be limited in frequency to $l_{max}$. For the sake of clarity, this limitation will not be indicated in the following equations. \\
Equation \eqref{eq:imagmodeldirect} can be recast in the spherical harmonic domain:
\begin{equation}
\mathbf{\hat{X}}^{l,m}_\nu = \mathbf{\hat{H}}^l_\nu  \mathbf{A}_\nu \mathbf{\hat{S}}^{l,m} + \mathbf{\hat{N}}^{l,m}_\nu,
\end{equation}
where the superscript $(l, m)$ refers to the column of the matrix with degree $l$ and mode $m$, in accordance with Eq.~\eqref{eq:SHproj}. The convolution kernels being isotropic, $\mathbf{\hat{H}}$ does not depend on $m$. The mixing model is finally rewritten by grouping the channels, yielding:
\begin{equation}
\label{eq:imagmodel}
\mathbf{\hat{X}}^{l,m} = \diag\left({\mathbf{\hat{H}}^{l}}\right) \mathbf{A} \mathbf{\hat{S}}^{l,m} + \mathbf{\hat{N}}^{l,m}.
\end{equation}

To build an estimator for the sources $\bS$ and the mixing matrix $\bA$, the following assumptions are made:
\begin{itemize}
\item{\it Data-fidelity term: } Thereafter, the noise is assumed to be Gaussian; the data-fidelity term is then defined as the Euclidean distance between the observations and the forward model as follows: 
\begin{equation}
	\label{eq:datafid}
	h \left(\mathbf{A}, \mathbf{S}\right) =  \sum\limits_{(l,m) \in \mathcal{D}}  \norm{\mathbf{\hat{X}}^{l,m} - \diag\left({\mathbf{\hat{H}}^{l}}\right) \mathbf{A} \mathbf{\hat{S}}^{l,m}}^2_2.
\end{equation}

\item{\it Source regularization: } The sources are assumed to be sparse in a given dictionary $\mathbf{\Phi}$. This sparsity can be enforced by a $\ell_1$-penalization in the transformed domain, hence the constraint term on $\mathbf{S}$:
\begin{equation}
\label{eq:consS}
\mathcal{G}\left( \mathbf{S}\right) = \norm{\mathbf{\Lambda} \odot \left( \mathbf{S} \mathbf{\Phi}^\top\right)}_{\ell_1},
\end{equation}
where $\mathbf{\mathbf{\Lambda}}$ is a matrix that stores the thresholding parameters, which will be described in detail in Section~\ref{sec:threshparam}. For the sake of simplicity, we will assume that the sparse domain is orthogonal, \textit{ie.} that $\mathbf{\Phi}$ is a square matrix verifying $\mathbf{\Phi}^\top \mathbf{\Phi} = \mathbf{I}$, with $\mathbf{I}$ the identity matrix. However, as mentioned in \cite{AMCA15}, the forthcoming mathematical derivations also apply to tight frames at first order approximation \cite{Elad06}.

In some applications, the sources are expected to be non-negative. In this case, one can also impose a non-negativity constraint on $\bf S$, yielding:
\begin{equation}
\label{eq:consSPos}
\mathcal{G}\left( \mathbf{S}\right) = \norm{\mathbf{\Lambda} \odot \left( \mathbf{S} \mathbf{\Phi}^\top\right)}_{\ell_1} + \chi_\mathcal{K_S}\left(\bS\right),
\end{equation}
where $\chi_\mathcal{K_S}$ is the characteristic function of the positive orthant for the sources $\mathcal{K_S} = \{ \mathbf{S} \in \mathbb{R}^{N_s \times N_p},~\bS \geq 0 \}$.\\

\item{\it Mixing matrix regularization: } To mitigate the scale indeterminacy of the product $\mathbf{A \hat{S}}$, the columns of $\mathbf{A}$ are enforced to belong to the $\ell_2$-hypersphere, which is defined as the set of vectors with unit $\ell_2$ norm:
\begin{equation}
	\label{eq:consA}
	\mathcal{J}\left( \mathbf{A}\right) = \chi_\mathcal{O}\left(\mathbf{A}\right),
\end{equation}
where $\chi_\mathcal{O}$ is the characteristic function of the oblique ensemble $\mathcal{O} = \{ \mathbf{A} \in \mathbb{R}^{N_c \times N_s}, \forall n \in \left[1, N_s\right], \norm{\mathbf{A}^n}_2 = 1 \}$.

In addition, it is possible to impose the non-negativity of the mixing matrix; in this case:
\begin{equation}
\label{eq:consAPos}
\mathcal{J}\left( \mathbf{A}\right) = \chi_\mathcal{O}\left(\mathbf{A}\right) + \chi_\mathcal{K_A}\left(\bA\right),
\end{equation}
where $\chi_\mathcal{K_A}$ is the characteristic function of the positive orthant for the mixing matrices.\\
\end{itemize}
To sum up, the mixing matrix and the sources will be estimated by looking for a stationary point of the following cost function:
\begin{equation}
\label{eq:sdecg_cost}
\min_{\bA,\hat{\bS}} \; \left\|{\bf \Lambda} \odot \left(\hat{\bS} {\mathcal{F}^\dagger} {\bf \Phi}^\top\right) \right\|_{\ell_1} + \chi_\mathcal{O}\left(\mathbf{A}\right) + \sum\limits_{(l,m) \in \mathcal{D}}  \norm{\mathbf{\hat{X}}^{l,m} - \diag\left({\mathbf{\hat{H}}^{l}}\right) \mathbf{A} \mathbf{\hat{S}}^{l,m}}^2_2,
\end{equation} 
plus the characteristic functions of the positive orthants, if the sources and/or the mixing matrix are non-negative. $\mathcal{F}^\dagger$ is the inverse spherical harmonic transform. 

Being an unsupervised matrix factorization task, the problem in Equation~\eqref{eq:sdecg_cost} is a challenging non-convex problem. More precisely, it is multi-convex \cite{Xu_13_BlockCoordinateDescent} since it is convex with respect to $\bS$ ({\it resp.} $\bA$) when $\bA$ ({\it resp.} $\bS$) is fixed. In this context, most standard algorithms perform with alternating minimization steps, with respect to the mixing matrix $\bA$ and the sources $\bS$ in a sequential manner.\\
When it comes to DBSS, a major challenge is the need for a robust and yet effective algorithm, but with a reasonable computational burden. As the above optimization combines non-differentiable terms, such as the sparsity-enforcing $\ell_1$ regularization term or the characteristic function of the Oblique ensemble $\chi_\mathcal{O}$, it would make perfect sense to design a minimizer based on plain proximal algorithms \cite{Boyd_Proximal14}. However, this would be lead to algorithms with very high computational cost, especially if one considers that DBSS problems are generally ill-conditioned. Moreover, when BSS only is considered, tuning the regularization parameters, which has a paramount impact on the separation performances, is particularly complex with standard proximal algorithms \cite{KervazoPALM19}.\\
In contrast, projected Alternating Least-Squares (pALS), which were first introduced in the field of Non-Negative Matrix Factorization \cite{Paatero_94_Positivematrixfactorization}, have been proved to be particularly effective at providing fast sparse BSS algorithms. Furthermore, they come with almost automatic strategies to fix the regularization parameters in a robust way \cite{AMCA15,KervazoPALM19}. In a nutshell, when either $\bA$ or $\bS$ is updated, a first least-squares estimate is computed by minimizing the data-fidelity term. This step is then followed by the application of the proximal operator of the corresponding regularization term\footnote{We refer to \cite{Boyd_Proximal14} for insightful details about proximal calculus and algorithms.}. These two steps are detailed in the following sections.

\subsection{Update of \texorpdfstring{$\mathbf{A}$}{A}}
\label{sec:UpdateA}

Updating the mixing matrix $\mathbf{A}$ when the sources $\mathbf{S}$ are fixed requires solving the least-squares problem $\argmin\limits_{\mathbf{A}} h\left(\mathbf{A}, \mathbf{S}\right)$, which yields for all channel $\nu \in [1, N_c]$:
\begin{equation}
\mathbf{A}_\nu \leftarrow \left( \sum\limits_{(l,m) \in \mathcal{D}}  \mathbf{\hat{X}}_\nu^{l,m} \mathbf{\hat{H}}_\nu^{l} {{}\mathbf{\hat{S}}^{l,m}}^\dagger \right) \left( \sum\limits_{(l,m) \in \mathcal{D}}  {{}\mathbf{\hat{H}}_\nu^l}^2 \mathbf{\hat{S}}^{l,m} {{}\mathbf{\hat{S}}^{l,m}}^\dagger\right)^{-1}.
\end{equation}
In this equation, the estimate is defined with sums over all multipoles, which are significantly larger than the number of sources $N_s$. Consequently, the matrix $\left( \sum\limits_{(l,m) \in \mathcal{D}}  {{}\mathbf{\hat{H}}_\nu^l}^2 \mathbf{\hat{S}}^{l,m} {{}\mathbf{\hat{S}}^{l,m}}^\dagger\right)$ is generally invertible if not well conditioned.\\ 
In a second phase, the proximal operator of $\mathcal{J}$ is applied on the solution. If there is no constraint on the sign of $\bA$, it is the projection on the multidimensional $\ell_2$-hypersphere $\mathcal{O}$:
\begin{equation}
\mathbf{A}^n \leftarrow \Pi_{\mathcal{O}}(\bA) = \frac{\mathbf{A}^n }{\norm{\mathbf{A}^n}_2}.
\end{equation}
If $\bA$ is non-negative, the proximal operator of $\mathcal{J}$ is the composition of the proximal operators of the oblique and non-negativity constraints:
\begin{equation}
\mathbf{A} \leftarrow \Pi_{\mathcal{O}}\left(\Pi_{\mathcal{K_A}}(\bA)\right).
\end{equation}

\subsection{Update of \texorpdfstring{$\bS$}S}\label{sec:updateS}
\label{sec:UpdateS}
Updating $\bS$ assuming $\bA$ is fixed leads to the following optimization problem:
\begin{equation}
\label{eq:sdecg_cost_S}
\min_{\hat{\bS}} \; \left\|{\bf \Lambda} \odot \left(\hat{\bS} {\mathcal{F}^H} {\bf \Phi}^\top \right)\right\|_{\ell_1} + \sum\limits_{(l,m) \in \mathcal{D}}  \norm{\mathbf{\hat{X}}^{l,m} - \diag\left({\mathbf{\hat{H}}^{l}}\right) \mathbf{A} \mathbf{\hat{S}}^{l,m}}^2_2,
\end{equation} 
to which the term $\chi_\mathcal{K_S}\left(\bS\right)$ is added if the sources are non-negative. \\
According to the pALS minimization scheme, the solution of the above minimization problem is first approximated with a least-squares estimate by finding a solution to:
\begin{equation}
\label{eq:sdecg_cost_leastsquaresS}
\min_{\hat{\bS}} \; \sum\limits_{(l,m) \in \mathcal{D}} \norm{\mathbf{\hat{X}}^{l,m} - \diag\left({\mathbf{\hat{H}}^{l}}\right) \mathbf{A} \mathbf{\hat{S}}^{l,m}}^2_2.
\end{equation} 
Solving the least-squares problem $\argmin\limits_{\mathbf{S}} h\left(\mathbf{A}, \mathbf{S}\right)$ yields for all harmonic coefficient $(l,m) \in \mathcal{D}$:
\begin{equation}
\label{eq:updateS}
\mathbf{\hat{S}}^{l,m} \leftarrow \mathbf{M}[l]^{-1} \mathbf{A}^\top \diag\left({\mathbf{\hat{H}}^{l}}\right) \mathbf{\hat{X}}^{l,m},
\end{equation}
with $\mathbf{M}[l] = \mathbf{A}^\top \diag\left(\mathbf{\hat{H}}^l\right)^2 \mathbf{A}$.\\
It is important to notice that, in contrast to standard BSS problems, the above solution is not necessarily stable with respect to noise as the matrices $\mathbf{M}[l]$ may be ill-conditioned. Additionally, it is not always unique as $\mathbf{M}[l]$ may not be invertible; this might occur when the convolution kernels vanish for some spherical harmonics multipoles.\\
Borrowing ideas from sparsity enforcing deconvolution methods such as ForWard \cite{rest:neelamani01}, we proposed in \cite{Jiang_2017} to add an extra Tikhonov regularization \cite{ima:bertero98} to the least-squares problem: 
\begin{equation}
    \min_{\hat{\bS}} \; \dfrac{1}{2} \sum\limits_{n=1}^{N_s} \sum\limits_{(l,m) \in \mathcal{D}} \varepsilon_{n,l}  \left| \mathbf{\hat{S}}^{l,m}_n \right|^2 + \norm{\mathbf{\hat{X}}^{l,m} - \diag\left({\mathbf{\hat{H}}^{l}}\right) \mathbf{A} \mathbf{\hat{S}}^{l,m}}^2_2.
\end{equation}
The set $\{\varepsilon_{n,l} \geq 0\text{, } n\in[1,N_s]\text{, }l\in\mathbb{N}\}$ are the Tikhonov regularization coefficients. They depend on the frequency $l$ and on the source $n$. For all $(l,m) \in \mathcal{D}$, updating the sources can now be recast as:
\begin{equation}
	\label{eq:updateS2}
	\mathbf{\hat{S}}^{l,m} \leftarrow \left( \mathbf{M}[l] + \diag\limits_{n\in[1,N_s]} \left(\varepsilon_{n,l}\right) \right)^{-1} \mathbf{A}^\top \diag\left({\mathbf{\hat{H}}^{l}}\right) \mathbf{\hat{X}}^{l,m}.
\end{equation}
Similarly to standard deconvolution problems \cite{rest:neelamani01}, the way the Tikhonov regularization coefficients are fixed has a dramatic impact on the quality of the regularized least-squares solution, and eventually on the whole separation process. This is discussed in depth in the following section.\\

In a second step, the proximal operator of $\mathcal{G}$ is applied to the above least-squares solution. If there is no constraint on the sign of $\bS$, this amounts to applying the soft-thresholding operator $\mathcal{S}_\mathbf{\Lambda}$, with thresholds $\mathbf{\Lambda}$:
\begin{equation}
\mathbf{S} \leftarrow \mathcal{S}_{\mathbf{\Lambda}}\left(\mathbf{S} \mathbf{\Phi}^\top \right) \mathbf{\Phi}.
\end{equation}
Under the non-negativity constraint, the proximal operator of $\mathcal{G}$ has no analytical form. It is approximated as the composition of the proximal operator of the sparsity constraint and the proximal operator of the non-negativity constraint, which is the projection on the non-negative orthant $\Pi_\mathcal{K_S}$:
\begin{equation}
\mathbf{S} \leftarrow \Pi_\mathcal{K_S}\left(\mathcal{S}_{\mathbf{\Lambda}}\left(\mathbf{S} \mathbf{\Phi}^\top \right) \mathbf{\Phi}\right).
\end{equation}

\subsubsection{Regularization strategies}
\label{sec:regstrat}

In \cite{Jiang_2017}, the regularization strategy is defined quite arbitrarily with hyperparameters that are fixed to an {\it ad hoc} small value ({\it e.g. $10^{-3}$}). In this section, we will particularly highlight that this regularization significantly impacts the estimation precision. We further investigate different strategies allowing more efficient and adaptive way of tuning these key parameters.\\
Let $c$ be a positive number, that will be called the regularization hyperparameter. We will further investigate four different regularization strategies:

\begin{itemize}
\item{\bf Strategy \#1} Let us first consider the naive strategy where the regularization parameters are chosen independently of the source $n$ and the frequency $l$:
\begin{equation}
	\label{eq:str1}
	\varepsilon_{n,l} = c \text{.}
\end{equation}

\item{\bf Strategy \#2} The strategy presented in \cite{Jiang_2017} is also considered:
\begin{equation}
\label{eq:str2}
\varepsilon_{n,l} = c\, \lambda_{\text{max}}(\mathbf{M}[l]),
\end{equation}
where $\lambda_{\text{max}}(\cdot)$ returns the greatest eigenvalue. In \cite{Jiang_2017}, the main motivation of this choice was to set a regularization parameters that scales with the sources.

\item{\bf Strategy \#3} In Eq.~\eqref{eq:updateS}, the errors that contaminate the observations are amplified by the inverse of the smallest eigenvalue of $\mathbf{M}[l]$, denoted $\lambda_\text{min}(\mathbf{M}[l])^{-1}$. Limiting the noise amplification to $c$ amounts to choosing $\varepsilon_{n,l}$ such that  $\varepsilon_{n,l} + \lambda_\text{min}(\mathbf{M}[l]) \leq c \iff \varepsilon_{n,l}  \leq c - \lambda_\text{min}(\mathbf{M}[l])$. Bearing in mind that $\varepsilon_{n,l} \geq 0$, it is possible to set $\varepsilon_{n,l} = \max \left(0, c - \lambda_{\text{min}} \left(\mathbf{M}[l]\right)\right)$. A change of variable is finally operated to facilitate the interpretation of the hyperparameter $c$, yielding for all source $n\in[1,N_s]$ and frequency $l\in\mathbb{N}$:
\begin{equation}
	\label{eq:str3}
	\varepsilon_{n,l} = \max\left(0, c-\frac{\lambda_{\text{min}}(\mathbf{M}[l])}{\lambda_{\text{min}}(\mathbf{A}^\top\mathbf{A})+ \epsilon }\right),
\end{equation}
with $\epsilon = 1\mathrm{e}{-2}$, to prevent numerical issues. Since the sequence $(\lambda_\text{min}(\mathbf{M}[l]))_{l\in\mathbb{N}}$ decreases, the sequence $(\varepsilon_{n,l})_{l\in\mathbb{N}}$ increases. Consequently, the higher frequencies are more penalized, while the lower ones are preserved. This is advantageous because most of the information from a physical source generally lies in the lower frequencies. Moreover, the maximum operator in \eqref{eq:str3} allows to have $\varepsilon_{n,l} = 0$ for the smaller frequencies, provided that $c$ is small enough. This allows to keep the smaller frequencies unbiased.

\item{\bf Strategy \#4} As in the Euclidean case, it is possible to define an angular power spectrum, which describes the power distribution along the degrees $l$. It is likewise deduced from the harmonic decomposition: 
\begin{equation}
\mathbf{c_x}_l = \frac{1}{2l+1} \sum_{m=-l}^{l} \left|\mathbf{\hat{x}}_{l,m}\right|^2.
\end{equation} 
If the angular power spectra of the sources $\{\mathbf{c_{S_\mathnormal{n}}}\text{, }n\in[1,N_s]\}$ and the noise $\mathbf{c_N}$ (which is assumed observation independent) are known, the optimal strategy for the unpenalized least-squares problem is $\varepsilon_{n,l} = \mathbf{c_N}_l/\mathbf{c_{S_\mathnormal{n}}}_l$. This is reminiscent of a Wiener deconvolution filter. Yet, it has to be reminded that the update of $\mathbf{S}$ is further followed by a thresholding step, which likely alters the properties of the regularization strategy. Therefore, the regularization strategy is adapted by adding a hyperparameter $c$, yielding for all sources $n\in[1,N_s]$ and frequencies $l\in\mathbb{N}$:
\begin{equation}
	\label{eq:str4}
	\varepsilon_{n,l} = c\, \frac{\mathbf{c_N}_l}{\mathbf{c_{S_\mathnormal{n}}}_l}.
\end{equation}
$c$ tips the balance between the Tikhonov regularization and the sparsity regularization. 
\end{itemize}

\subsubsection{Numerical comparisons of the regularization strategies} 
\label{sec:nonblindpb}

In this paragraph, we propose illustrating the impact of the above regularization strategies with numerical experiments. For that purpose, we consider the non-blind separation case, \textit{ie.}~with the ground truth mixing matrix $\mathbf{A^*}$ known.\\

In these experiments, the data are sampled on the sphere using the HEALPix pixelization, with parameters $N_{side}=128$ and $l_{max}=384$. The observations are unmixed and deconvolved at the resolution of the best-resolved channel. This amounts to replacing $\mathbf{\hat{H}}_\nu^l$ by $\mathbf{\hat{H}}_\nu^l/\mathbf{\hat{H}}_{\nu_b}^l$, where $\nu_b$ is the index of the best-resolved channel. 

\begin{figure}
	\centering
	\includegraphics[width=0.4\linewidth]{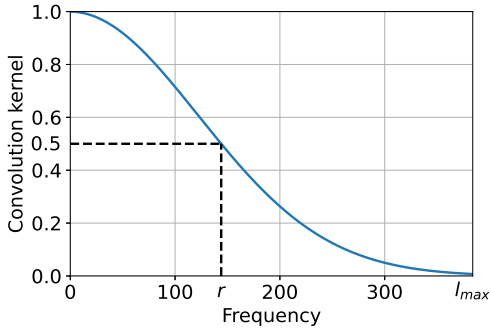}
	\caption{Example of a Gaussian filter $\mathbf{\hat{H}}_\nu$, with a resolution $r = 144$}
	\label{fig:filt_example}
\end{figure}

In these comparisons, the data are generated as follows:
\begin{itemize}
	\item the sources $\bS$ are random non-negative signals that are sparse in the spherical starlet (isotropic undecimated wavelets) domain \cite{Starck05}; they are also band-limited, with a frequency cut-off of $l_{max}/6 = 64$. See example in Fig.~\ref{fig:S1^*}.
	\item the mixing matrices $\bA$ are random non-negative matrices with a given condition number.
	\item the convolution kernels $\mathbf{\hat{H}}^l_{\nu} = \exp\left(-\frac{l(l+1)}{r(\nu)(r(\nu)+1)} \log 2\right)$ are Gaussian-shaped, with resolutions $r$ evenly spread between the minimum resolution $r_{min}$, which is a parameter to be set, and $l_{max}$. The resolution is defined as the full width at half maximum (FWHM) in the spherical harmonic domain of the convolution kernel (see example in Fig.~\ref{fig:filt_example}).
\end{itemize}
Throughout this section, we consider $N_s = 4$ sources.  Moreover, we define four parameters that characterize the observations:
\begin{itemize}
	\item number of observations $N_c$ ($N_c = 8$ by default),
	\item mixing matrix condition number $\cond(\mathbf{A})$ ($\cond(\mathbf{A}) = 2$ by default),
	\item minimum resolution $r_{min}$ ($l_{max}/8 = 48$ by default),
	\item overall signal-to-noise ratio (10~dB by default).
\end{itemize}

The joint deconvolution and non-blind separation problem is solved by performing the update of $\mathbf{S}$ proposed in \ref{sec:updateS}, using the ground truth mixing matrix $\mathbf{A^*}$. The update comprises a Tikhonov-penalized least square followed by a soft-thresholding. For strategy \#4, the regularization parameters are calculated with the angular power spectra of the ground truth sources. The sparsity of the sources is enforced in the spherical starlet domain. The thresholds are set in the same way than during the finale iteration of SDecGMCA; the strategy for the choice of the thresholds in SDecGMCA is discussed in detail in Section \ref{sec:threshparam}. 

The results are assessed with the normalized mean square error (NMSE) on the sources, defined by:
\begin{equation} \label{eq:nmse}
	\text{NMSE} = -10 \log_{10}\left(\frac{\norm{\mathbf{S}^*-\mathbf{S}}_{\ell_2}^2}{\norm{\mathbf{S}^*}_{\ell_2}^2}\right),
\end{equation}
with $\mathbf{S}^*$ the ground truth sources and $\mathbf{S}$ the estimated sources.\\

We test how the four regularization strategies we defined in the previous section behave with respect to the number of observations, the mixing matrix condition number, the minimum resolution of the convolution kernel and the SNR. Each experiment is composed of 100 trials, with varying random sources, mixing matrices and noise realizations. For each point, the optimal regularization hyperparameter is fine-tuned using a grid-search. The performances of each strategy is evaluated based on the NMSE of the estimated sources.\\
The results are reported in detail in \ref{app:resnonblind}. Strategy \#4 clearly provides the best reconstruction qualities, and  does so in various observation scenarios (varying number of observations, mixing matrix condition number, minimum resolution and SNR). Moreover, the optimal regularization hyperparameter of strategy \#4 is remarkably insensitive to the observation parameters; typically $c_{opt} \sim 0.5$. Yet, strategy \#4 requires to know the spectra of the sought sources, or at least an estimation of these. Among the other regularization strategies, which do not necessitate prior information on the sources, strategy \#3 achieves better results both in terms of reconstruction quality and hyperparameter sensitivity. \\

\section{Implementation of the SDecGMCA algorithm}
\label{sec:algo}

We shedded light on how critical the choice of the Tikhonov regularization strategy is. In this section, the details of implementation of SDecGMCA are described. In particular, a novel two-stage algorithm which exploits the proposed regularization strategies is introduced.\\

SDecGMCA is summarized in Algorithm \ref{alg:SDecGMCA}. The mixing matrix $\bA$ is initialized with a Principal Component Analysis (PCA) performed on the observations. For this step only, the data are first re-convolved so as they share a common resolution. To avoid noise amplification, such resolution is the one of the worse-resolved channel. SDecGMCA is comprised of two stages: the first stage estimates a first guess of the mixing matrix and the sources (\textbf{warm-up}); it is required to provide robustness with respect to the initial point. The second stage refines the separation by employing a more precise regularization strategy (\textbf{refinement}). The values of the parameters of SDecGMCA according to the stage are summarized in Table \ref{tab:paramSDecGMCA}.

\begin{algorithm}
	\caption{SDecGMCA}
	\label{alg:SDecGMCA}
	\footnotesize
	\textbf{Inputs:} $\mathbf{X}$, $\mathbf{\hat{H}}$, $N_s$, $c_{wu}$, $c_{ref}$, $\sigma^2$, $\mathbf{\Phi}$, $k=3$, $K_{max}=0.5$
	\begin{algorithmic}
		\STATE $i \leftarrow 0$
		\STATE PCA initialization of $\mathbf{A}^{(0)}$

		\STATE stage $\leftarrow$ warm-up \inlinecomment{Stage 1: warm-up}
		\WHILE{convergence not reached} 
			\STATE $i \leftarrow i+1$
			\STATE Update $c$ and $K$ according to Table \ref{tab:paramSDecGMCA} 
			\STATE Estimate $\mathbf{\hat{S}}$ with $\mathbf{A}$ fixed: $\mathbf{\hat{S}}^{(i)} \leftarrow \text{UpdateS}\left(\mathbf{\hat{X}},~ \mathbf{\hat{H}},~\mathbf{A}^{(i-1)},~\text{stage},~c,~\sigma^2,~\mathbf{\Phi},~k,~K\right)$

			\STATE Estimate $\mathbf{A}$ with $\mathbf{\hat{S}}$ fixed: $\mathbf{A}^{(i)} \leftarrow \text{UpdateA}\left(\mathbf{\hat{X}},~\mathbf{\hat{H}},~\mathbf{\hat{S}}^{(i)}\right)$
		\ENDWHILE

		\STATE stage $\leftarrow$ refinement \inlinecomment{Stage 2: refinement}
		\WHILE{convergence not reached} 
		\STATE $i \leftarrow i+1$
		\STATE Estimate $\mathbf{\hat{S}}$ with $\mathbf{A}$ fixed: $\mathbf{\hat{S}}^{(i)} \leftarrow \text{UpdateS}\left(\mathbf{\hat{X}},~ \mathbf{\hat{H}},~\mathbf{A}^{(i-1)},~\text{stage},~c_{ref},~\sigma^2,~\mathbf{\Phi},~k,~K_{max}\right)$

		\STATE Estimate $\mathbf{A}$ with $\mathbf{\hat{S}}$ fixed: $\mathbf{A}^{(i)} \leftarrow \text{UpdateA}\left(\mathbf{\hat{X}},~\mathbf{\hat{H}},~\mathbf{\hat{S}}^{(i)}\right)$
		\ENDWHILE
				
		\RETURN $\mathbf{A}^{(i)},~\mathbf{S}^{(i)}$
		
	\end{algorithmic}
\end{algorithm}

\begin{algorithm}
	\caption{UpdateS}
	\label{alg:UpdateS}
	\footnotesize
	\textbf{Inputs:} $\mathbf{\hat{X}}$, $\mathbf{\hat{H}}$, $\mathbf{A}$, stage, $c$, $\sigma^2$, $\mathbf{\Phi}$, $k$, $K$
	\begin{algorithmic}
		
		\STATE Calculate the regularization parameters $\{\varepsilon_{n,l}\}$ according to the current stage
		\FOR{$l = 0, 1, ..., l_{max}$ and $m = -l, -l+1, ..., l$} 
		\STATE Tikhonov-penalized least-squares update of $\mathbf{\hat{S}}$ :\\
		\hspace{1em}$\mathbf{\hat{S}}^{l,m} \leftarrow \left(  \mathbf{A}^\top \diag\left(\mathbf{\hat{H}}^l\right)^2 \mathbf{A}+ \diag\limits_{n\in[1,N_s]} \left(\varepsilon_{n,l}\right) \right)^{-1} \mathbf{A}^\top \diag\left({\mathbf{\hat{H}}^{l}}\right) \mathbf{\hat{X}}^{l,m}$
		\ENDFOR
		
		\STATE Calculate the thresholds $\mathbf{\Lambda}$, with a support-based strategy and a $\ell_1$-reweighting if need be, according to the current stage
		\STATE Soft-threshold $\mathbf{S}$:
		$\mathbf{S} \leftarrow \mathcal{S}_{\mathbf{\Lambda}}\left(\mathbf{S} \mathbf{\Phi}^\top \right) \mathbf{\Phi}$
		\STATE Project $\bS$ on the positive orthant (if non-negativity constraint): $\mathbf{S} \leftarrow \max\left(\mathbf{S},~\mathbf{0}\right)$ 
		
		\RETURN $\mathbf{\hat{S}}$
		
	\end{algorithmic}
\end{algorithm}

\begin{algorithm}
	\caption{UpdateA}
	\label{alg:UpdateA}
	\footnotesize
	\textbf{Inputs:} $\mathbf{\hat{X}}$, $\mathbf{\hat{H}}$, $\mathbf{\hat{S}}$
	\begin{algorithmic}

		\FOR{$\nu = 1, 2, ..., N_p$}
		\STATE Least-squares update of $\mathbf{A}$:\\
		\hspace{1em} $\mathbf{A}_\nu \leftarrow \left( \sum\limits_{(l,m) \in \mathcal{D}}  \mathbf{\hat{X}}_\nu^{l,m} \mathbf{\hat{H}}_\nu^{l} {{}{{}\mathbf{\hat{S}}}^{l,m}}^\dagger \right) \left( \sum\limits_{(l,m) \in \mathcal{D}}  {{}\mathbf{\hat{H}}_\nu^l}^2 {{}\mathbf{\hat{S}}}^{l,m} {{}{{}\mathbf{\hat{S}}}^{l,m}}^\dagger\right)^{-1}$
		\ENDFOR
		\STATE 
		\FOR{$n =  1, 2, ..., N_s$}
		\STATE Normalization of $\mathbf{A}^n$:  $\mathbf{A}^n \leftarrow \dfrac{\mathbf{A}^n}{\norm{\mathbf{A}^n}}_2$
		\ENDFOR
		\STATE Project $\bA$ on the positive orthant (if non-negativity constraint): $\mathbf{A} \leftarrow \max\left(\mathbf{A},~\mathbf{0}\right)$ 
		
		\RETURN $\mathbf{A}$
		
	\end{algorithmic}
\end{algorithm}

\begin{table}
	\centering
	\small
	\begin{tabular}{@{}lp{6em}p{6em}@{}}
		\toprule
		Stage  & 1: warm-up  & 2: refinement \\ \midrule
		Regularization strategy  & \#3   & \#4  \\
		Regularization hyperparameter $c$  & $10 \,c_{wu} \rightarrow c_{wu}$ & $c_{ref}$  \\
		Active source support $K$  & $0 \rightarrow K_{max}$  & $K_{max}$ \\
		$\ell_1$-reweighting   & No & Yes \\ 
		Non-negativity constraint on $\bS$ & No& Yes\\\bottomrule
	\end{tabular}
	\caption{Parameters of SDecGMCA for each stage}
	\label{tab:paramSDecGMCA}
\end{table}

\subsection{Choice of the thresholding parameters} \label{sec:threshparam}

As emphasized in \cite{KervazoPALM19}, the thresholding parameters play a central role for the robustness of the minimization scheme with respect to spurious local minima and the precision of the estimates. To that respect, this role is twofold:
\begin{itemize}
\item{Robustness with respect to noise: } The strength of GMCA lies in proposing an adapted thresholding strategy. The thresholds $\mathbf{\Lambda}$ are calculated at each iteration, for each source $n\in[1,N_s]$ by:
\begin{equation}
\mathbf{\Lambda}_{n} = k\,\sigma_{\bS_n}\, \mathbf{1}, 
\end{equation}
where $\sigma_{\bS_n}$ is the standard deviation of the noise which contaminates $(\mathbf{S\Phi}^\top)_{n}$ and $k$ is a parameter adjusting the sparsity intensity. In practice, $k=3$ in order to reject samples that are likely to be noise-related. When the noise level that contaminates the data is known, $\sigma_{\bS_n}$ can be evaluated analytically; this is described in details in \ref{app:multires}. This is however not always the case in most applications in BSS; when the noise level remains unknown, it can be computed empirically with the Median Absolute Deviation applied to the sources in the sparse domain. We refer the interested reader to \cite{KervazoPALM19} for more details. 

\item{Decreasing thresholding strategy: } It has long been showed empirically that applying a decreasing threshold strategy dramatically improves the robustness of the separation process with respect to local spurious minima \cite{Bobin_07_Sparsityandmorphological,AMCA15,KervazoPALM19}. In the present article, we adapt this strategy to the context of DBSS. To that end, the thresholds are set so as to keep only a fixed amount of samples, say the $K$~\% samples with the greatest amplitudes, above the noise-related threshold $k\,\sigma_{\bS_n}$. The percentage $K$ linearly increases along the iterations from 0 to $K_{max}$. This procedure (i) enforces the decrease of the thresholds, thus improving the robustness of the separation, and (ii) provides an additional implicit regularization, by selecting only the most significant samples. $K_{max}$ sets the intensity of the two aforementioned phenomena. More specifically, the smaller the $K_{max}$, the slower the decrease of the thresholds and the greater the regularization. In practice, $K_{max}$ can be set to $0.5$.\\ 
Assuming that the coefficients of $(\mathbf{S\Phi}^\top)_n$ are sorted in order of descending modulus, this procedure is implemented for all source $n\in[1,N_s]$ by:
\begin{equation}
\label{eq:perdecthr}
\mathbf{\Lambda}_{n} = \left|(\mathbf{S\Phi}^\top)_n^{p_0}\right| \mathbf{1},
\end{equation}
with $p_0 = \left\lfloor K \card\left(\left|(\mathbf{S\Phi}^\top)_n\right| \geq k\,\sigma_{\bS_n}\right) \right\rfloor$, $K \in \left]0,1\right]$ the active source support and $\mathbf{1}$ a vector of ones.  
\end{itemize}

The soft-thresholding induced by the $\ell_1$-penalization introduces a bias in the estimation of the sources. Resorting to a $\ell_1$-reweighting scheme \cite{CandesIRL108, KervazoPALM19} allows to reduce this phenomenon. It consists in deriving sample-wise thresholds based on the samples' value in the sparse domain; the greater the sample's amplitude, the smaller the threshold and thus the smaller the bias.
Assuming that $\lambda_{n}^{(i)}$ is the initial non-sample-dependent threshold of source $n\in[1,N_s]$ at iteration $i$, the $\ell_1$-reweighted threshold for sample $p\in[1,N_p]$ is given by:
\begin{equation}
	{{}\mathbf{\Lambda}_{n}^{p}}^{(i)} = \frac{\lambda_{n}^{(i)}}{1+\frac{\left|{{}(\mathbf{S\Phi}^\top)_n^p}^{(i-1)}\right|}{\lambda_{n}^{(i)}}}.
\end{equation}

\subsection{A two-stage minimization approach}

We highlighted in Section~\ref{sec:nonblindpb} that the regularization strategy \#4 provides significantly better results in the non-blind source separation case. Since, this strategy is defined based on some estimate of the sources, we propose proceeding with a two-step approach: i) a warm-up step, whose goal is to provide a quick rough estimate of $\bA$ and $\bS$, with increased robustness with respect to the initialization and spurious local minima, and ii) a refinement step that makes use of regularization strategy \#4.\\

\paragraph*{Warm-up stage}
During the warm-up, regularization strategy \#3 is employed. Indeed, as seen in Section~\ref{sec:nonblindpb}, it provides the best results among the three first regularization strategies, which require no additional information about the sources. During this stage, the estimates of the sources are likely to be dominated by the artifacts due to the noise amplification, especially if the regularization hyperparameter at warm-up $c_{wu}$ is too small. The estimation of $\mathbf{A}$ is then erroneous and the algorithm might get stuck in a local minimum. This phenomenon can be alleviated by over-regularizing the sources during the first iterations, \textit{i.e.}~by overestimating $c_{wu}$ \cite{Jiang_2017}. Doing so, the noise contamination is reduced and only the major features of the sources are kept. In the spirit of \cite{Jiang_2017}, the regularization hyperparameter is then progressively decreased along the iterations to the input value, in order to refine the estimations of $\mathbf{S}$ and consequently $\mathbf{A}$. The decrease of the warm-up regularization hyperparameter markedly improves the robustness of the separation in terms of convergence. Moreover, the earlier mentioned support-based strategy is achieved during the warm-up. Since the starting point of this step is likely to be quite far from the sought-after sources, no reweighted $\ell_1$ is applied as it would tend to favor spurious solutions. For the same reason, the non-negativity constraint is likewise not applied on the sources during this stage (but it is for $\bA$).
The warm-up ends when the decrease of $c_{wu}$ and the increase of $K$ are completed, and when the estimations of the sources have converged, that is when $	{||{\mathbf{\hat{S}}^{(i)} - \mathbf{\hat{S}}^{(i-1)}}||_{\ell_2}}/{||{\mathbf{\hat{S}}^{(i)}}||_{\ell_2}} \leq \epsilon_{wu}$. In practice, $\epsilon_{wu}$ does not require to be very small (\textit{e.g.}~$\epsilon_{wu}  = 1\mathrm{e}{-2}$).

\paragraph*{Refinement stage}
During the second stage, the estimations are refined by using the more precise regularization strategy \#4 introduced in \ref{sec:regstrat}. The regularization parameters are calculated with the angular power spectra of the sources estimated at the previous iteration, which are assumed to be close enough to the ground truth ones:
\begin{equation}
\varepsilon_{n,l}^{(i)} = c_{ref} \frac{\mathbf{c_N}_l}{\mathbf{c_{S^\mathnormal{(i-1)}_\mathnormal{n}}}_l},
\end{equation}
where $c_{ref}$ is the regularization hyperparameter at refinement. $\mathbf{c_N}_l$ is deduced from the SNR of the observations. Concerning the choice of the thresholds, the active support is kept constant at the same $K_{max}$ as the previous stage. The $\ell_1$-reweighting of the sources we described above is applied during this stage, as well as the non-negativity constraint on the sources. The refinement ends when the estimations of the sources have converged, that is when $	{||{\mathbf{\hat{S}}^{(i)} - \mathbf{\hat{S}}^{(i-1)}}||_{\ell_2}}/{||{\mathbf{\hat{S}}^{(i)}}||_{\ell_2}} \leq \epsilon_{ref}$ (for instance $\epsilon_{ref}  = 1\mathrm{e}{-5}$).\\
Setting $K_{max}$ to a different value than 1 allows to keep the major features of the sources to improve the separation. However, it biases the estimation of the sources. Consequently, after the refinement stage, a final estimation of the sources with $\bA$ fixed and with $K = 1$ is performed. This final step does not appear in Algorithm \ref{alg:SDecGMCA} for the sake of clarity; it is however described in \ref{app:multires}.

\subsection{Convergence proprieties}
\begin{figure}
	\centering
	\includegraphics[width=0.4\linewidth]{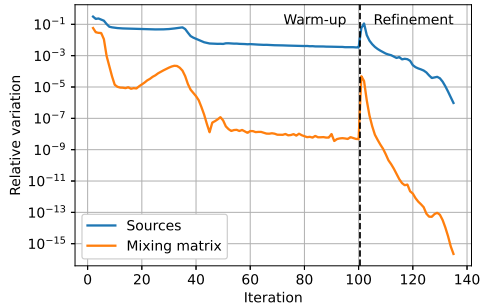}
	\caption{Empirical stabilization. The peak at iteration 101 corresponds to the switch from the warm-up stage to the refinement stage.}
	\label{fig:stab}
\end{figure}

DBSS requires solving a multi-convex optimization problem. No method can guarantee finding in general the global minimum. At best, it is possible to guarantee a convergence toward a local minimum (\textit{e.g.}~BCD \cite{Tseng_01_ConvergenceBlockCoordinate}, PALM \cite{Bolte_13_Proximalalternatinglinearized}).
SDecGMCA is built upon a projected ALS scheme. To the best of our knowledge, is has not been theoretically proved that projected ALS algorithms converge. However, we empirically show that the regularization parameters tend to stabilize along the iteration, as well as the estimates of $\bA$ and $\bS$ (see Fig.~\ref{fig:stab}).  

\section{Numerical experiments}
\label{sec:numerics}

In this section, we investigate the performances of the SDecGMCA algorithm on synthetic toy examples that allow performing Monte-Carlo trials, as well as realistic simulations with partial sky coverage. The code that is used is open source (see \ref{app:code}).\\
For all these experiments, we make use of the HEALPix pixelisation on the sphere. \\
Comparison criteria are first based on the NMSE (see Eq.~\eqref{eq:nmse}), which measures the reconstruction quality of the sources. As well, we make use of the mixing matrix criterion \cite{AMCA15} to assess the quality of the estimated mixing matrices:
\begin{equation}
	\text{C\textsubscript{A}} = -10 \log_{10}\left(\text{mean}\left(\left|\mathbf{A}^+\mathbf{A}^*-\mathbf{I}\right|\right)\right),
\end{equation}
where the mean operator is an average over all entries, $\mathbf{A}^*$ is the ground truth mixing matrix and $\mathbf{A}^+$ the Moore–Penrose inverse of the estimated mixing matrix. The mixing matrix criterion is more appropriate to compare DBSS and BSS methods with different source regularization, since it only depends on $\bA$.\\
Let us define the oracle as the solution of the non-blind problem (\textit{ie.}~knowing the ground truth mixing matrix) with regularization strategy \#4 and the optimal regularization hyperparameter. It provides an upper bound of the NMSE that SDecGMCA can reach.

\subsection{Toy examples} \label{sec:toyex}

The same toy examples as in Section~\ref{sec:nonblindpb} are employed, with a similar parameterization. The input parameters of SDecGMCA are summarized in Table~\ref{tab:inputParamSDecGMCA}. 

\begin{table}
	\centering
	\small
	\begin{tabular}{@{}lll@{}}
		\toprule
		Parameter & Notation & Value \\
		\midrule
		Minimum number of iterations at warm-up & $N_{wu} $& 100 \\
		Number of detail scales & $J$ & 3 \\
		Thresholding parameter & $k$ & 3 \\
		Maximum source support & $K_{max}$ & 0.5 \\
		\bottomrule
	\end{tabular}
	\caption{Input parameters of SDecGMCA for the toy example}
	\label{tab:inputParamSDecGMCA}
\end{table}

\subsubsection{Impact of the Tikhonov regularization}
\begin{figure}
		\subfloat[Ground truth source $\mathbf{S}_1^*$]{
			\centering
			\includegraphics[width=.45\linewidth]{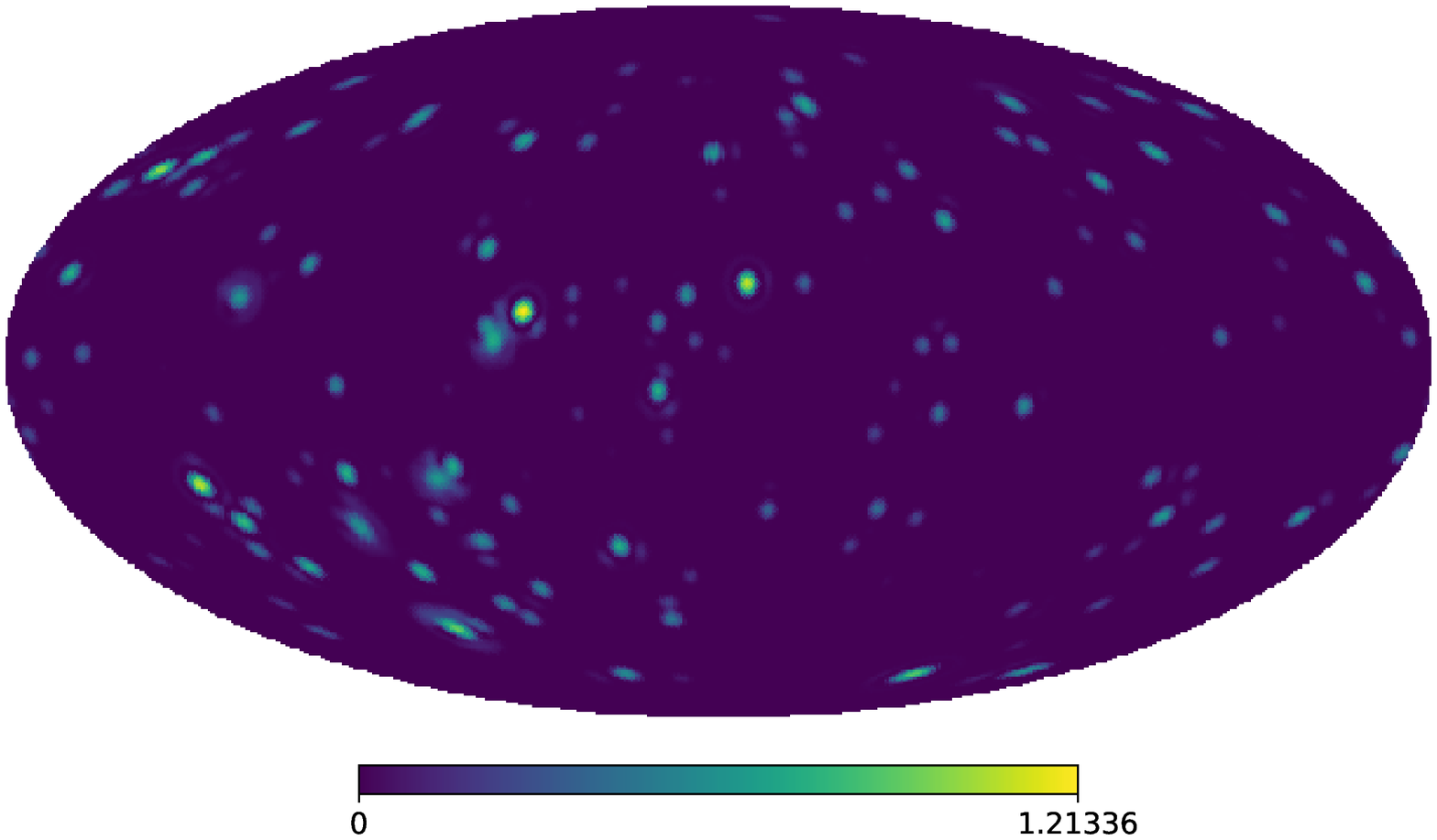}
			\label{fig:S1^*}
		} \qquad
	\subfloat[Convolution kernels $\{\mathbf{\hat{H}}_\nu, 1 \leq \nu \leq 8\}$]{
		\centering
		\includegraphics[width=.45\linewidth]{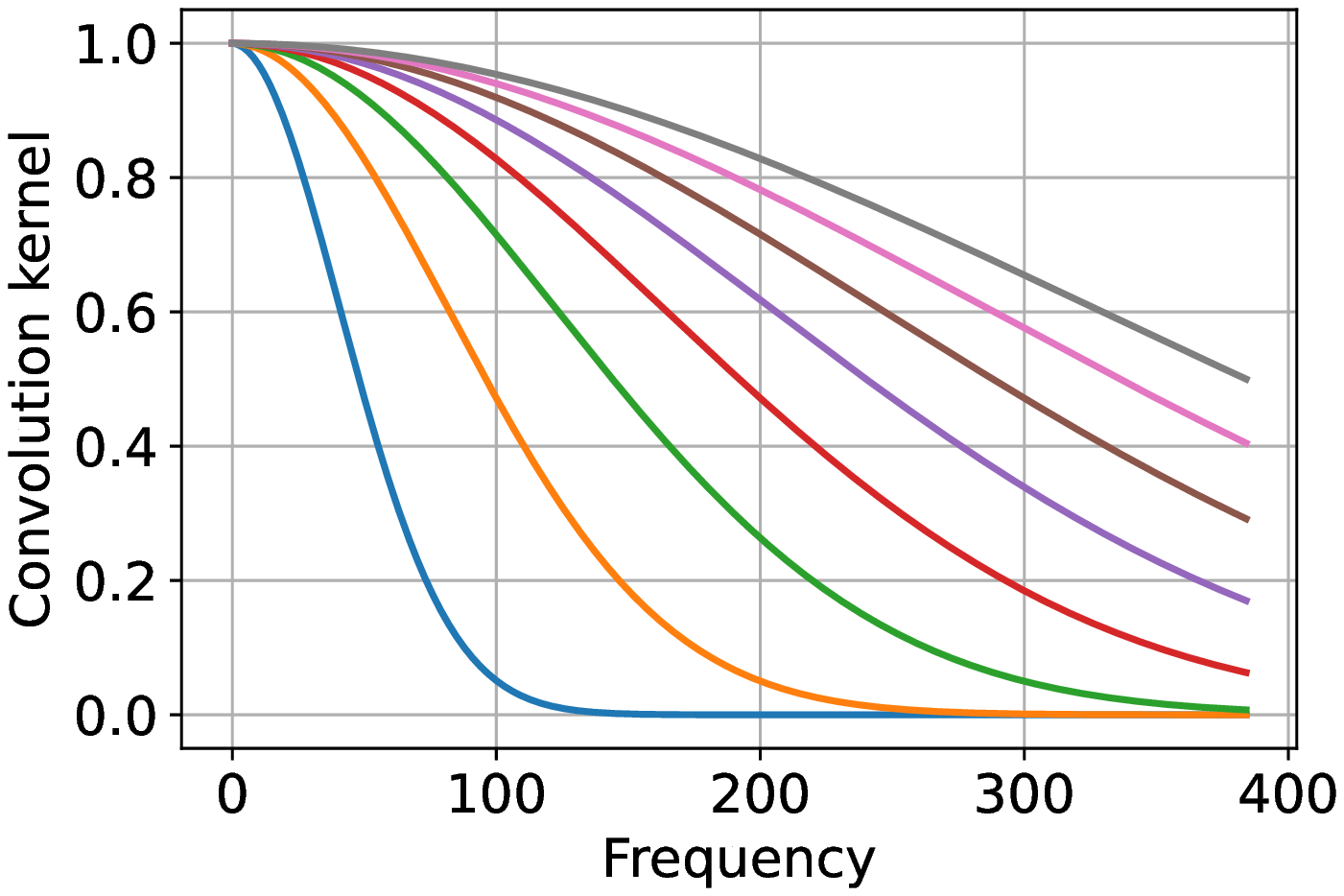}
		\label{Hl}
	} \\
	\subfloat[Worse-resolved observation $\mathbf{X}_1$]{
		\centering
		\includegraphics[width=.45\linewidth]{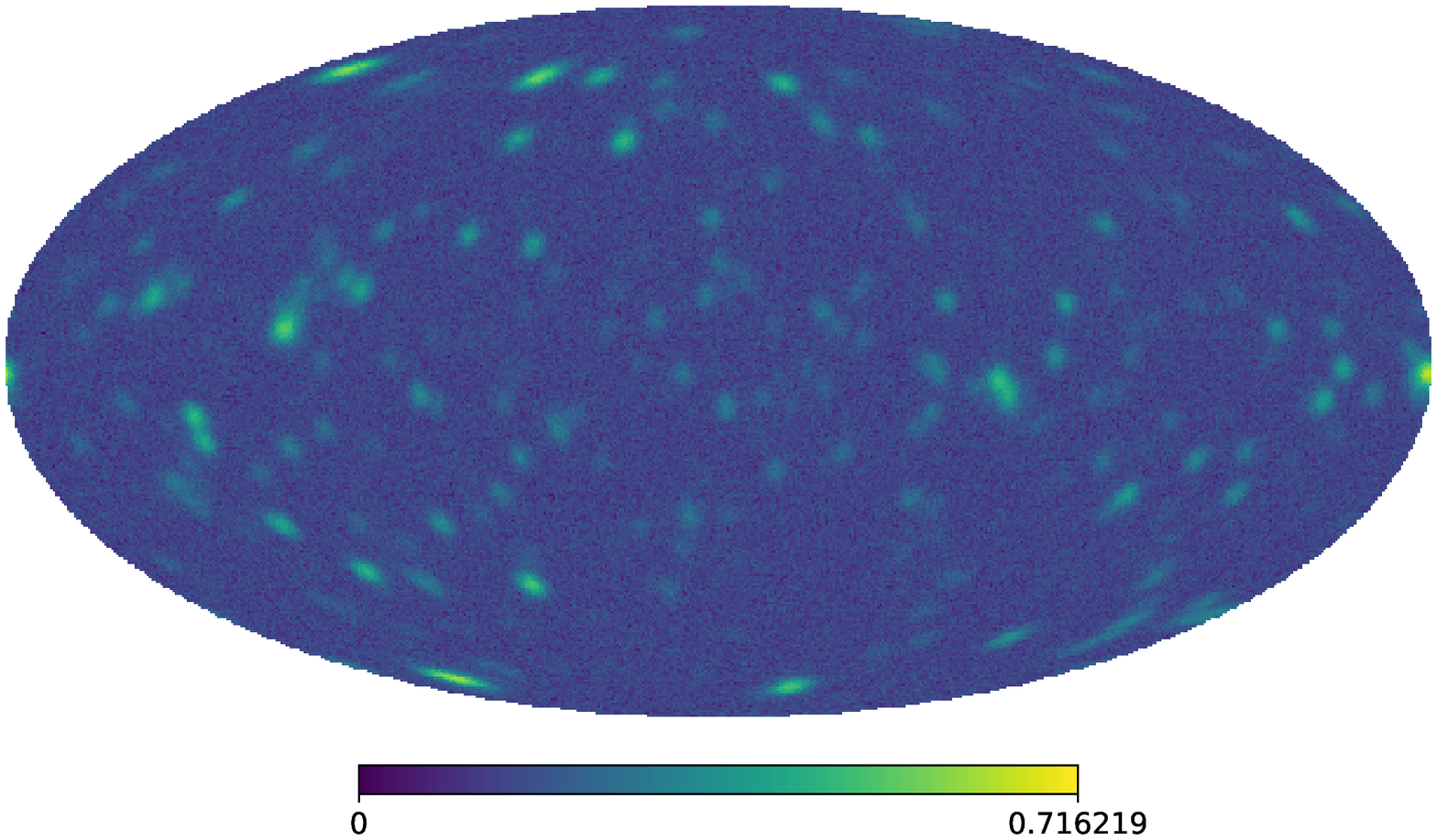}
	} \qquad
	\subfloat[Best-resolved observation $\mathbf{X}_8$]{
		\centering
		\includegraphics[width=.45\linewidth]{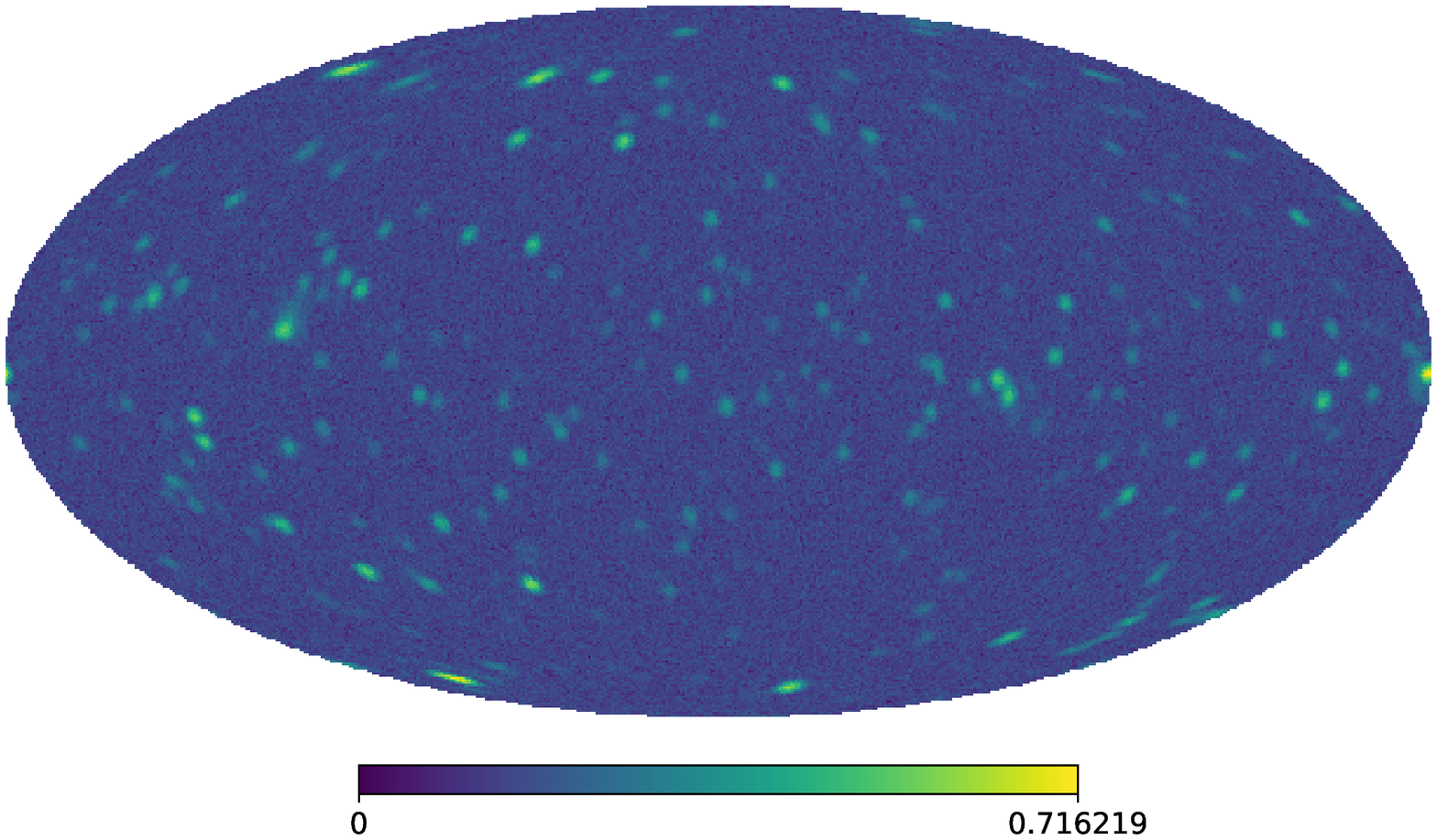}
	} \\
	\subfloat[Estimated source $\mathbf{S}_1$]{
		\centering
		\includegraphics[width=.45\linewidth]{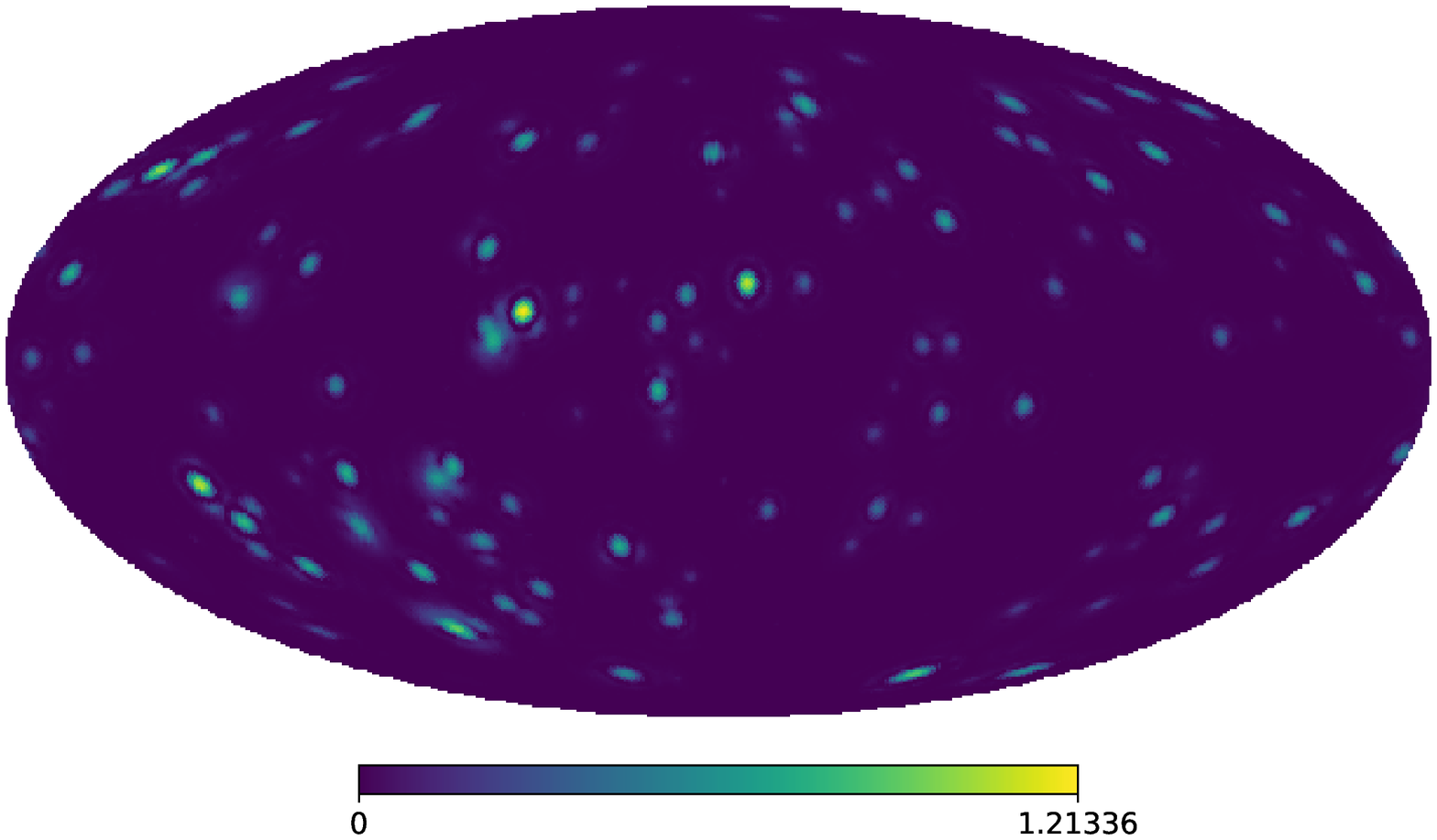}
	} \qquad
	\subfloat[Absolute error $|\mathbf{S}_1^*-\mathbf{S}_1|$]{
		\centering
		\includegraphics[width=.45\linewidth]{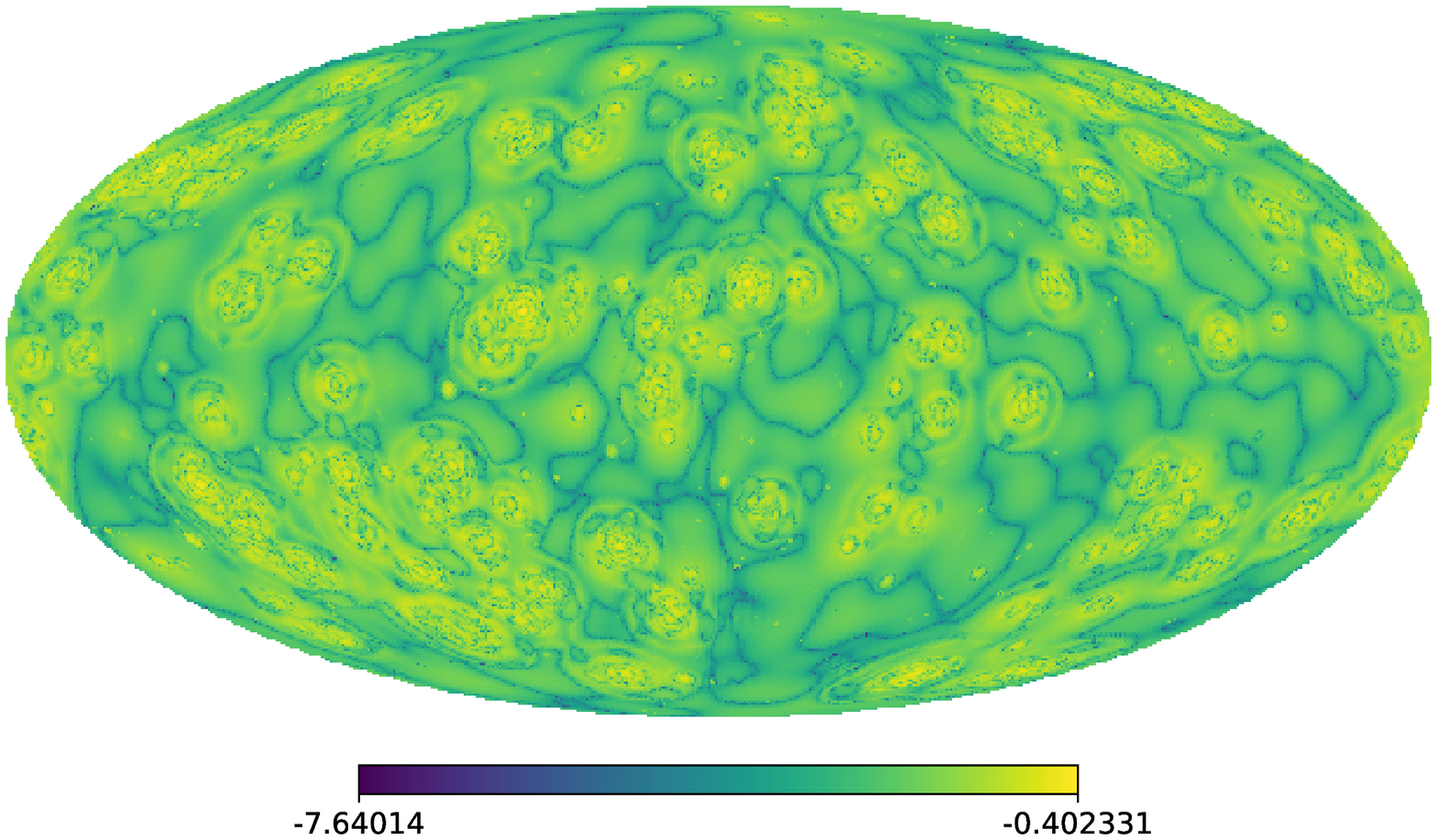}
		\label{fig:E0}
	}
	\vspace{0.2em}
	\caption{DBSS example with SDecGMCA (sources: arbitrary unit, logarithmic scale)}
	\label{fig:dbss_toyex}
\end{figure}

In standard BSS, the expected sources of error are the interferences ({\it i.e.}~the leakage between the sources due to the mis-estimation of the mixing matrix), the noise contamination ({\it i.e.} the backprojected noise of the observations which is not filtered out) and the artifacts (\textit{i.e.}~the remaining errors) \cite{Vincent_06_Performancemeasurementin}. Within the scope of the joint deconvolution and BSS, this linear decomposition is not suitable anymore. For instance, when the mixing matrix is known, the estimation of the sources on its own can generate interferences in addition to the deconvolution artifacts due to the biasing Tikhonov regularization. Generally speaking, we can consider that a separation is successful when the reconstructions errors are dominated by the deconvolution artifacts. \\
In this subsection, the data have the default observation parameters (see Section~\ref{sec:nonblindpb}). Figure \ref{fig:dbss_toyex} shows an example of a solution given by SDecGMCA. The reconstructions errors are dominated by the deconvolution artifacts (see Fig.~\ref{fig:E0}).

We first assess the impact of the regularization hyperparameters on the performances of SDecGMCA. To this end, we execute SDecGMCA with different warm-up and refinement regularization hyperparameters. Since the sources are updated similarly in the non-blind problem, we consider the mean optimal regularization hyperparameters found in Section \ref{sec:nonblindpb} as reference regularization hyperparameters.  

Each experiment is performed 100 times, with varying sources, mixing matrices and noise realizations. The mean NMSE and the mean \ca{} are reported in Table \ref{tab:perfregparam}. 
The choice of $c_{wu}$ for the warm-up stage has little impact on the performance metrics; indeed, the loss of NMSE and \ca{} due to a poor choice of $c_{wu}$ is respectively at most 0.42~dB and 0.77~dB in the tested range. On the contrary, the selection of $c_{ref}$ for the refinement stage may be more critical. It is however very interesting to highlight that, in a range of one order of magnitude around the optimal hyperparameter, the loss of NMSE and \ca{} remains quite limited ({\it i.e.}~respectively about $-2.21$~dB and $-1.05$~dB at most). It is noted that the oracle mean NMSE is 25.74~dB; thus, the extra estimation of the mixing matrix is only the origin of a 0.95~dB loss in the NMSE.\\ 

\begin{table}
	\centering
	\small
	\begin{tabular}{@{}lllllll@{}} 
		\toprule
		& & \multicolumn{5}{c}{$c_{ref}$ ($\times\, {c_{ref}}_{opt}$)} \\ 
		\cmidrule(rl){3-7}
		& & $10^{-1} $ & $10^{-0.5} $ & $\mathbf{10^{0}} $ & $10^{0.5} $& $10^{1} $  \\ 
		\midrule
		\multirow{10}{*}{\shortstack[c]{$c_{wu}$\\($\times\, {c_{wu}}_{opt}$)}} & $10^{0} \rightarrow 10^{-1}$  & 22.86 & 24.43 & 24.61 & 22.6 & 18.11\\
		& $10^{0.5} \rightarrow 10^{-0.5}$ & 22.99 & 24.58 & \textbf{24.79} & 22.83 &18.35 \\
		& $\mathbf{10^{1} \rightarrow 10^{0}}$    & 23.00 & 24.58 & 24.79 & 22.82 &18.34 \\
		& $10^{1.5} \rightarrow 10^{0.5}$ & 23.06 & 24.59 & 24.65 & 22.44 & 18.10  \\
		& $10^{2} \rightarrow 10^{1}$     & 23.06 & 24.23 & 24.4 & 22.43 & \textit{17.93} \\
		\cmidrule(rl){2-7}
		& $10^{0} \rightarrow 10^{-1}$  & $\mathbf{25.62}$ & 25.20 & 24.86 & 24.41 & 22.66\\
		& $10^{0.5} \rightarrow 10^{-0.5}$ & 25.60 & 25.19 & 24.86 & 24.46 & 22.81 \\
		& $\mathbf{10^{1} \rightarrow 10^{0}}$    & 25.57 & 25.15 & 24.81 & 24.39 & 22.80 \\
		& $10^{1.5} \rightarrow 10^{0.5}$ & 25.28 & 24.80  & 24.37 & 23.75 & 22.07  \\
		& $10^{2} \rightarrow 10^{1}$     & 25.28 & 24.56 & 24.25 & 23.84 & \textit{22.04} \\
		\bottomrule
	\end{tabular}
	\caption{Mean NMSE (top) and \ca{} (bottom) in dB, over 100 realizations, performed by SDecGMCA as a function of $c_{wu}$ and $c_{ref}$. These are given as multiples of ${c_{wu}}_{opt}$ and ${c_{ref}}_{opt}$, which are the mean optimal hyperparameters for the non-blind problem. It is noted that the oracle mean NMSE is 25.74~dB.} 
	\label{tab:perfregparam}
\end{table}

\subsubsection{Comparison with other blind source separation methods}
In this paragraph, compararisons with other blind source separation methods are carried out. Since few DBSS methods have been investigated so far, a natural comparison would be with DecGMCA. However, since it has not directly been designed for data sampled on the sphere, a direct comparison cannot be performed. We rather propose to substitute DecGCMA's regularization strategy \#2 within SDecGMCA to quantify the impact of the regularization strategy. In contrast to \cite{Jiang_2017}, where the regularization parameter is chosen in an {\it ad hoc} manner, we employe the optimal regularization hyperparameters; this method is therefore referred to as oDecGMCA (for optimized DecGMCA). Moreover, in order to highlight the benefit of combining the deconvolution to the BSS, we propose to compare SDecGMCA to three non-deconvolving BSS methods:
\begin{itemize}
	\item GMCA (including the decreasing thresholding strategy and the $\ell_1$-reweighting) 
	\item Hierarchical Alternate Least-Squares (HALS) \cite{Cichocki_07_HierarchicalALSAlgorithms, Gillis_12}: non-negative matrix factorization algorithm solving $\argmin\limits_{\bA  \geq 0,\,\bS \geq 0} \norm{\bX - \bA\bS}^2_{\ell_2}$. It is based on block coordinate descent (one column of $\bA$ and one row of $\bS$ is updated at each iteration), with multiplicative updates.
	\item Beta Sparse Non-negative Matrix Factorization ($\beta$-SNMF) \cite{Cherni20BNMF}: non-negative matrix factorization algorithm promoting the sparsity of $\bS$ in the direct domain, which solves $\argmin\limits_{\bA \geq 0, \,\bS \geq 0} \norm{\bX - \bA\bS}^2_{\ell_2} + \lambda \norm{\bS}_{\ell_1}$. The minimization is also based on block coordinate descent with multiplicative updates.
\end{itemize}
As these methods can only process observations with the same resolution, the observations are convolved beforehand to the resolution of the worse-resolved observation so as to avoid noise amplification. The NMSE is adapted to take account of the resolution loss:
\begin{equation}
\text{NMSE}_\text{w} = -10 \log_{10}\left(\frac{\norm{\mathbf{{H}}_{\nu_w}*\mathbf{S}^*-\mathbf{S}}_{\ell_2}^2}{\norm{\mathbf{{H}}_{\nu_w}*\mathbf{S}^*}_{\ell_2}^2}\right),
\end{equation}
where $\nu_w$ is the worse-resolved observation channel. The NMSE\textsubscript{w} of the DBSS methods can be calculated by deteriorating the estimated sources. \\
The performance metrics achieved by the different DBSS and BSS algorithms are reported in Table \ref{tab:comp}. Compared to oDecGMCA, SDecGMCA performs a significant gain in NMSE and a moderate increase in \ca{}. This result confirms that the choice of the regularization strategy is crucial for the estimation of the sources. Moreover, the BSS algorithms achieve poor results; indeed, high-frequency information, which is essential for the separation process, is lost when the data are deteriorated.

\begin{table}
	\centering
	\small
		\begin{tabular}{@{}llll@{}} 
			\toprule
			& $\text{C\textsubscript{A}}$ & NMSE\textsubscript{w}& NMSE \\
			\midrule
			SDecGMCA                & \textbf{24.81} & \textbf{27.08} & \textbf{24.79}  \\
			oDecGMCA & 23.01 & 20.94 & 15.03 \\
			GMCA                       & 21.98  & 19.35 & N/A \\
			HALS                        & 8.17 & 5.83  & N/A \\
			$\beta$-SNMF        & 8.43  & 7.01   &  N/A \\
			\bottomrule
		\end{tabular}
	\caption{Mean performance metrics in dB, over 100 realizations, achieved by different algorithms}
	\label{tab:comp}
\end{table}

\subsubsection{Varying observation parameters}
\begin{figure}
	\subfloat{
		\includegraphics[width=.34\linewidth]{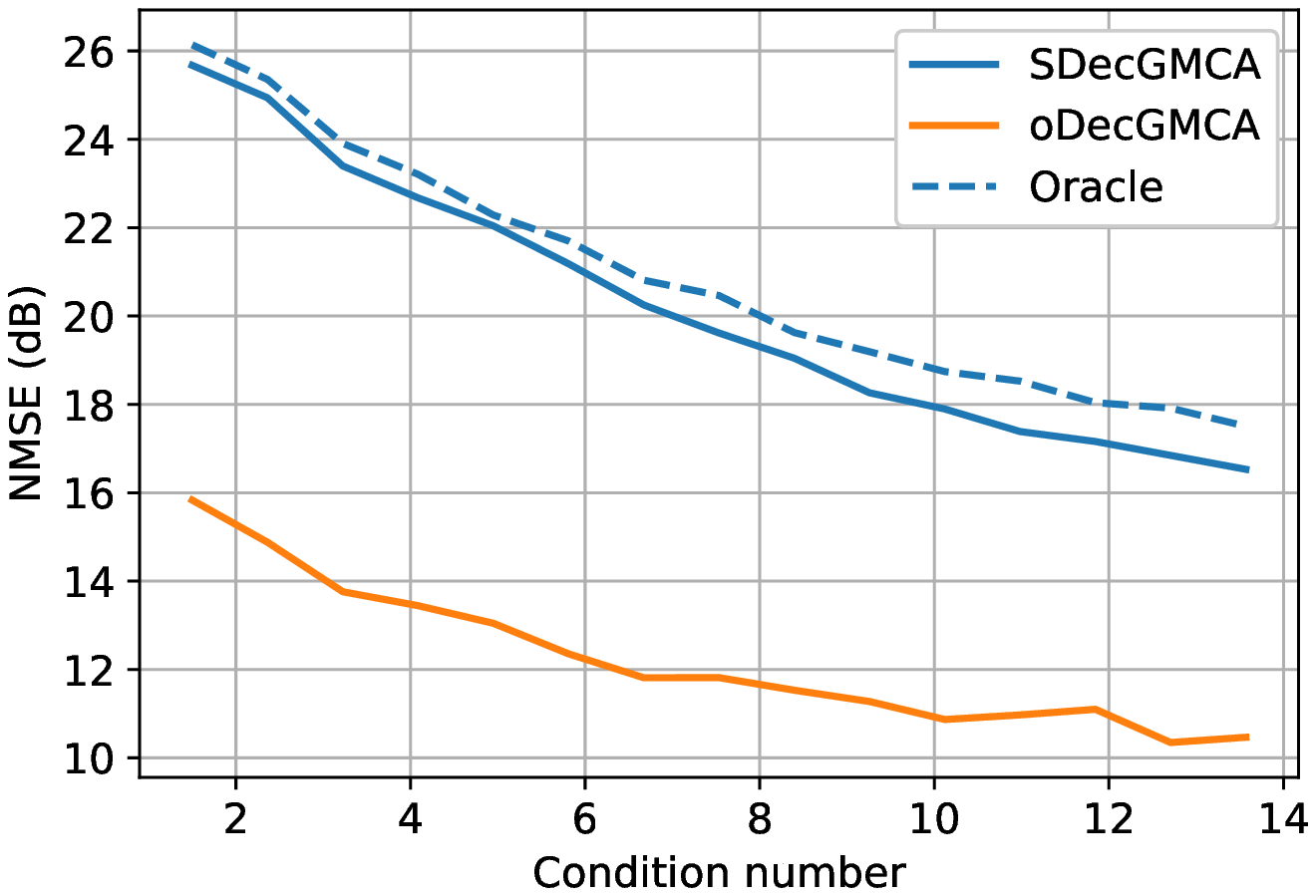}
	} 
	\hspace{-.7cm}
	\subfloat{
		\includegraphics[width=.34\linewidth]{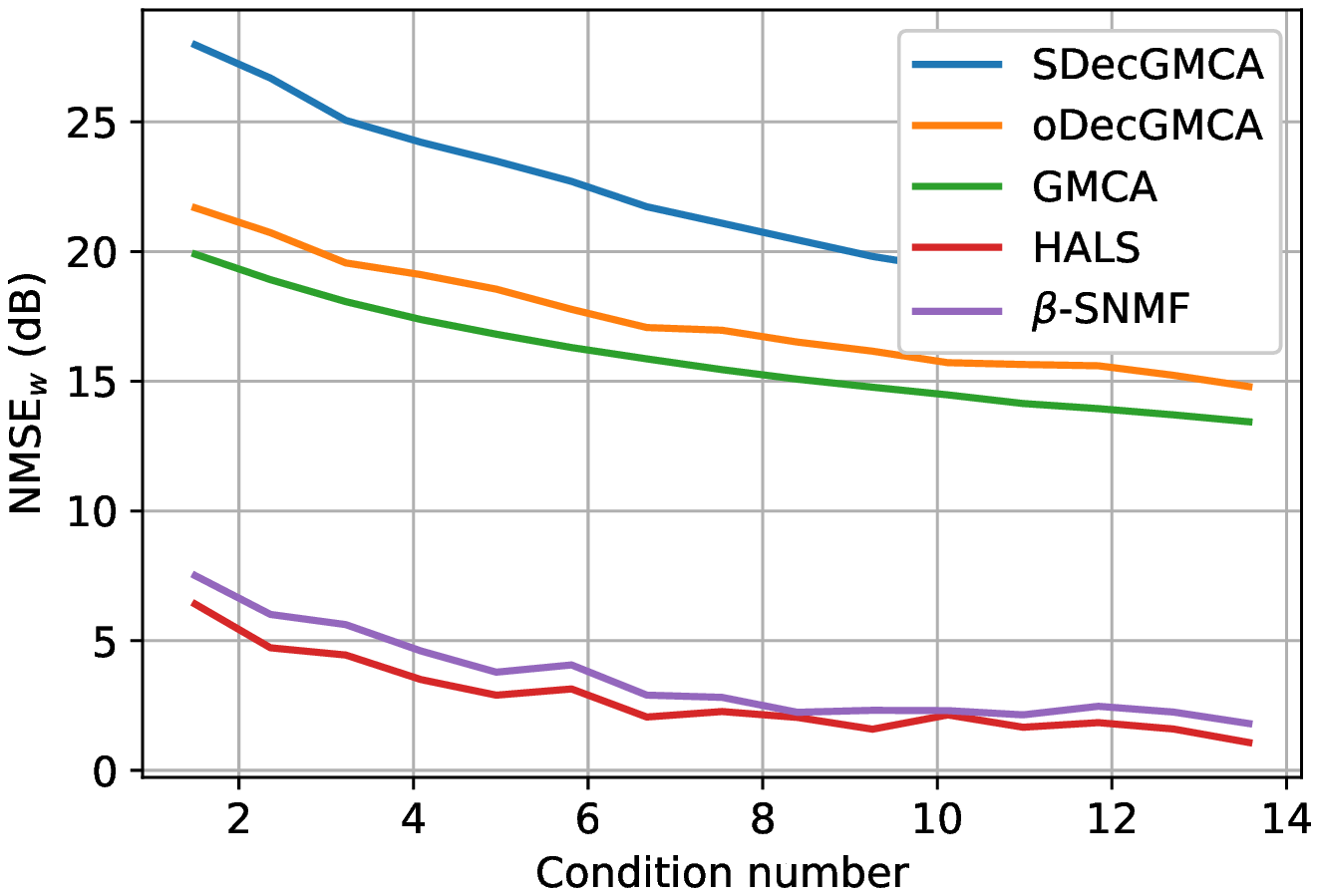}
	} 
	\hspace{-.7cm}
	\subfloat{
		\includegraphics[width=.34\linewidth]{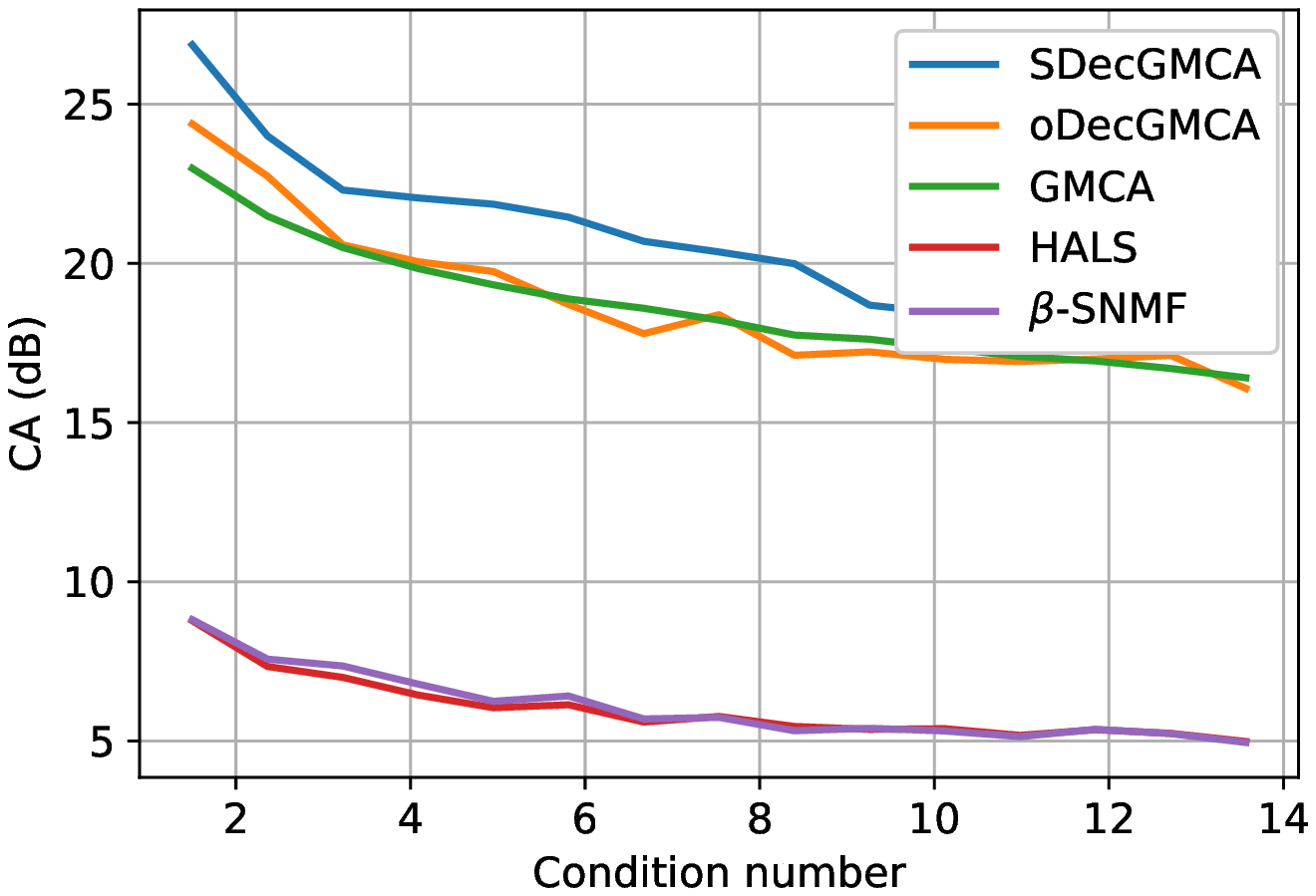}
	}\\[-.3cm]
	\subfloat{
		\includegraphics[width=.34\linewidth]{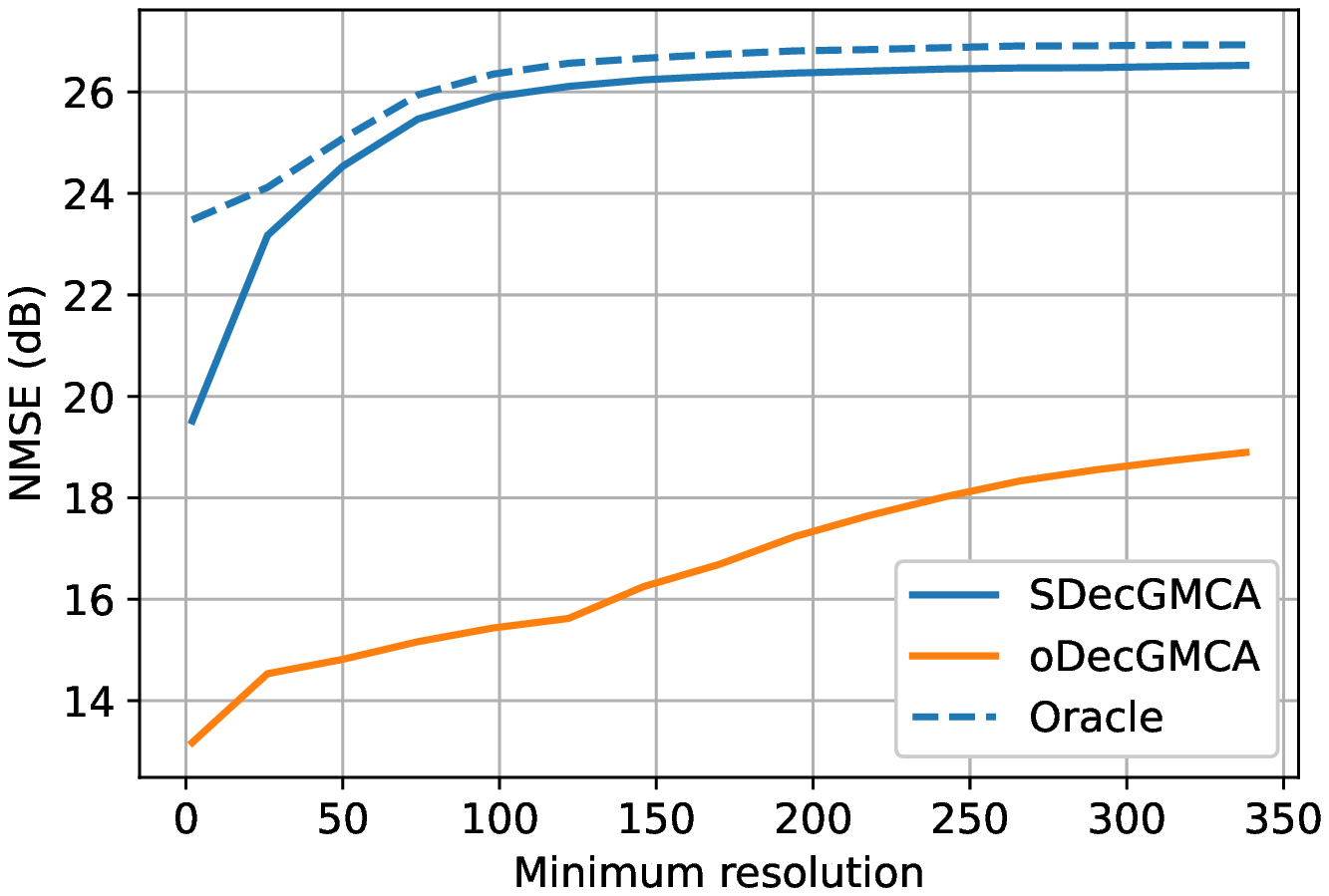}
	} 
	\hspace{-.7cm}
	\subfloat{
		\includegraphics[width=.34\linewidth]{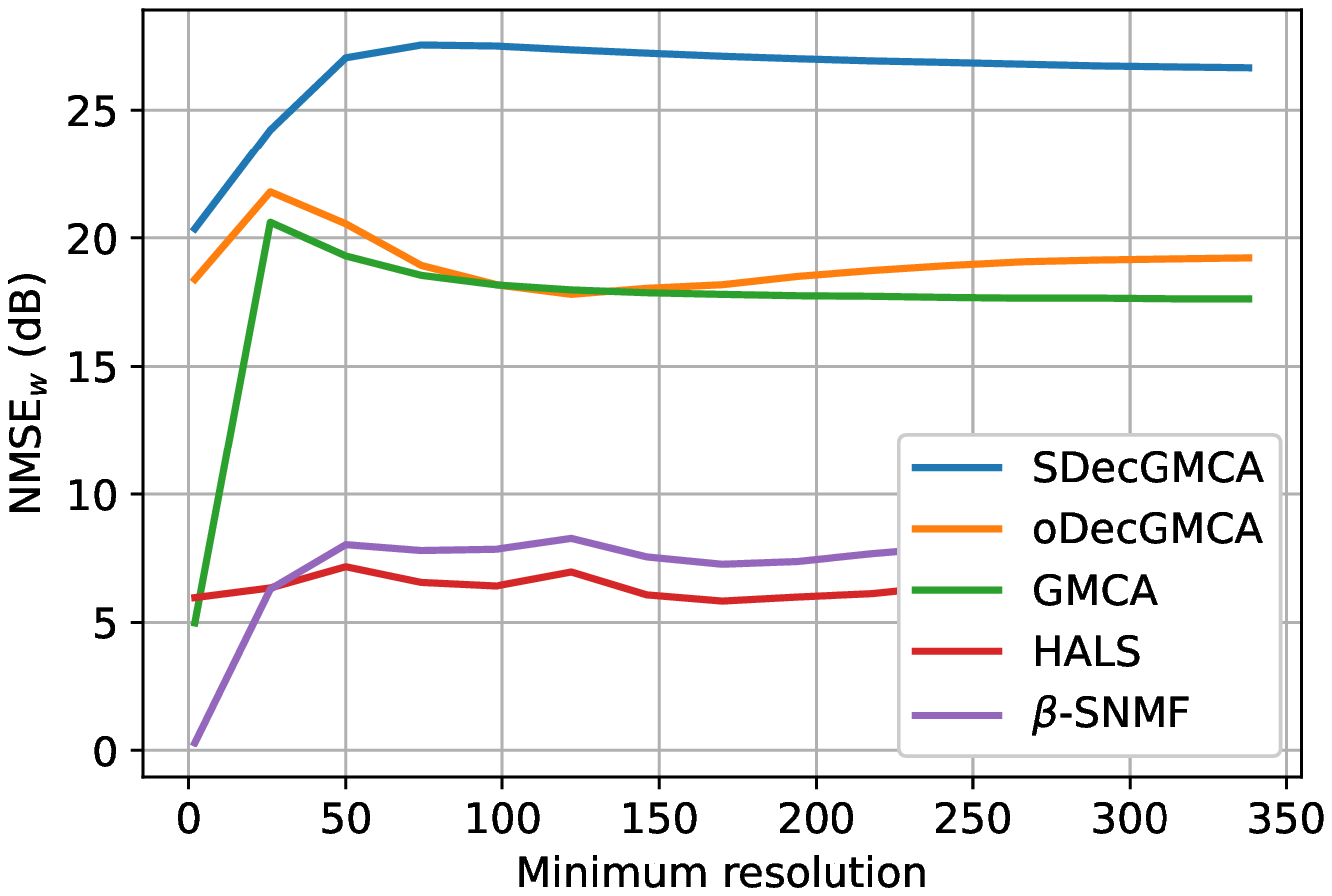}
	} 
	\hspace{-.7cm}
	\subfloat{
		\includegraphics[width=.34\linewidth]{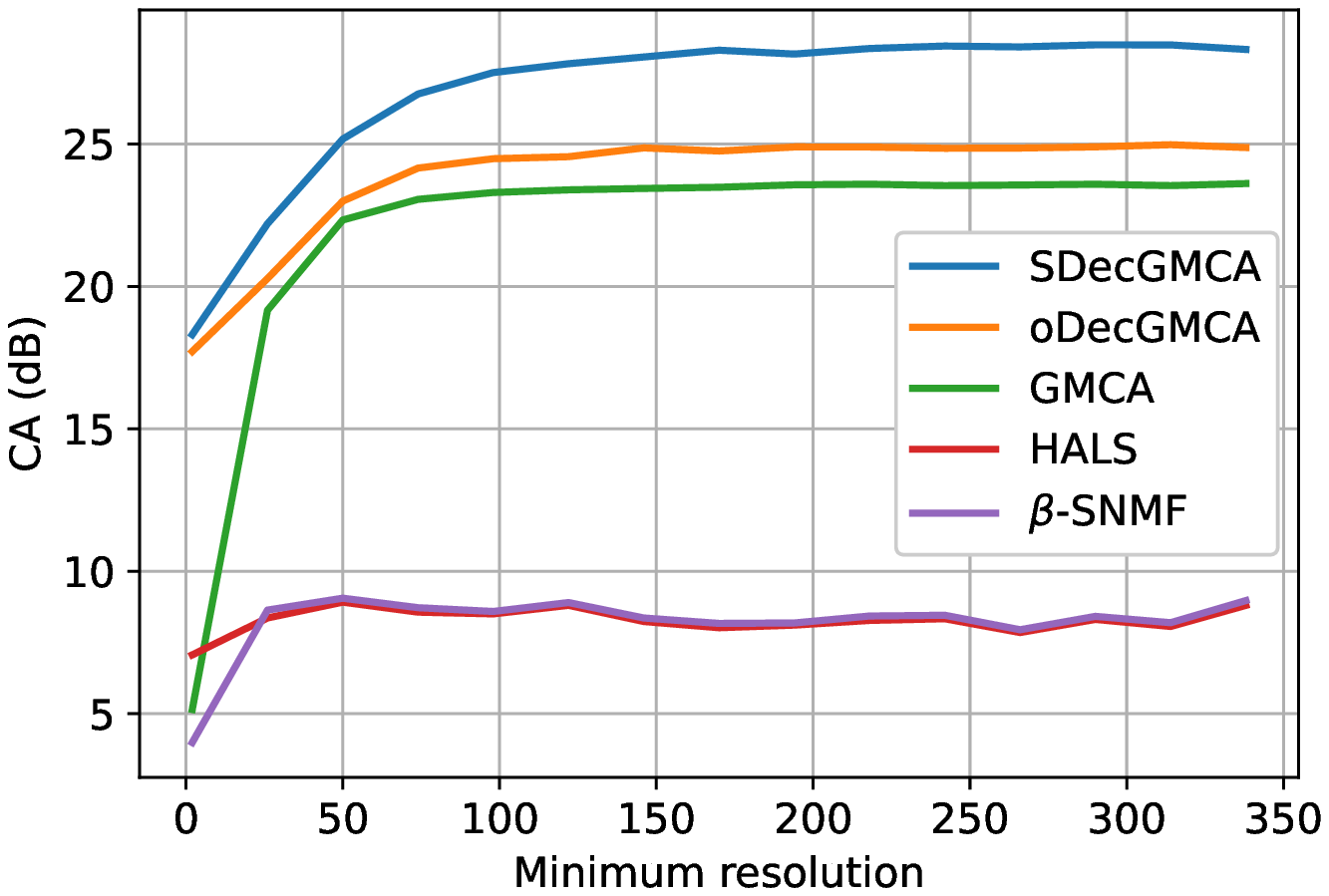}
	}\\[-.3cm]
	\subfloat{
		\includegraphics[width=.34\linewidth]{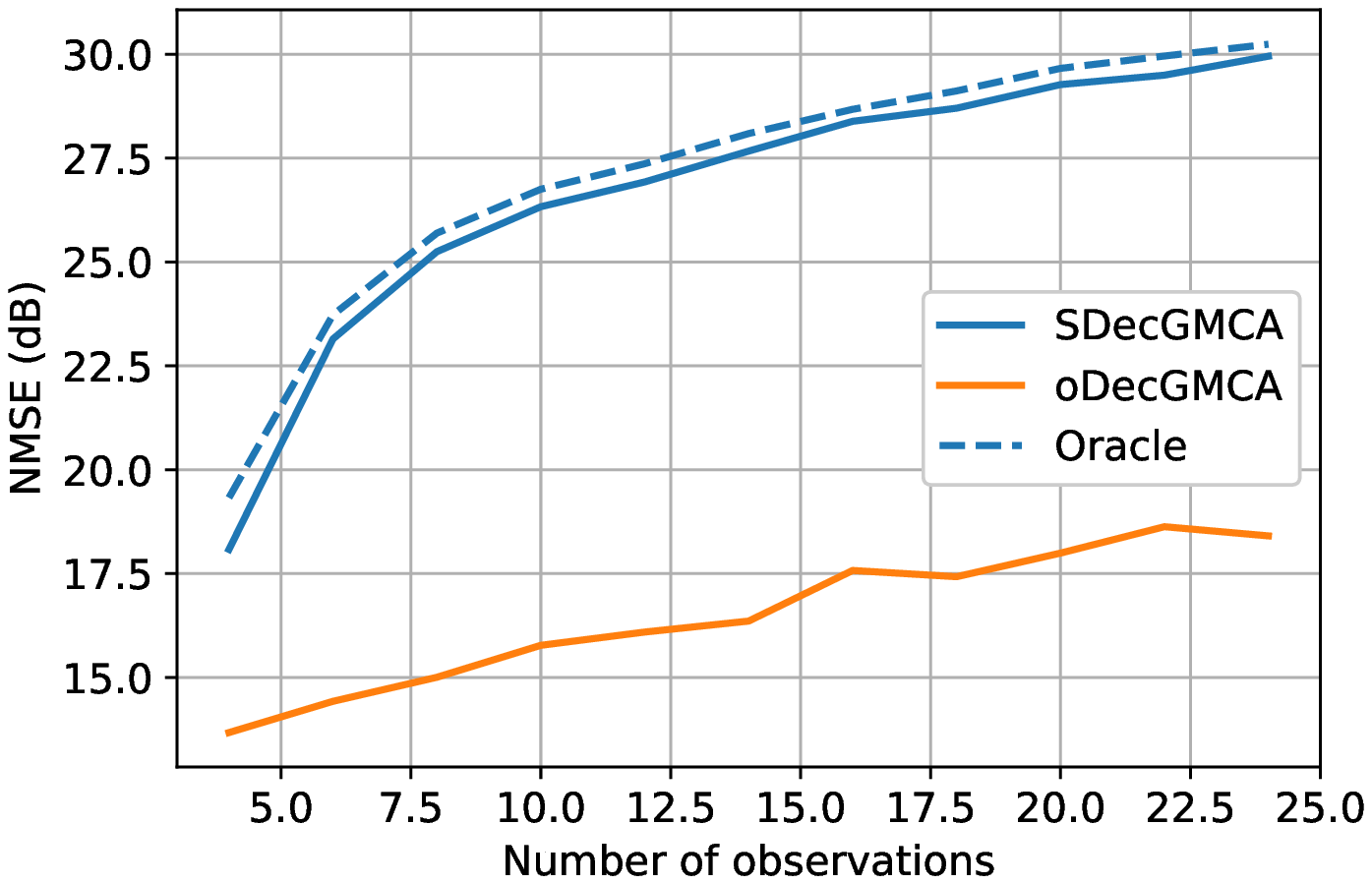}
	} 
	\hspace{-.7cm}
	\subfloat{
		\includegraphics[width=.34\linewidth]{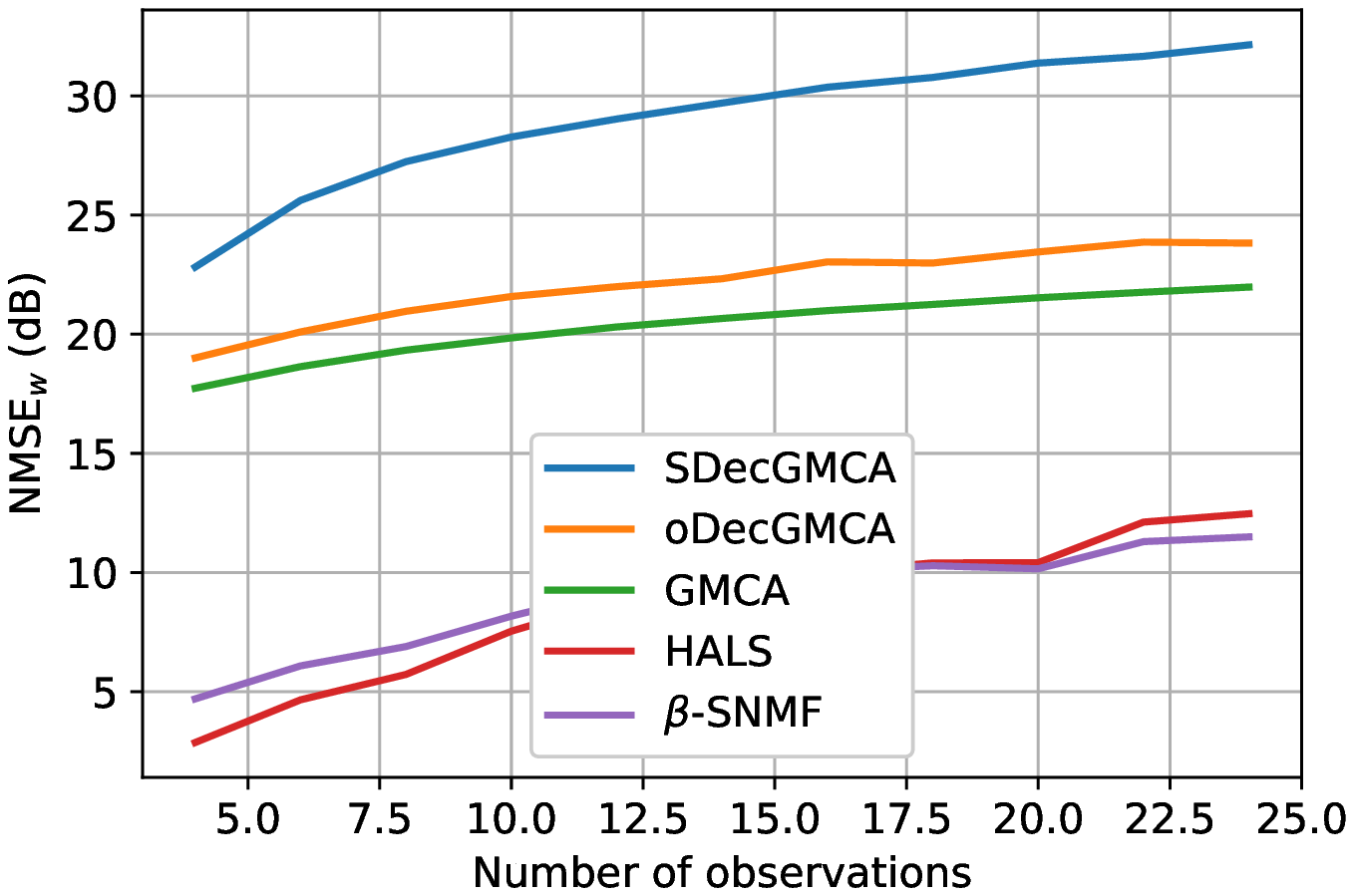}
	} 
	\hspace{-.7cm}
	\subfloat{
		\includegraphics[width=.34\linewidth]{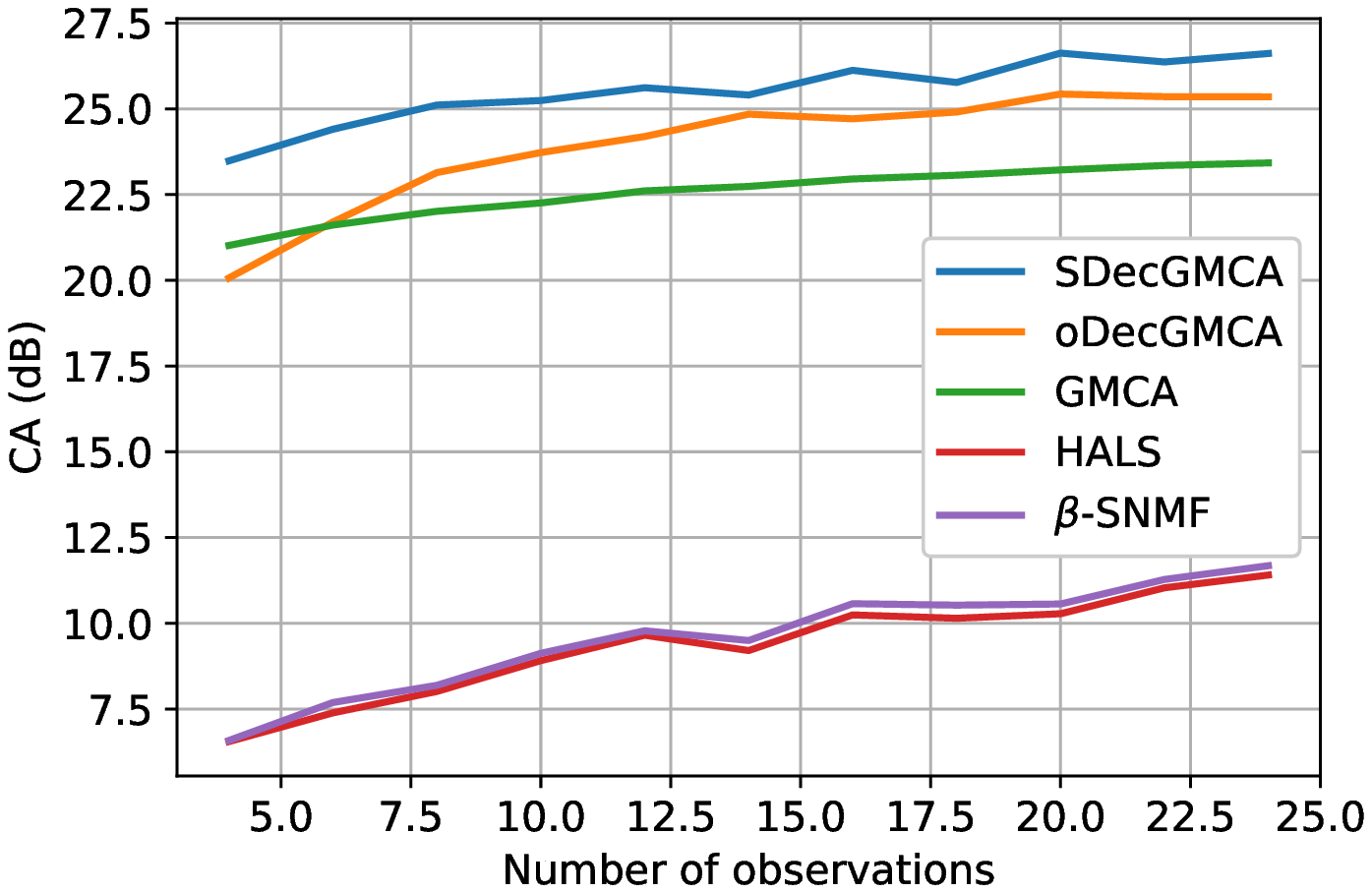}
	}\\[-.3cm]
		\subfloat{
	\includegraphics[width=.34\linewidth]{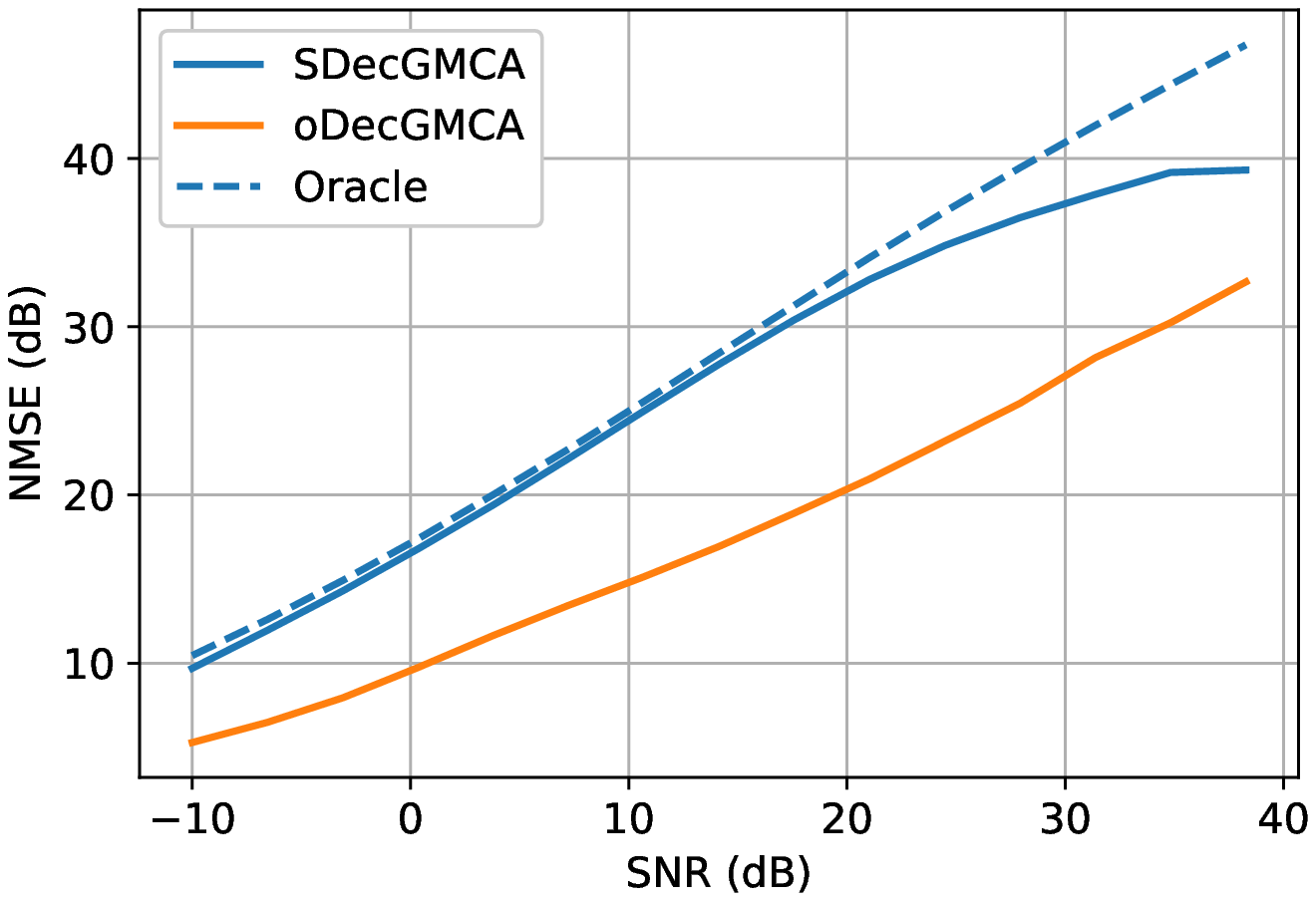}
	} 
	\hspace{-.7cm}
	\subfloat{
		\includegraphics[width=.34\linewidth]{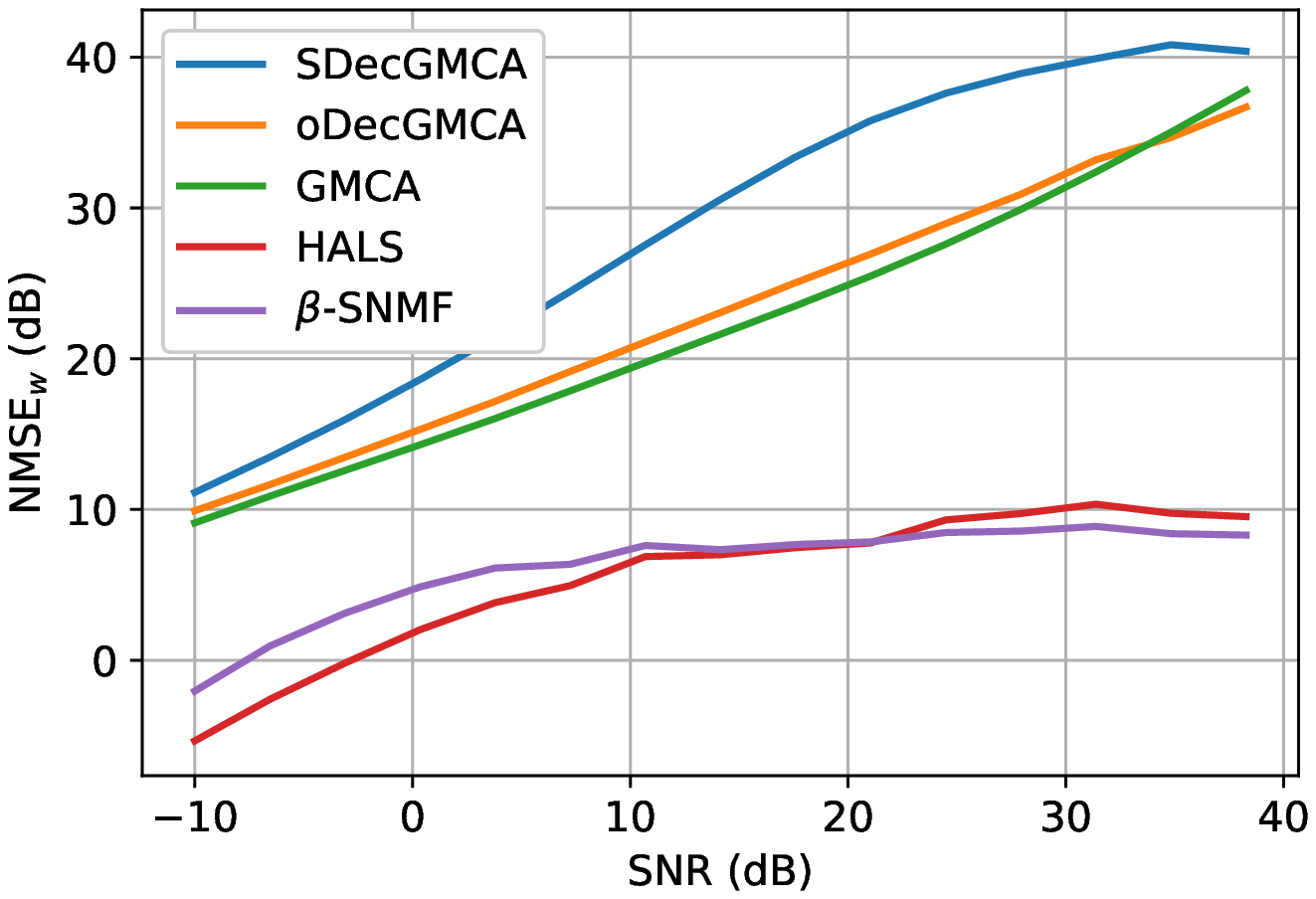}
	} 
	\hspace{-.7cm}
	\subfloat{
		\includegraphics[width=.34\linewidth]{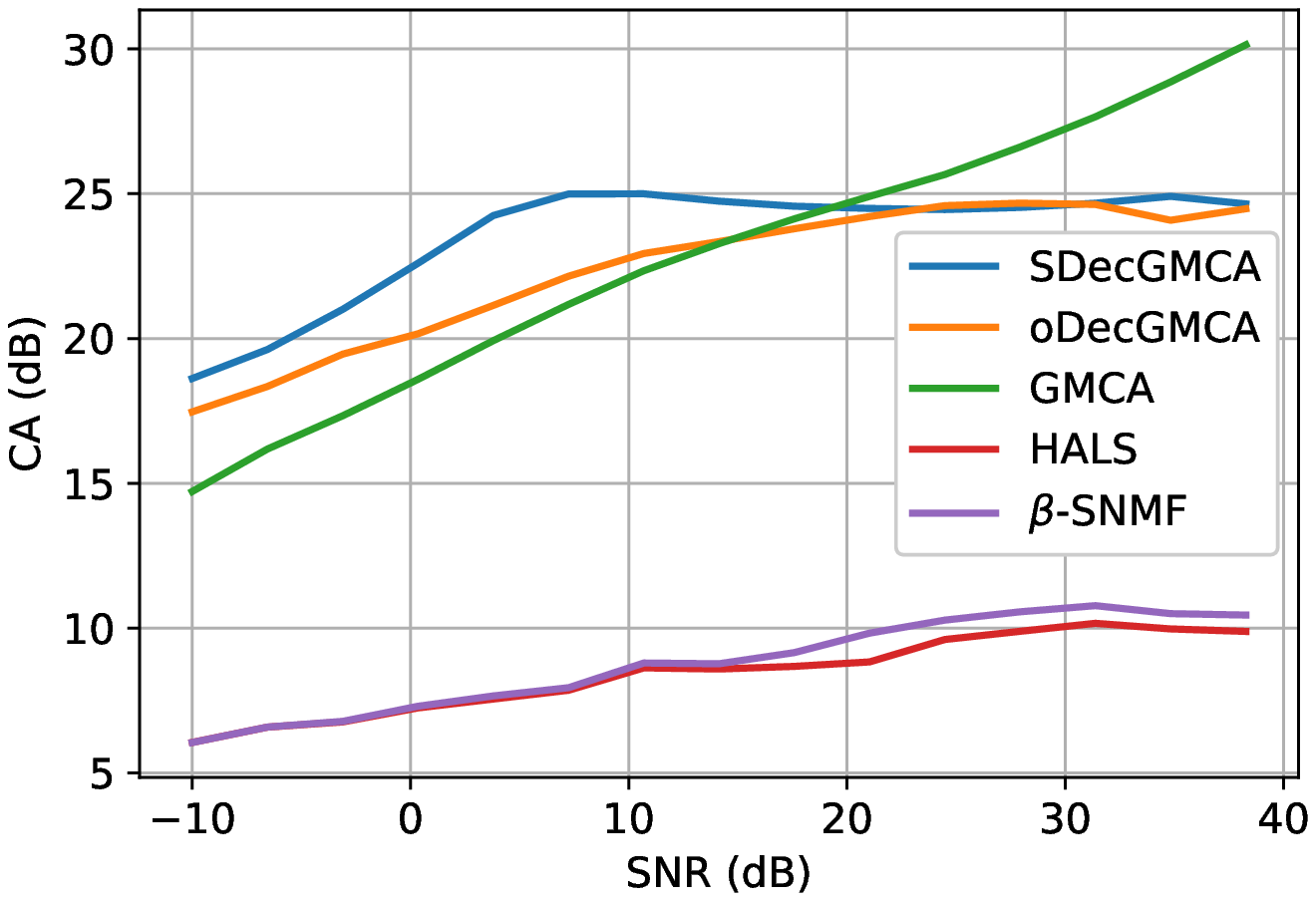}
	}
	\caption{Mean performance metrics over more than 30 realizations as a function of observation parameters. The default values of the parameters are summarized in Section~\ref{sec:nonblindpb}.}
	\label{fig:blind_metrics}
\end{figure}

Let us evaluate the sensitivity of the different source separation algorithms to the observation parameters, that is the mixing matrix condition number, the minimum resolution of the convolution kernels, the number of observations and the SNR. For SDecGMCA and oDecGMCA, the mean optimal regularization hyperparameter found above with the non-blind problem are used. At each point, the algorithms are executed 30 times with varying sources, mixing matrices and noise realization. The mean performance metrics are reported in Figure \ref{fig:blind_metrics}.\\ 
The NMSE achieved by SDecGMCA is close to the oracle; the loss is typically of 1~dB. In every scenario, SDecGMCA clearly outperforms oDecGMCA in terms of NMSE and \ca{}. Overall, the tendencies are consistent; the performance metrics increase with increasing minimum resolution, number of observations and SNR, while decrease with increasing mixing matrix condition number. The first notable exception is the NMSE\textsubscript{w} against the minimum resolution. It is due to the fact that the reference $\mathbf{{H}}_{\nu_w} * \textbf{S}$ in the definition of the NMSE\textsubscript{w} varies from one point to the other. The second exception concerns the SNR; the performance metrics stabilize or decrease when there is little noise. This is an effect of the implicit regularization provided by the noise. According to the proposed threshold tuning strategy, when the noise level is low, the thresholds are low and the sparsity constraint is loosened. On the contrary, a higher noise level yields higher thresholds that tend to select high amplitude coefficients, which better discriminate between the sources. 

\begin{figure}
	\subfloat{
		\includegraphics[width=.43\linewidth]{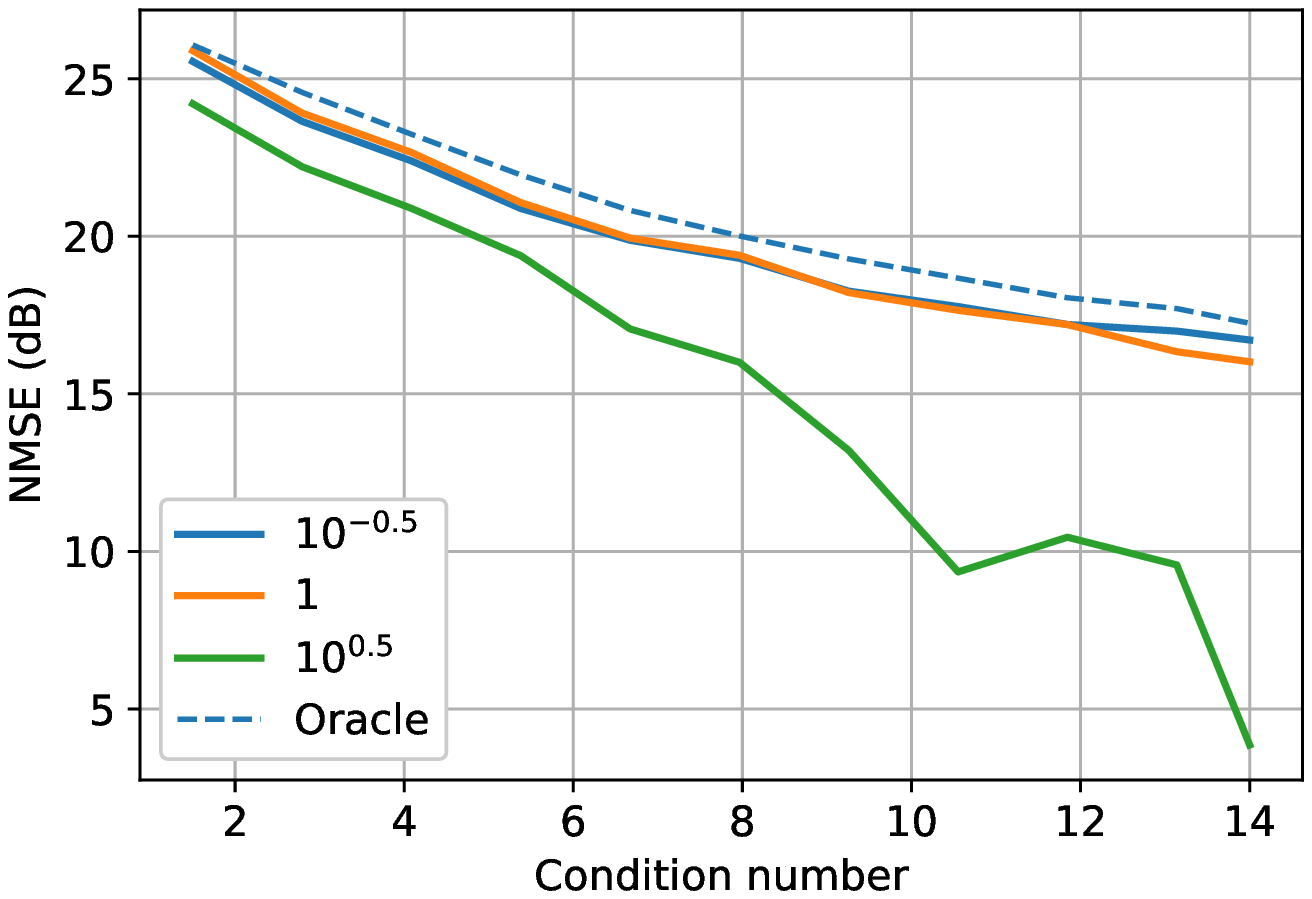}
	} 
	\hfill
	\subfloat{
		\includegraphics[width=.43\linewidth]{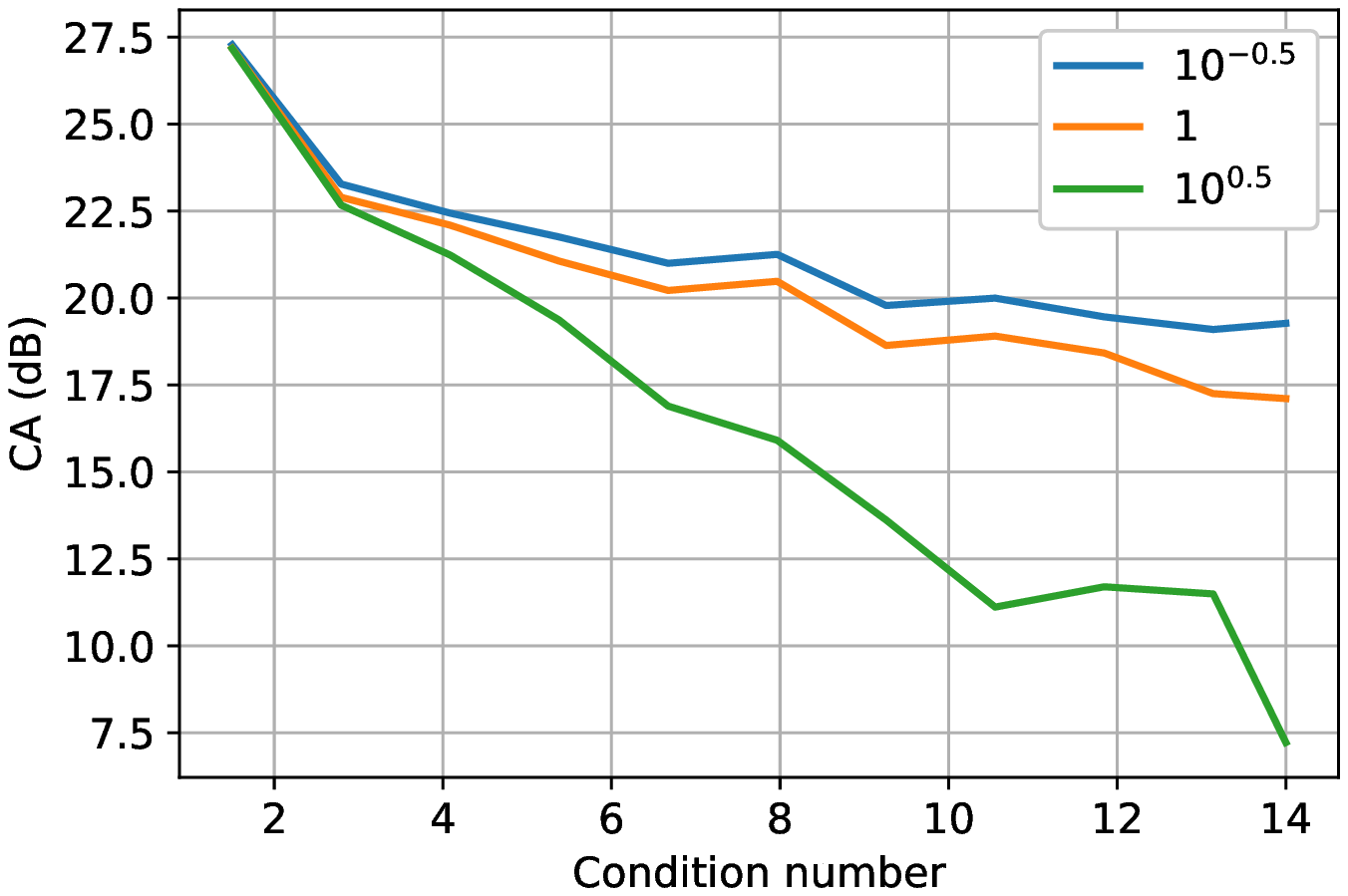}
	}\\[-.3cm]
	\subfloat{
		\includegraphics[width=.43\linewidth]{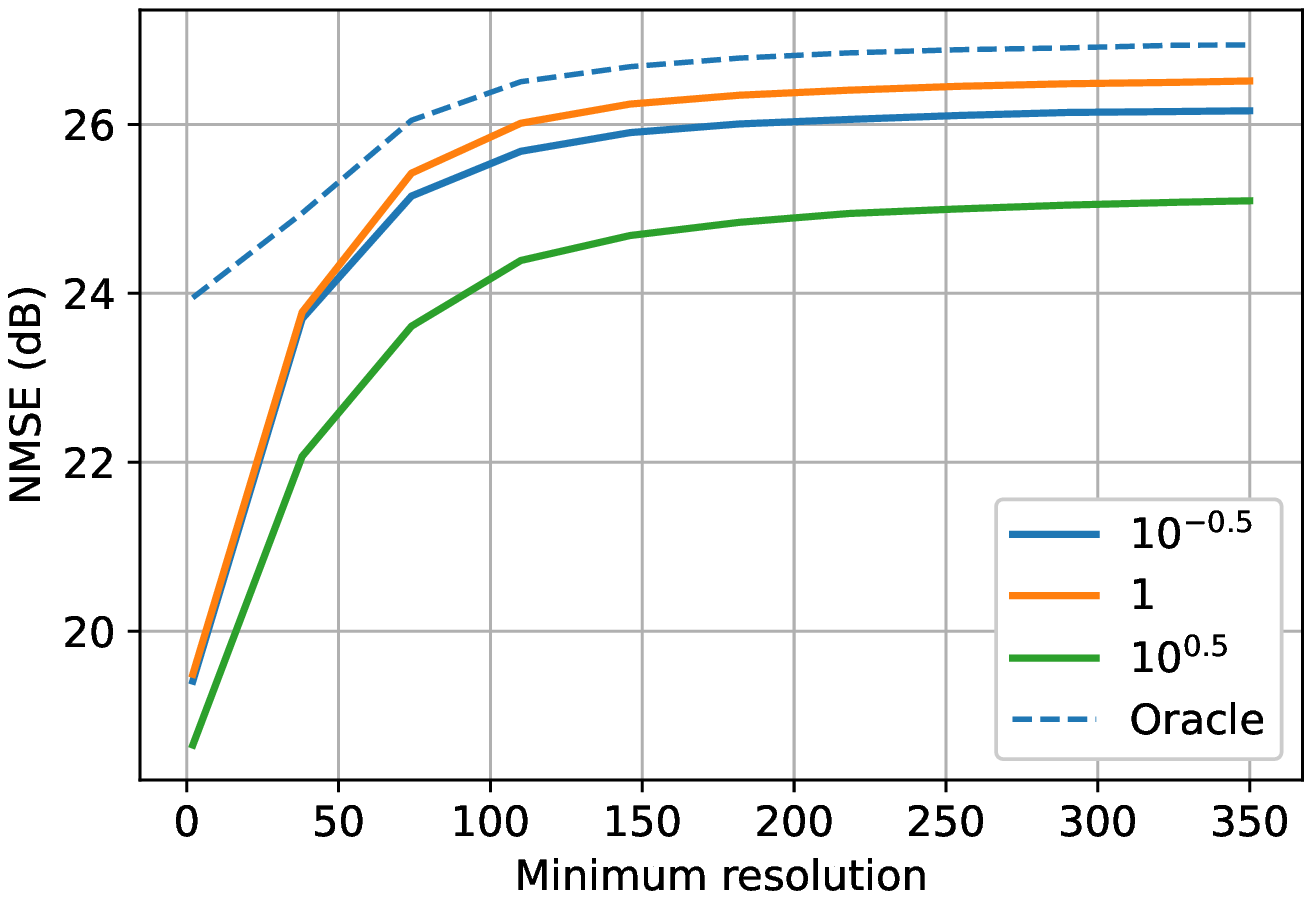}
	} 
	\hfill
	\subfloat{
		\includegraphics[width=.43\linewidth]{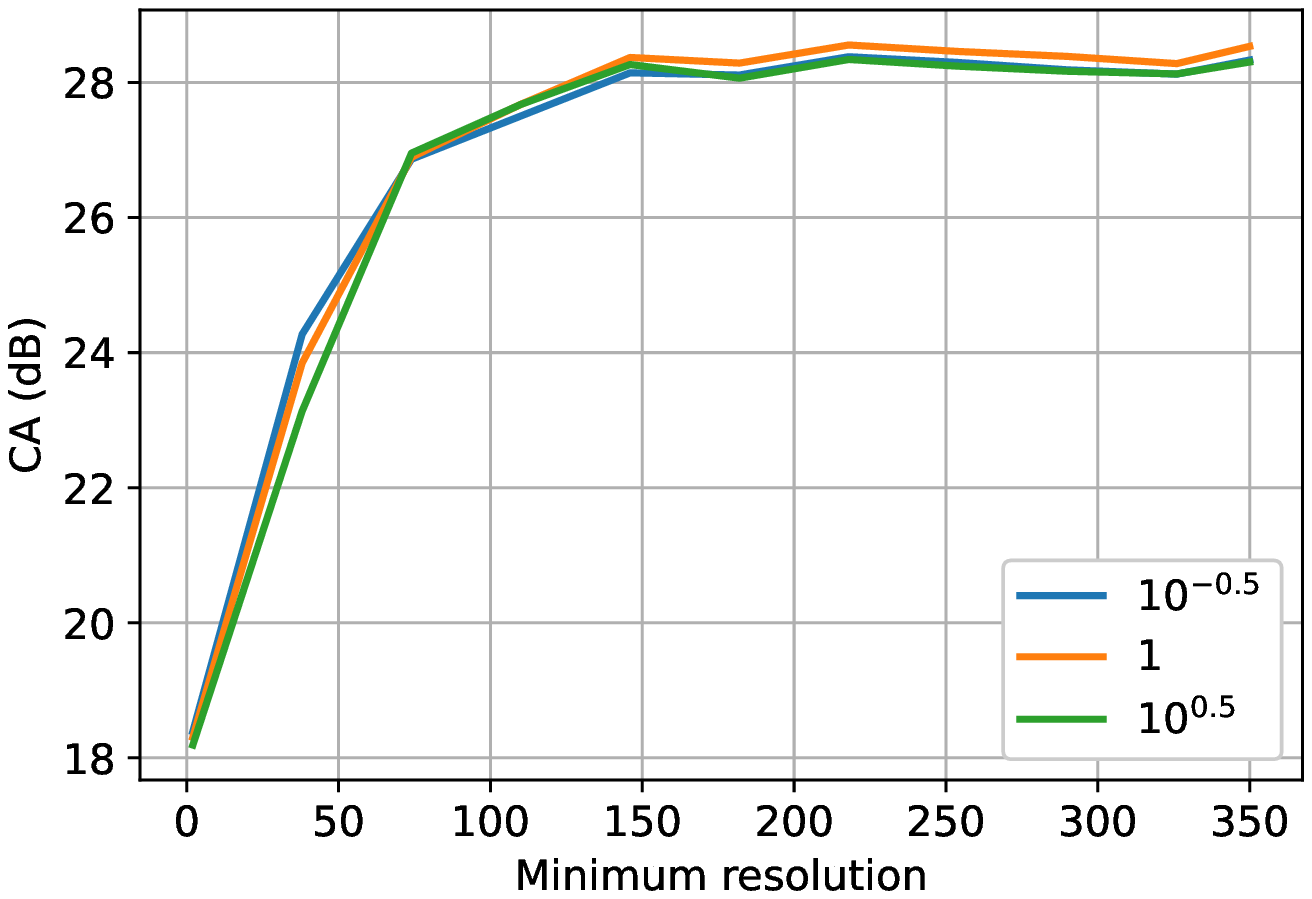}
	}\\[-.3cm]
	\subfloat{
		\includegraphics[width=.43\linewidth]{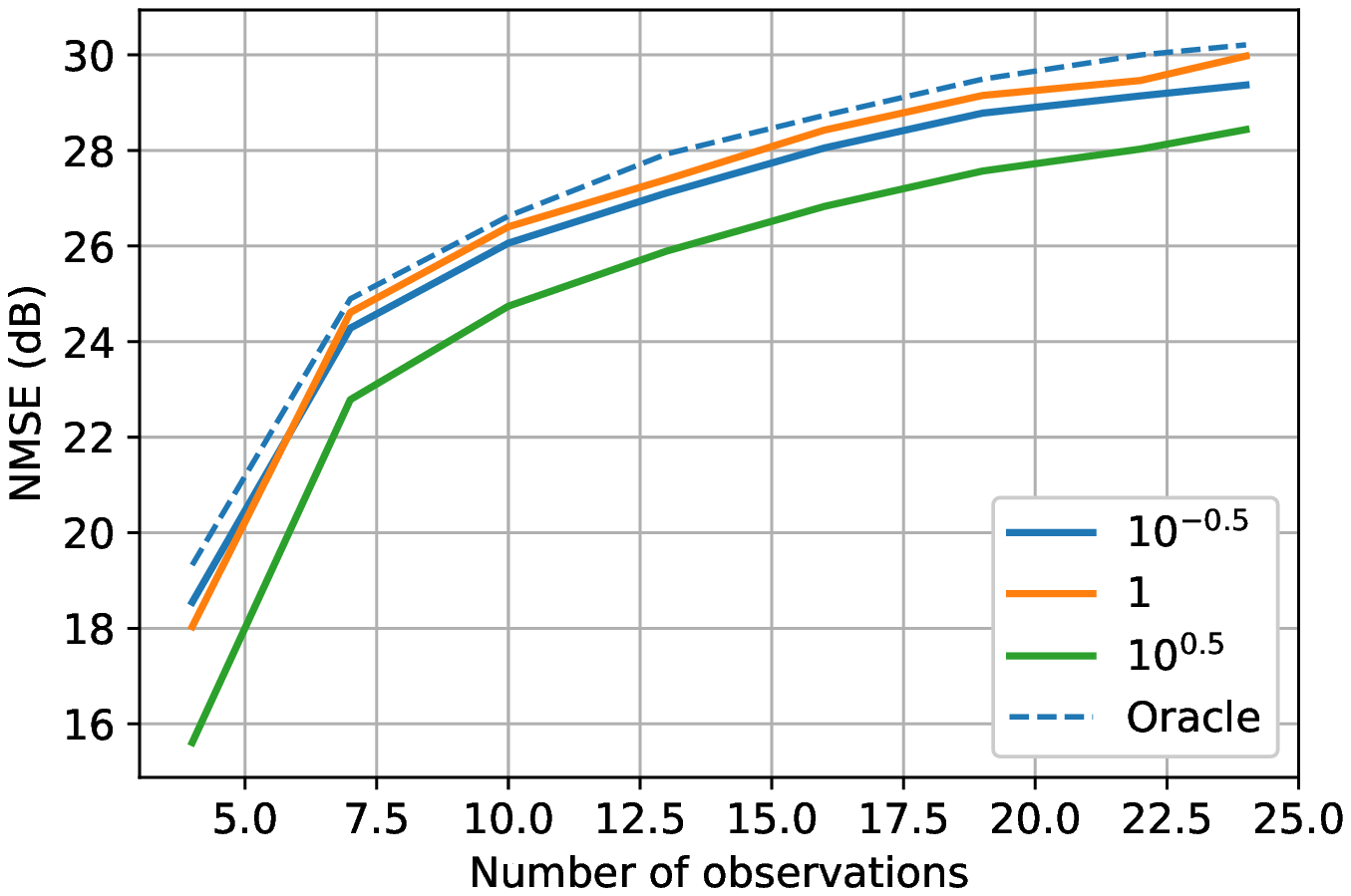}
	} 
	\hfill
	\subfloat{
		\includegraphics[width=.43\linewidth]{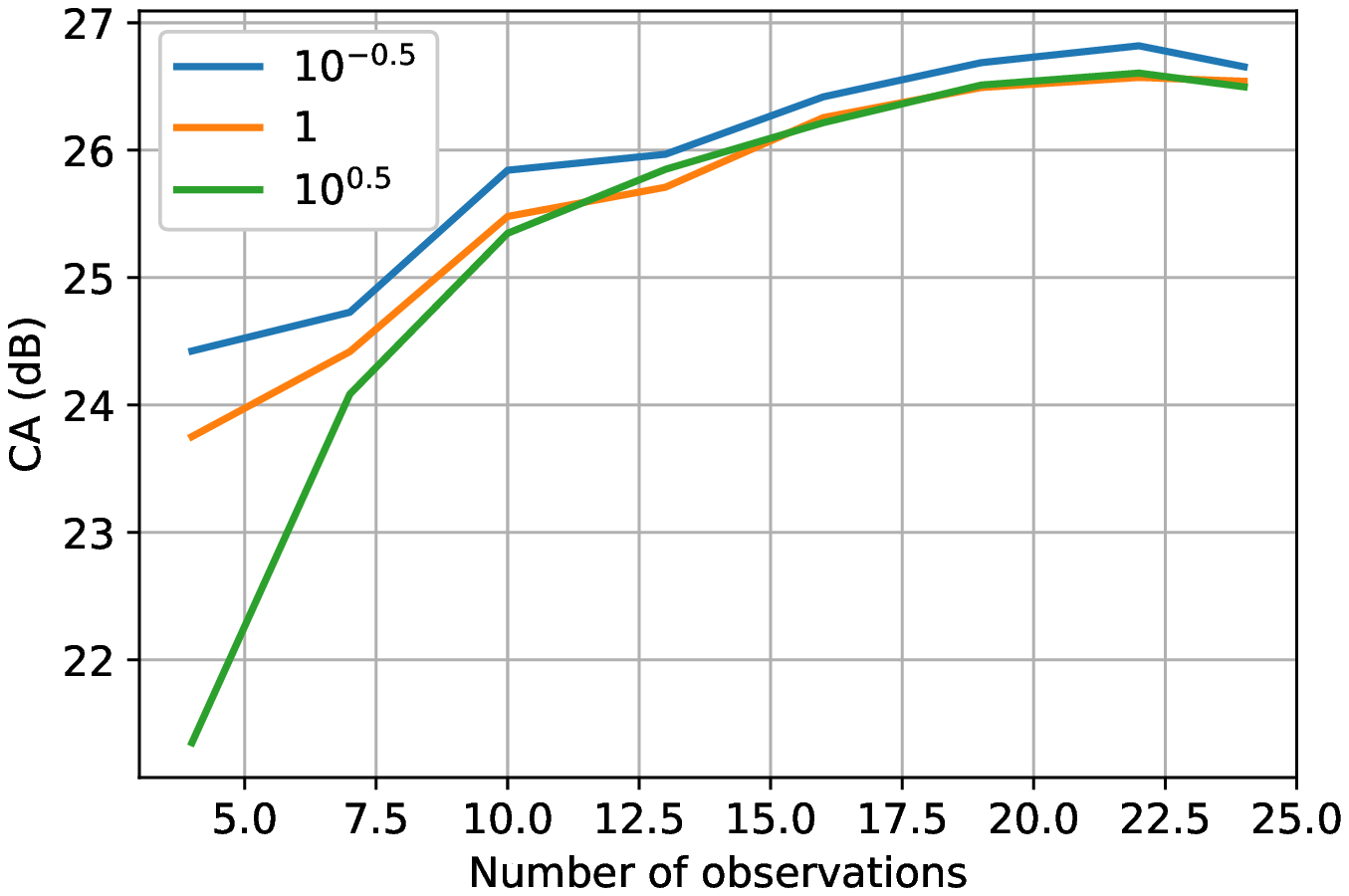}
	}\\[-.3cm]
	\subfloat{
		\includegraphics[width=.43\linewidth]{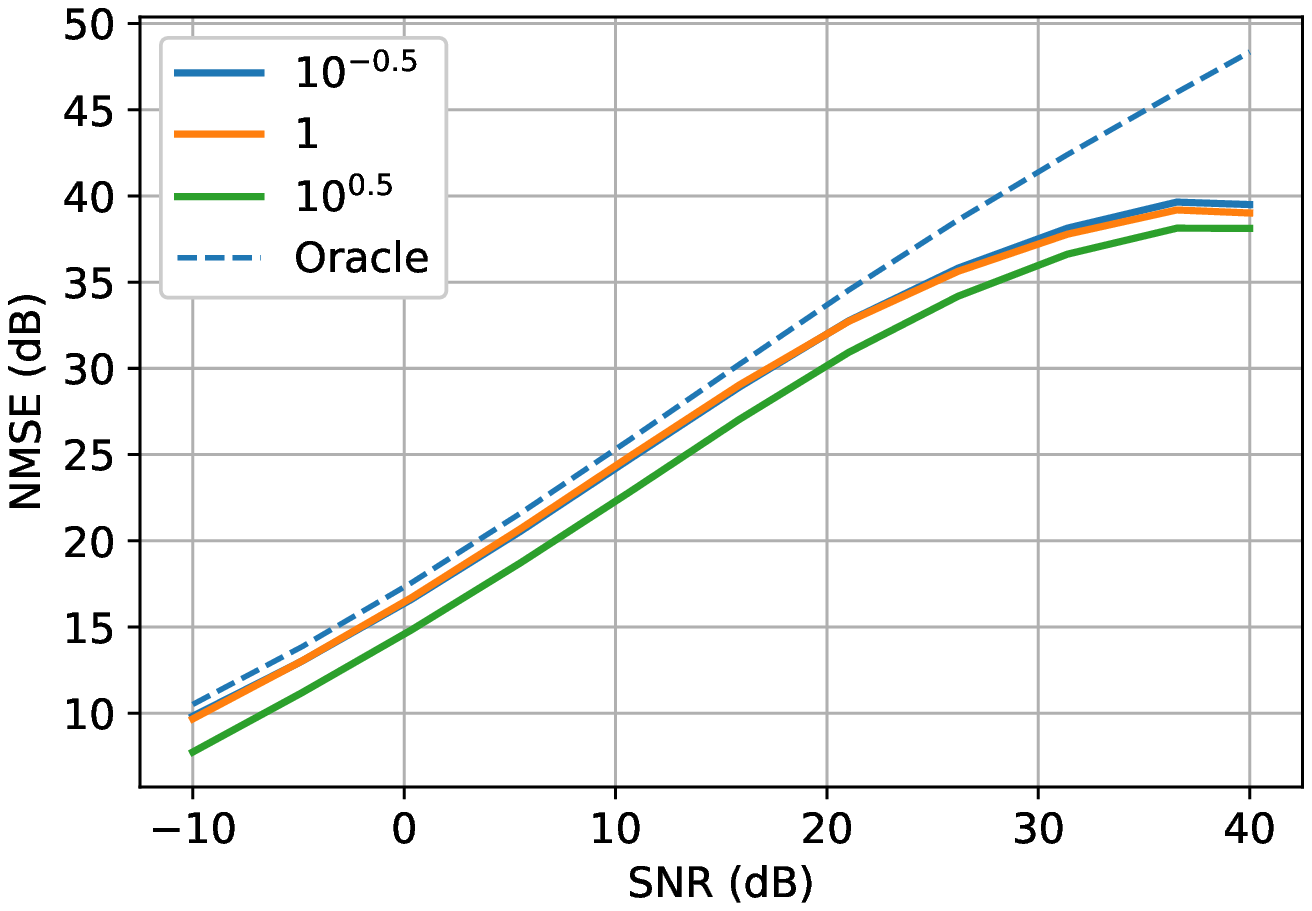}
	} 
	\hfill
	\subfloat{
		\includegraphics[width=.43\linewidth]{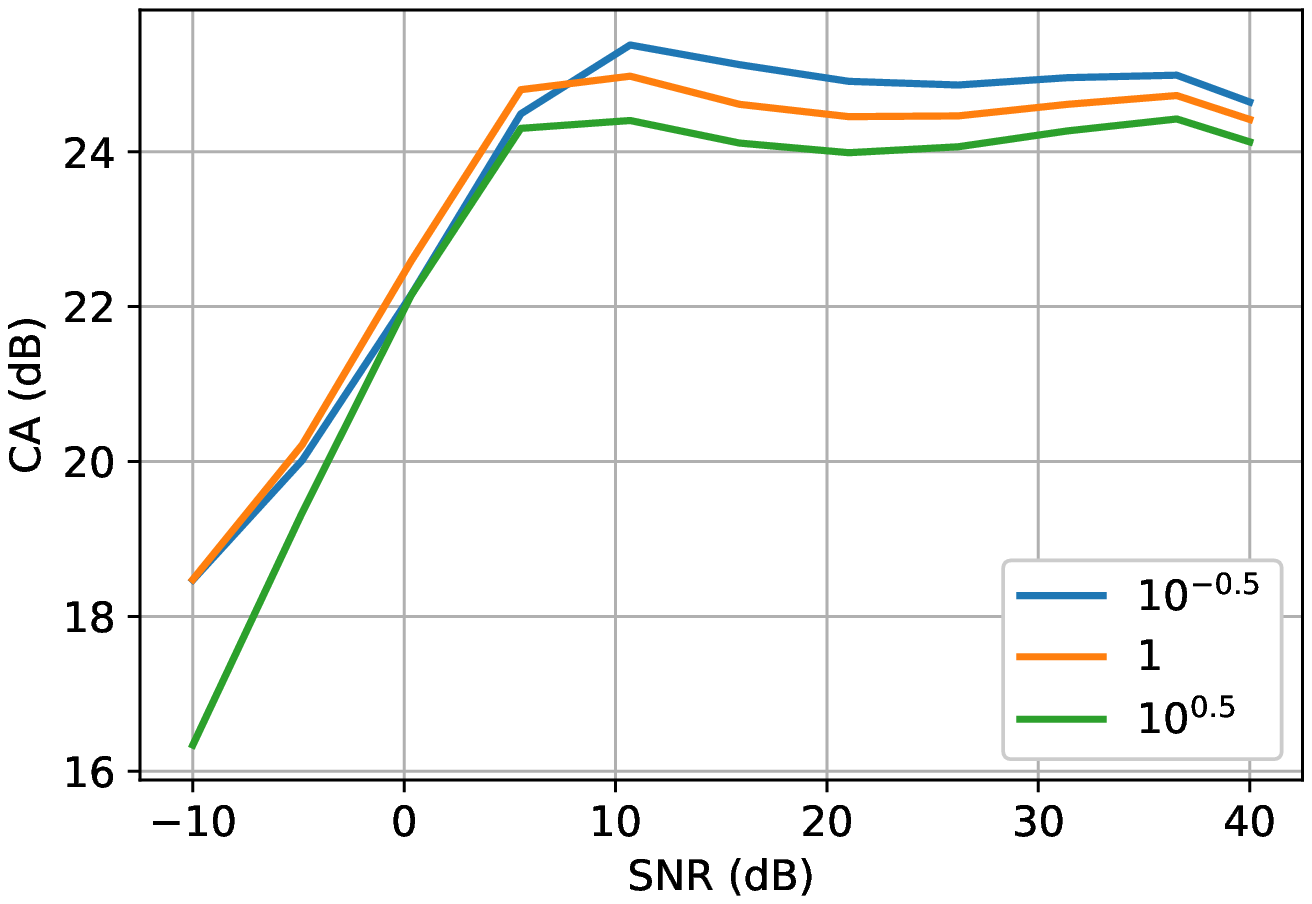}
	}
	\caption{Mean performance metrics over more 30 realizations performed by SDecGMCA with the optimal regularization hyperparameter at warm-up ${c_{wu}}_{opt}$ and with different regularization hyperparameters at refinement $c_{ref}$, which are indicated as multiples of ${c_{ref}}_{opt}$ in the legends.}
	\label{fig:cref}
\end{figure}

Finally, let us assess the sensitivity of SDecGMCA to the regularization hyperparameter at refinement as a function of the observation parameters (see Figure \ref{fig:cref}). Both NMSE and \ca{} losses are limited in a range of one order of magnitude around the optimal regularization hyperparameter (typically $-2~\si{dB}$). The noticeable exception is when $\mathbf{A}$ is ill-conditioned. The higher sensibility to the regularization hyperparameter may come from the induced ill-condition of the $(\mathbf{M}[l])_{l\in\mathbb{N}}$ (matrices which are inverted in Eq.~\eqref{eq:updateS2}).

\subsection{Application to realistic astrophysical data}

\begin{figure}
	\subfloat[][Synchrotron source $\bS_1^*$]{
		\centering
		\includegraphics[width=.495\linewidth]{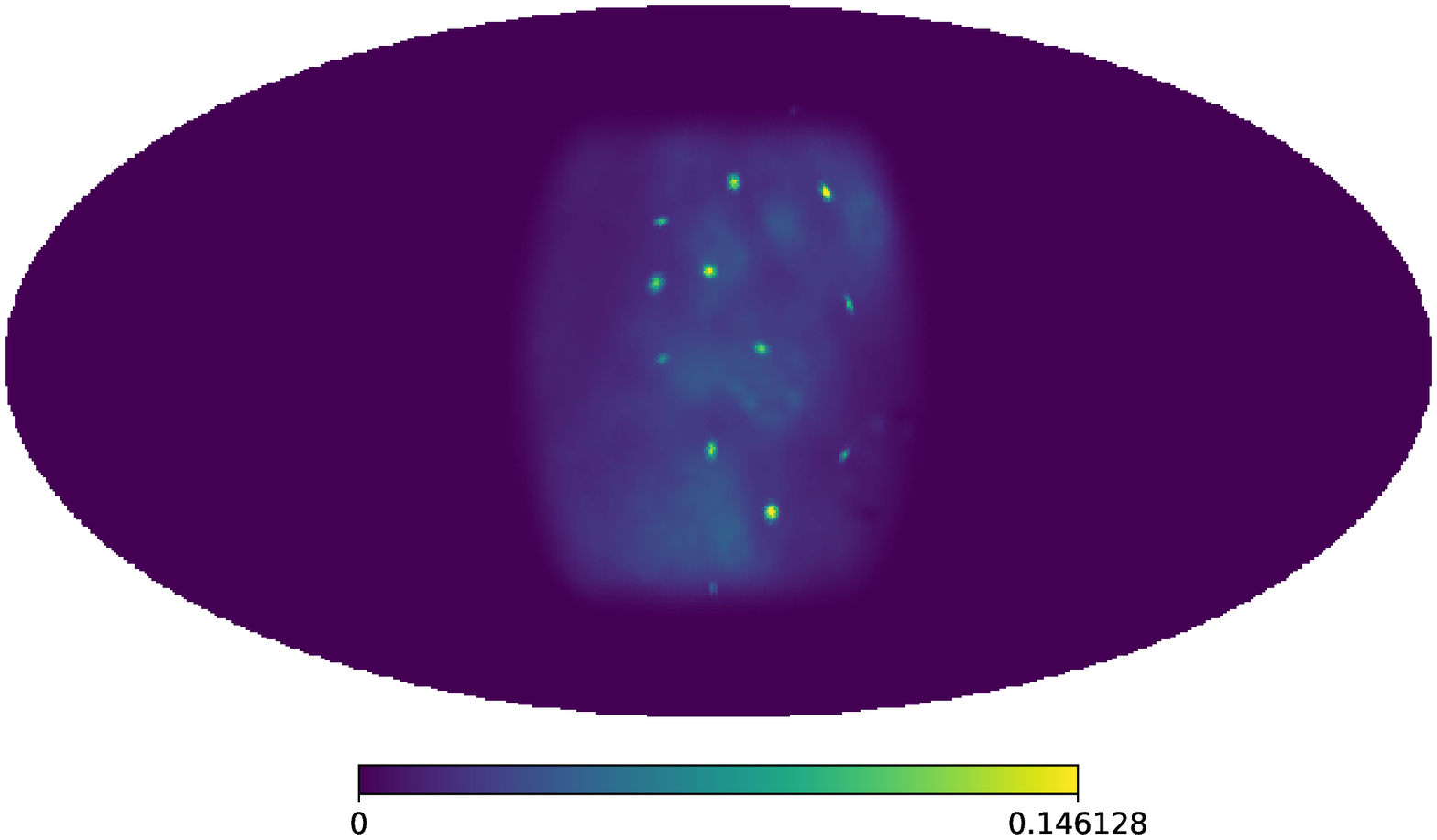}
		\label{fig:planckS0}
	} 
	\hfill
	\subfloat[][Thermal source $\bS_2^*$]{
		\centering
		\includegraphics[width=.47\linewidth]{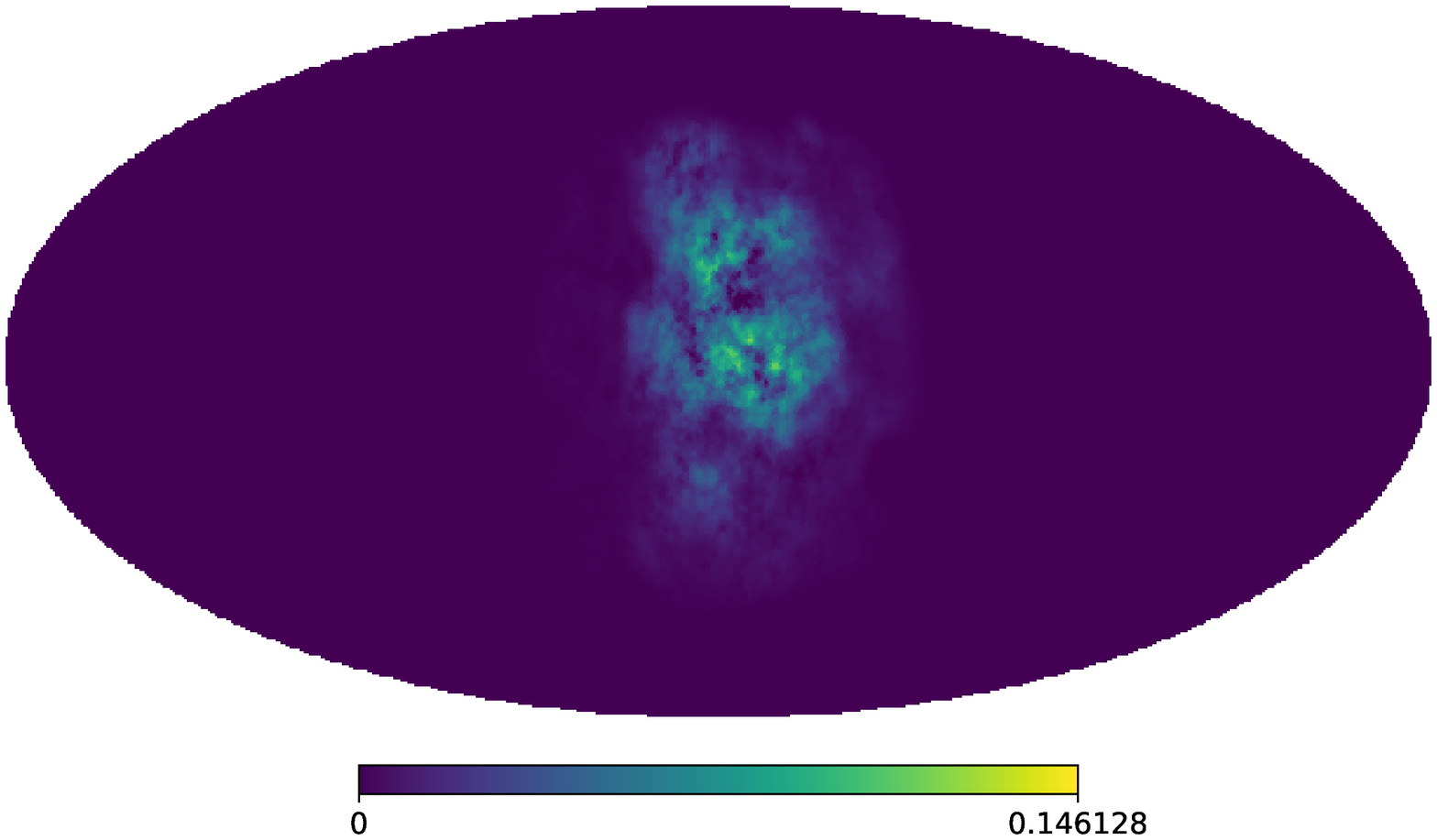}
		\label{fig:planckS1}
	}\\[-.3cm]
	\subfloat[][Emission line source $\bS_3^*$]{
		\centering
		\includegraphics[width=.47\linewidth]{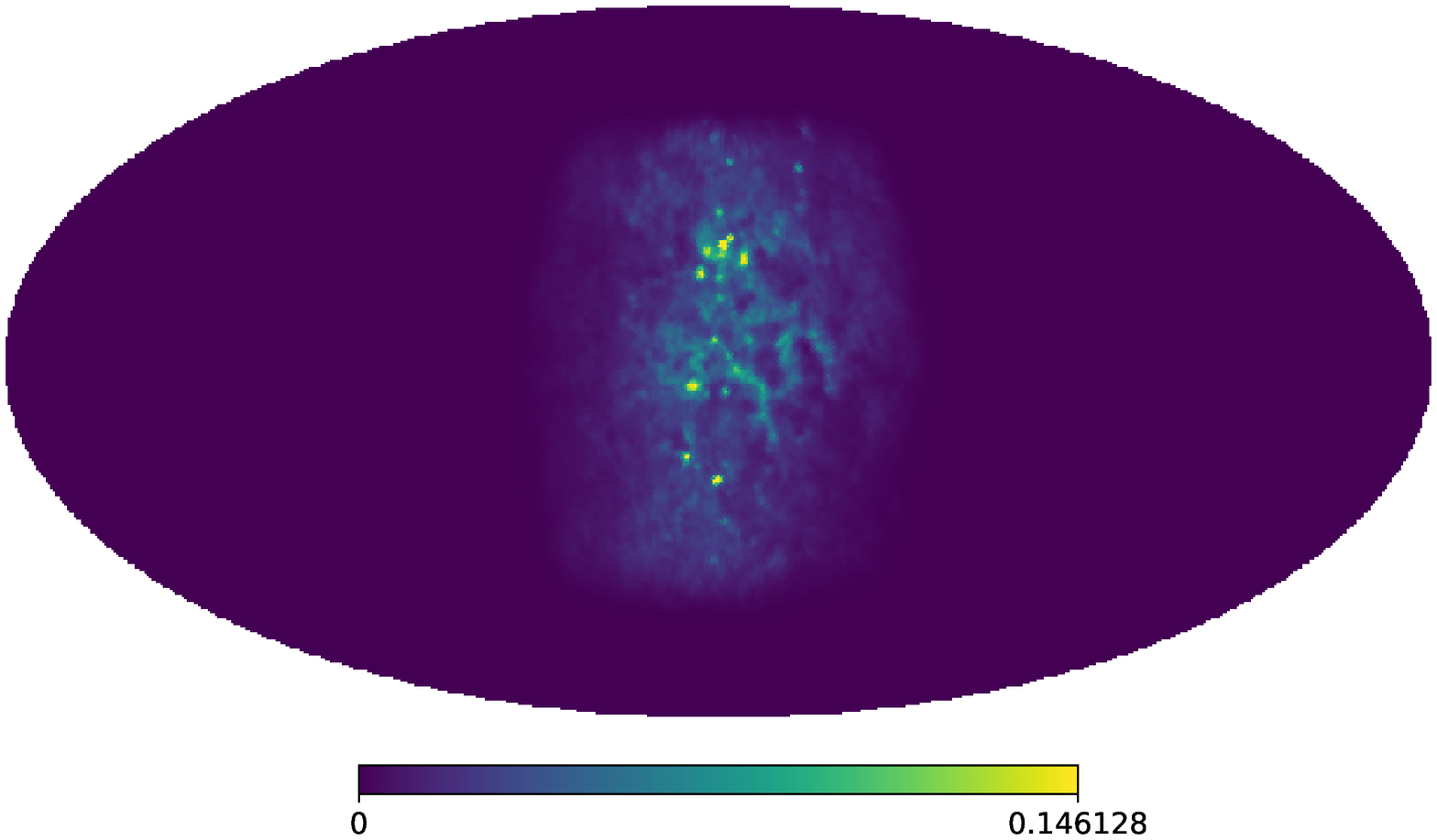}
		\label{fig:planckS2}
	} 
	\hfill
	\subfloat[][Mixing matrix $\bA^*$]{
		\centering
		\includegraphics[width=.47\linewidth]{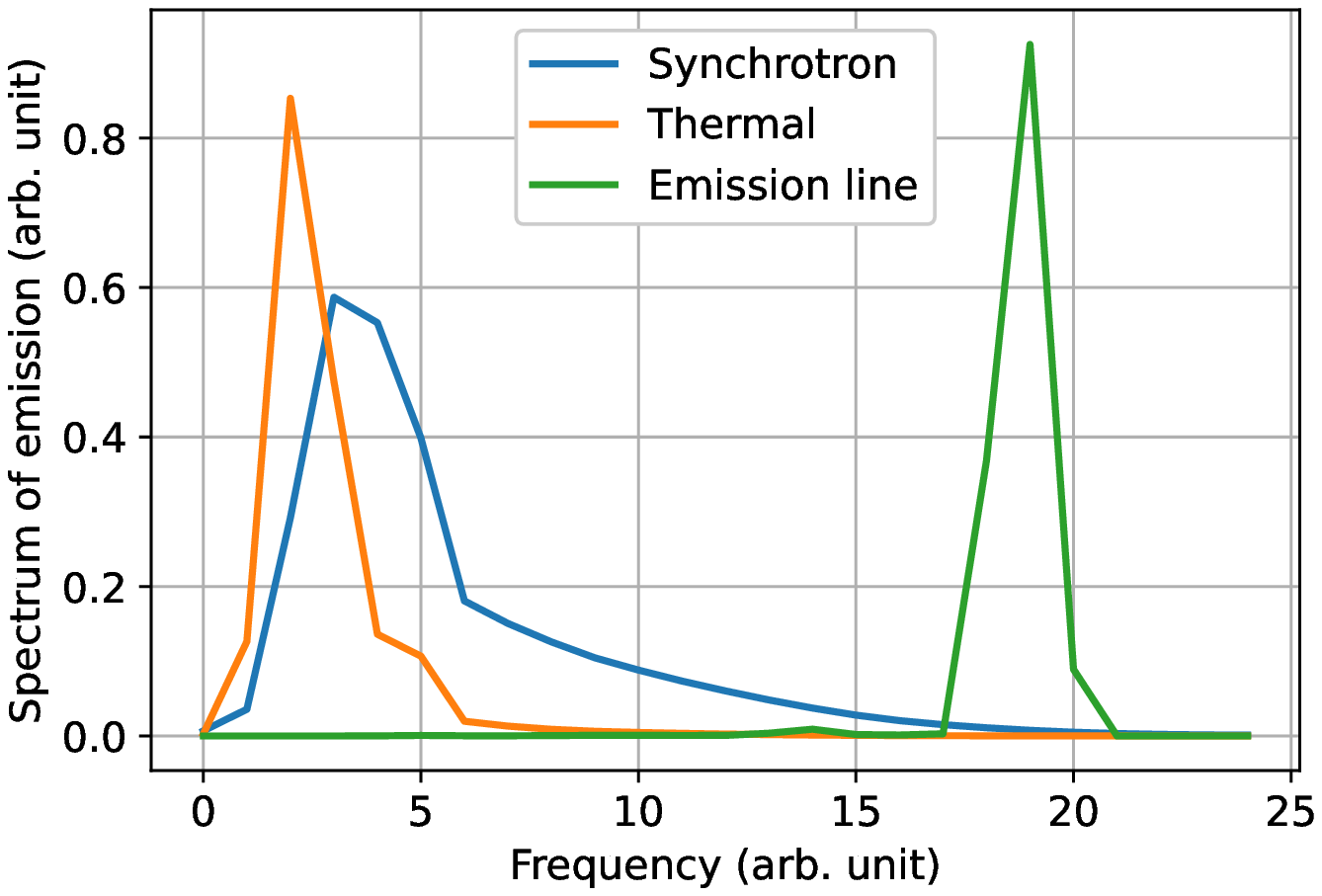}
		\label{fig:planckA}
	}
	\hfill
	\subfloat[][Low resolution observation $\bX_3$, with synchrotron and thermal emissions]{
		\centering
		\includegraphics[width=.47\linewidth]{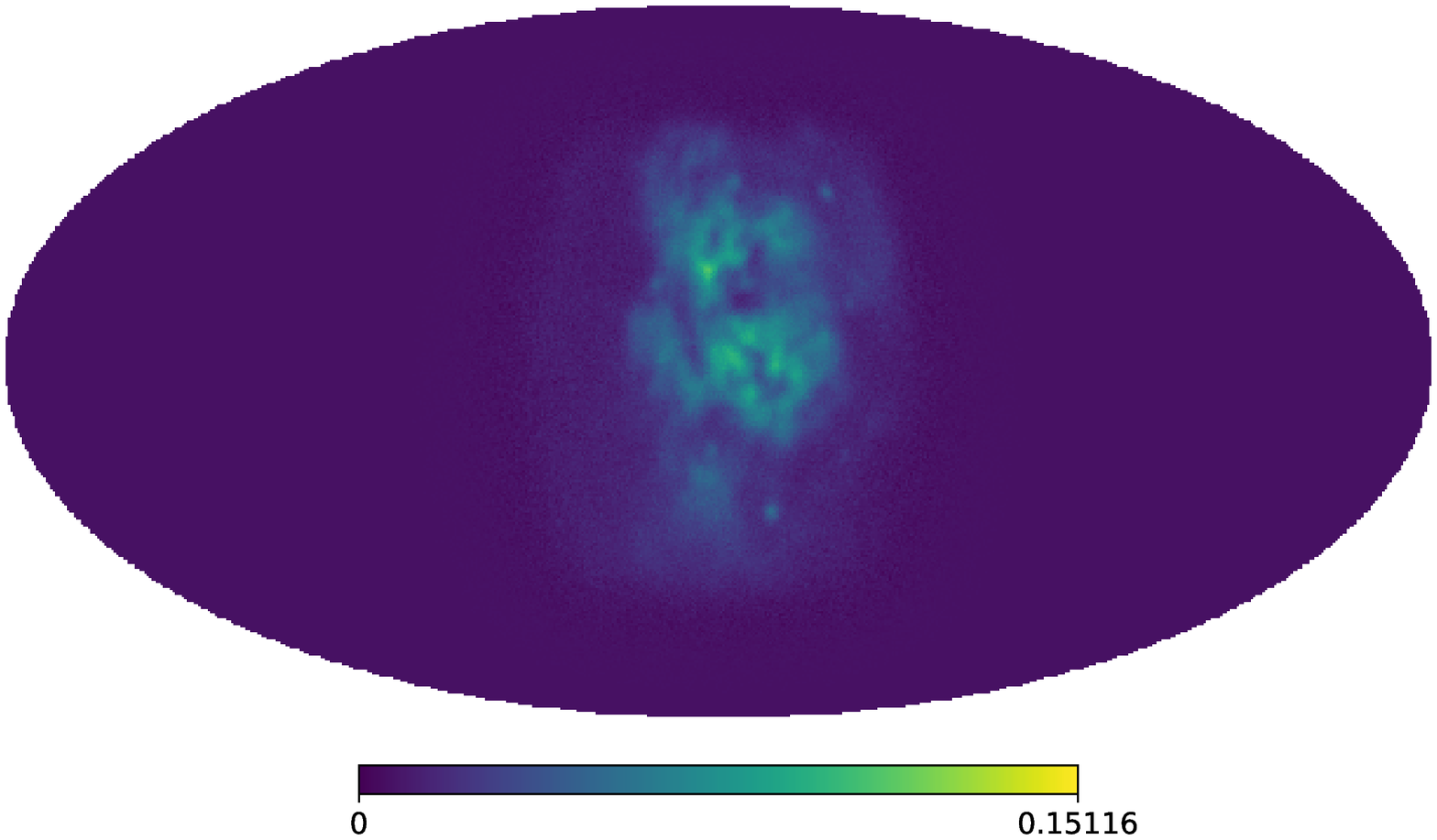}
		\label{fig:planckY2}
	} 
	\hfill
	\subfloat[][High resolution observation $\bX_{20}$, with mostly emission line source]{
		\centering
		\includegraphics[width=.47\linewidth]{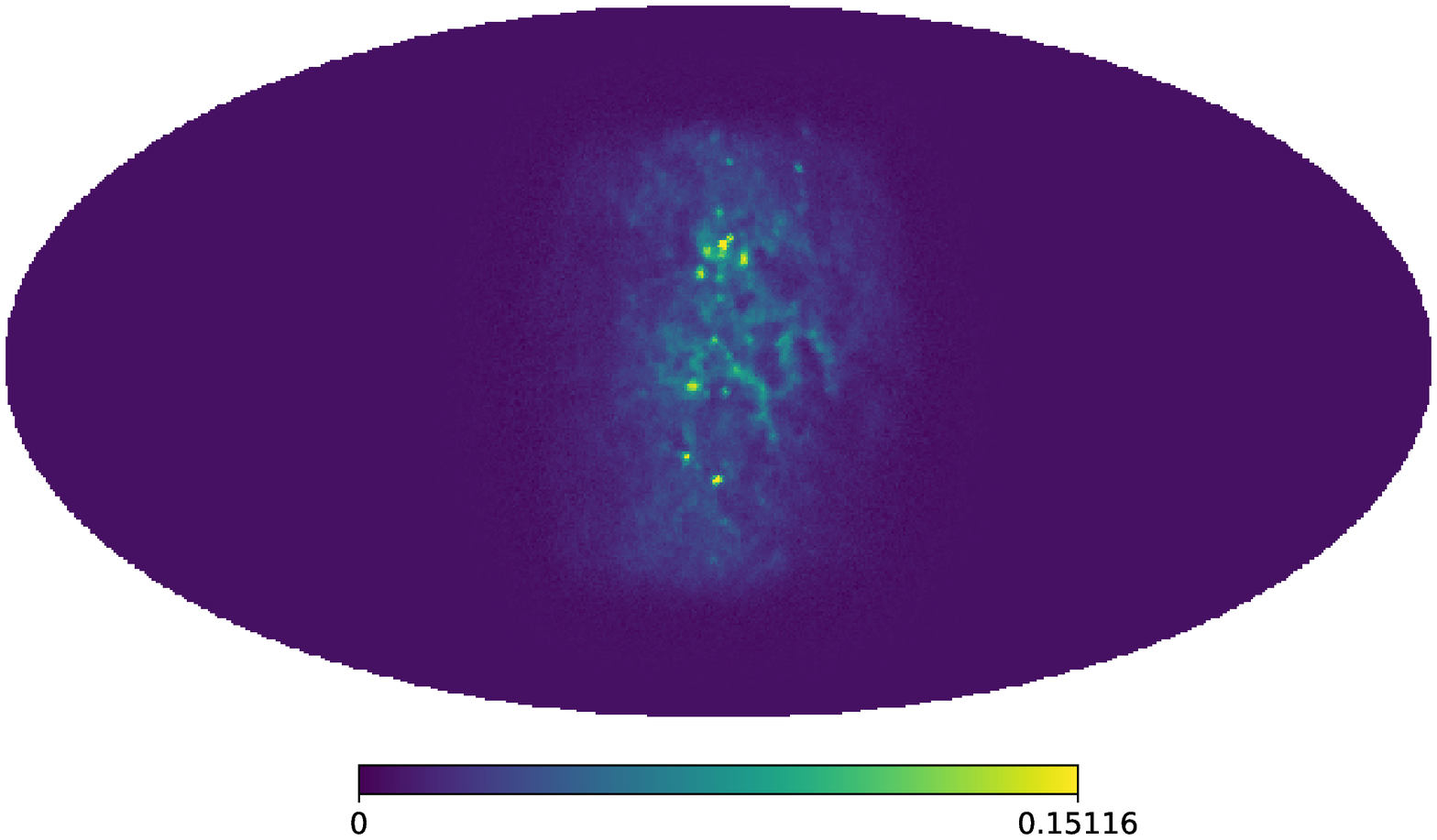}
		\label{fig:planckY19}
	}
	\caption{Realistic data and example of observations with a SNR of 10 dB.}
	\label{fig:planckdata}
\end{figure}

Joint deconvolution and blind source separation is now performed on realistic X-ray astrophysical simulations, whose spectra are representation of the emissions provided by the Chandra space telescope\footnote{\url{http://chandra.harvard.edu/}} in range of energy $2-6$ keV \cite{Picquenot19}. These data are composed of $25$ observations, which are built as mixtures of three sources. The latter are associated with a synchrotron, a thermal and an iron line emissions as displayed in Fig.~\ref{fig:planckdata}, whose spectra are displayed in the panel (d). It is commonplace in such an application that only a partial sky coverage is observed, which is simulated by projecting the sources on a limited portion of the sphere (see Fig.~\ref{fig:planckdata}). Similarly to the synthetic data, we set the pixelization parameters to $N_{side} = 128$ and $l_{max} = 384$. The 25 observations have resolutions evenly spread between $l_{max}/8$ and $l_{max}$. \\

\begin{figure}
	\subfloat[][Estimated source $\bS_1$]{
		\centering
		\includegraphics[width=.4\linewidth]{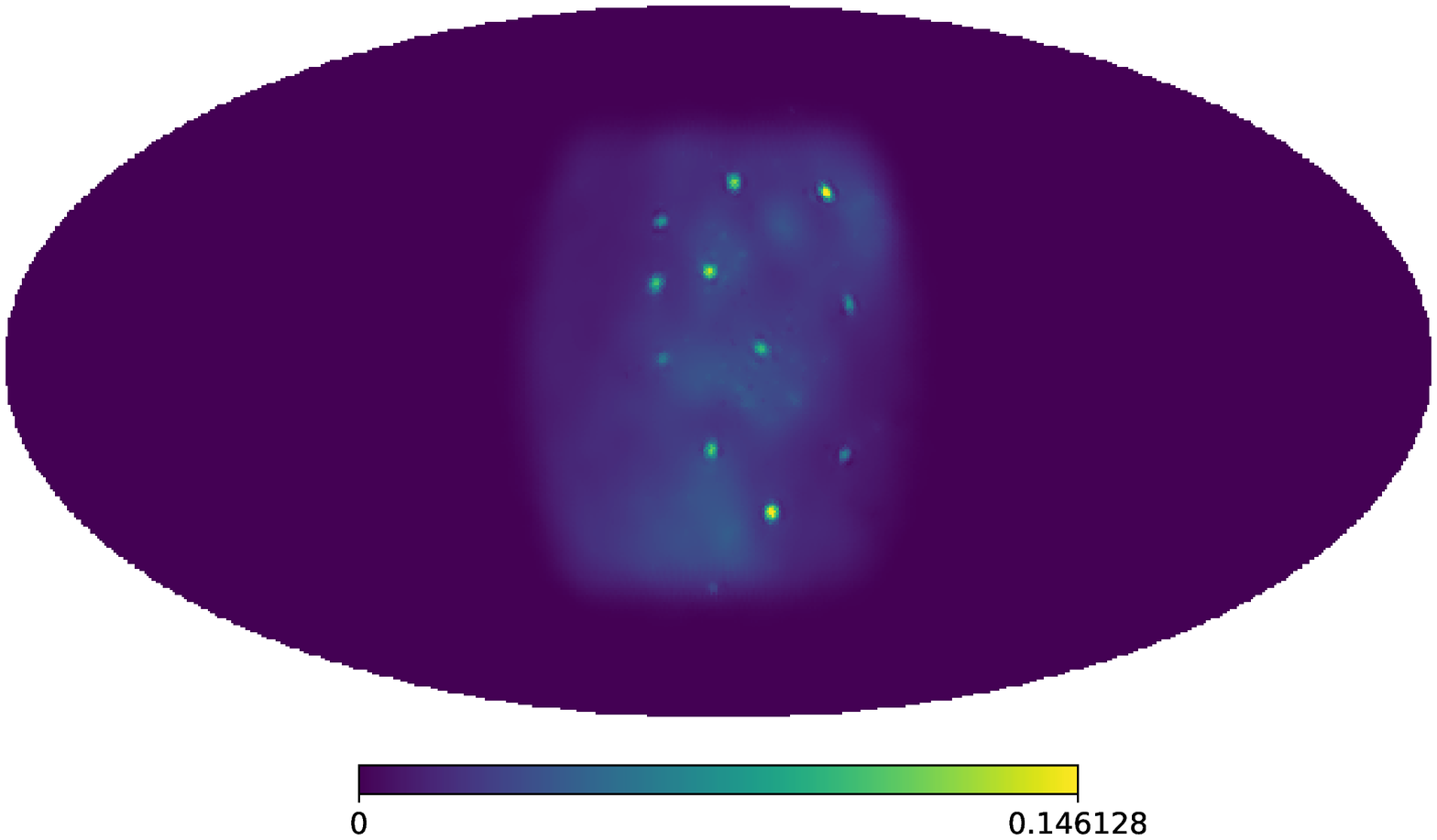}
		\label{fig:planckSe0}
	} 
	\hfill
	\subfloat[][Absolute error  $|\mathbf{S}_1^*-\mathbf{S}_1|$]{
		\centering
		\includegraphics[width=.4\linewidth]{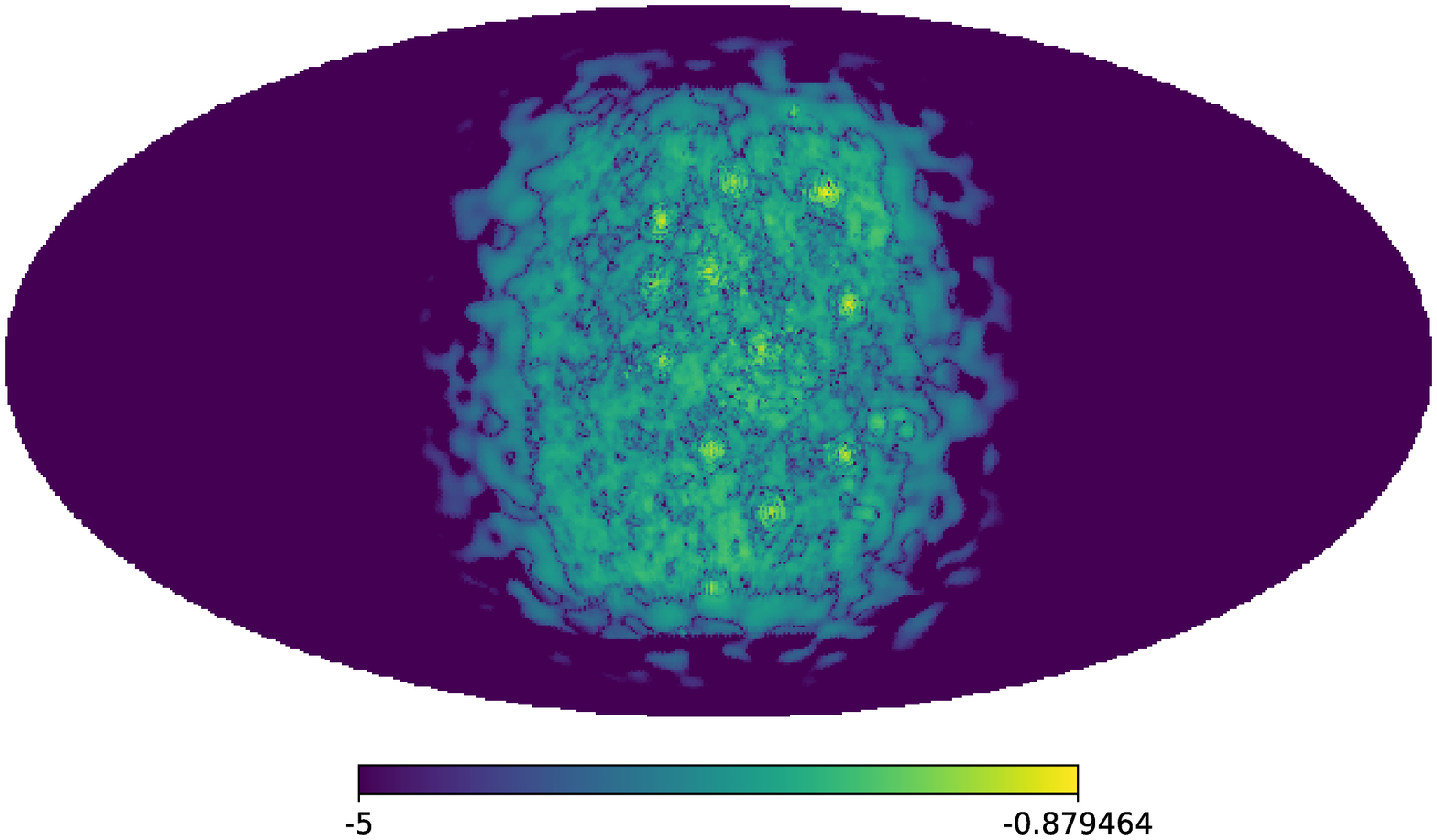}
		\label{fig:planckE0}
	} \\[-.3cm]
	\subfloat[][Estimated source $\bS_2$]{
		\centering
		\includegraphics[width=.4\linewidth]{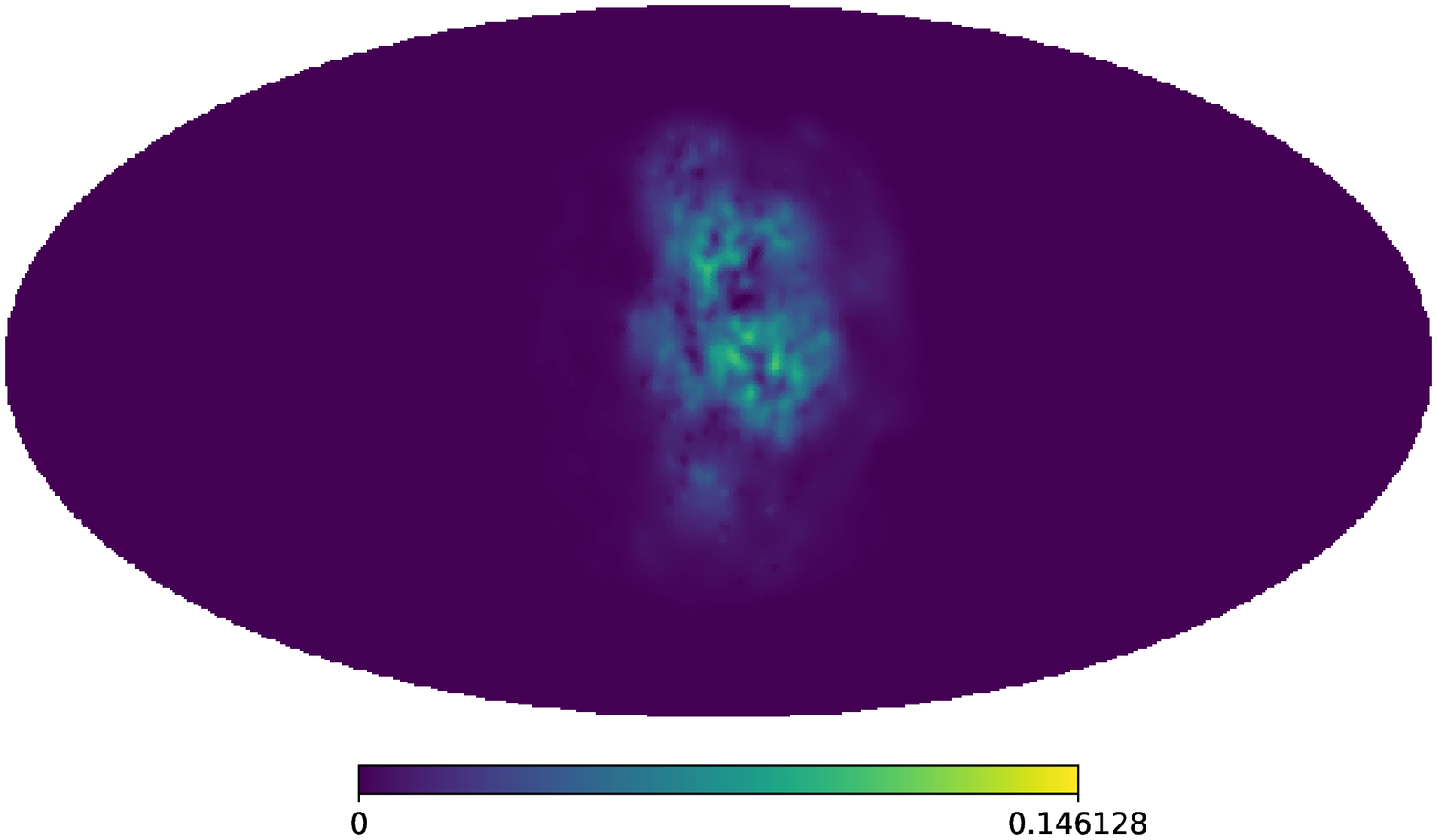}
		\label{fig:planckSe1}
	}
	\hfill
	\subfloat[][Absolute error  $|\mathbf{S}_2^*-\mathbf{S}_2|$]{ 
		\centering
		\includegraphics[width=.4\linewidth]{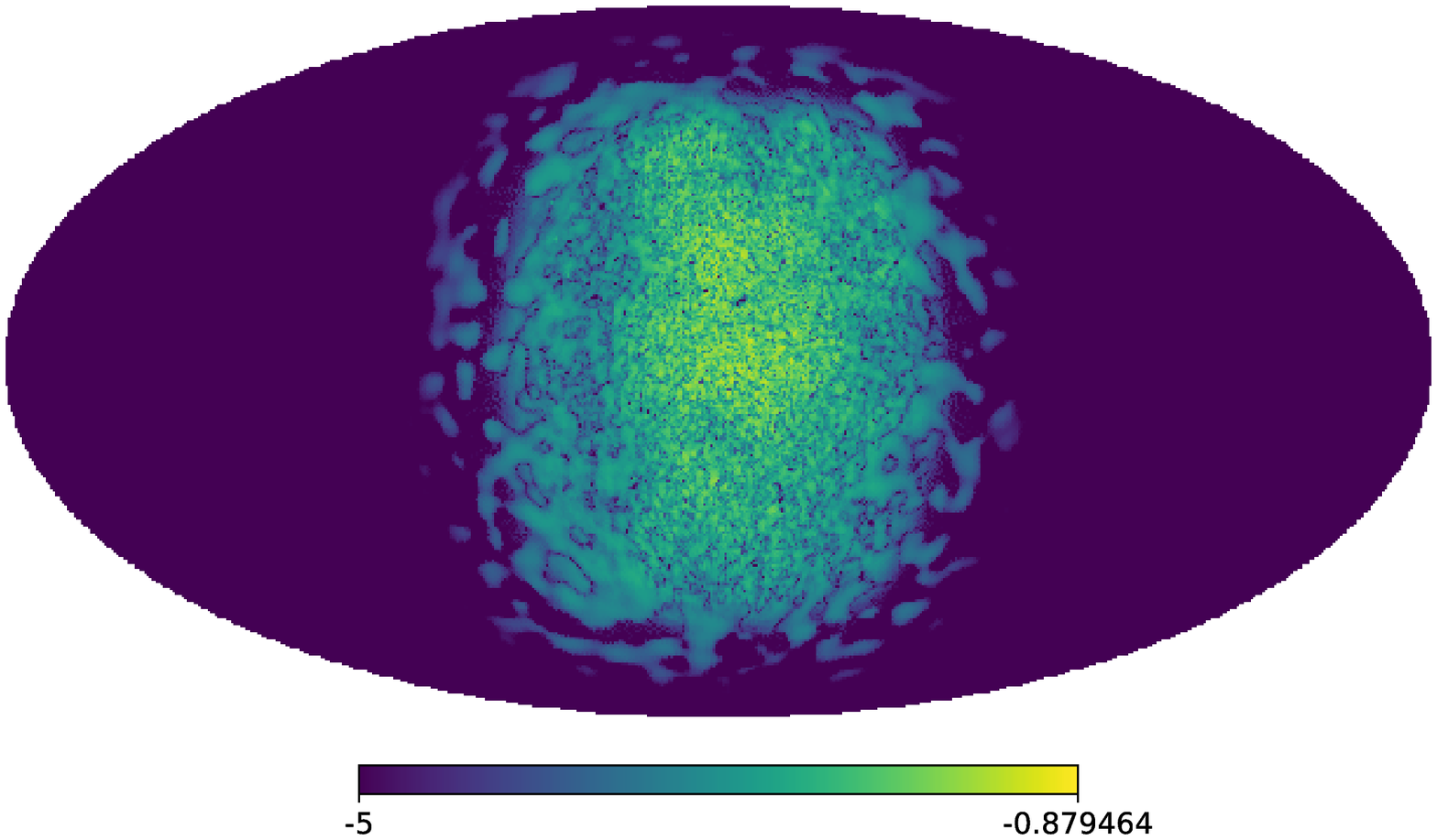}
		\label{fig:planckE1}
	}\\[-.3cm]
	\subfloat[][Estimated source $\bS_3$]{
		\centering
		\includegraphics[width=.4\linewidth]{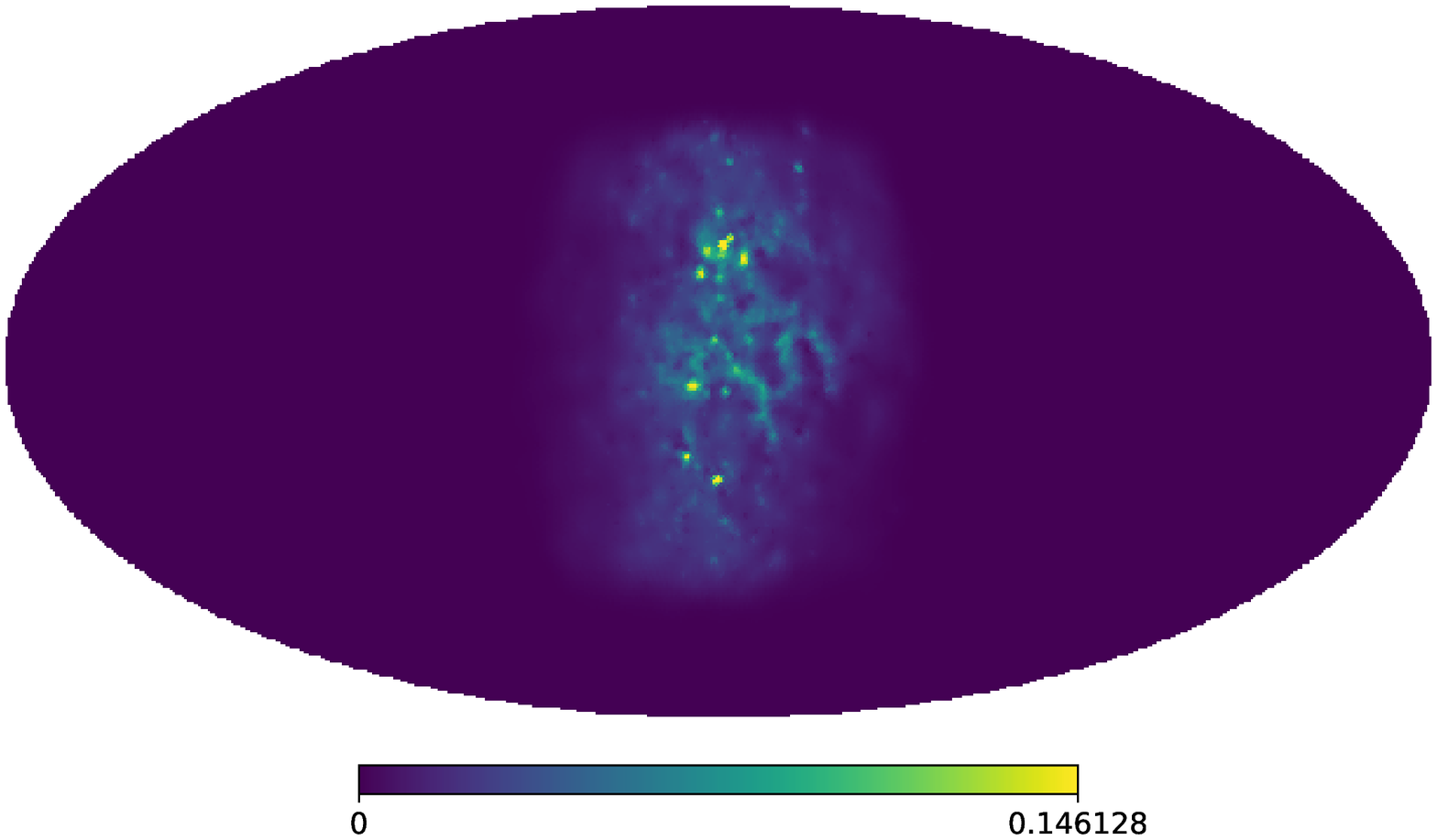}
		\label{fig:planckSe2}
	} 	
	\hfill
	\subfloat[][Absolute error $|\mathbf{S}_3^*-\mathbf{S}_3|$]{
	\centering
	\includegraphics[width=.4\linewidth]{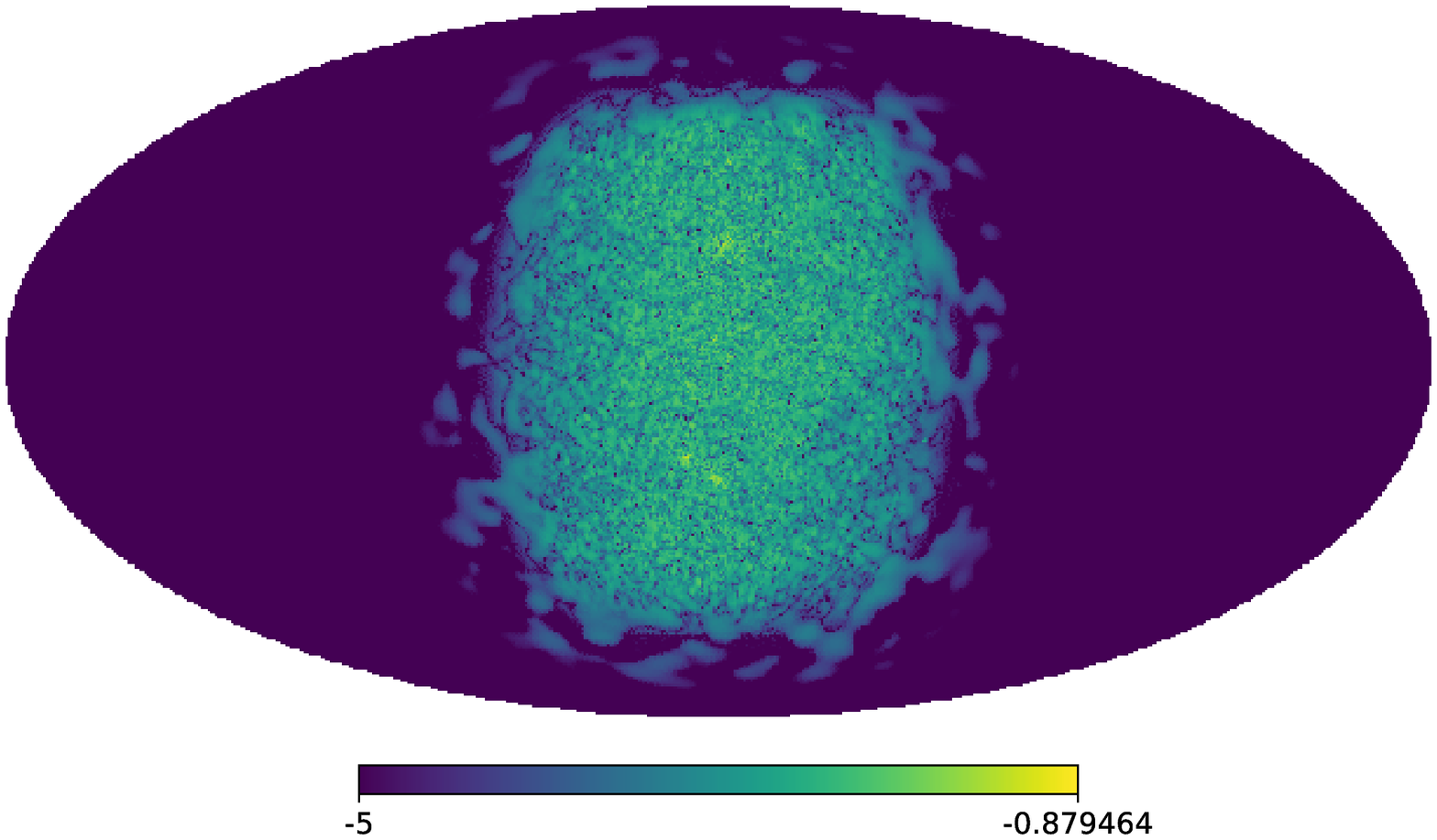}
	\label{fig:planckE2}
	}\\[-.3cm]
	\subfloat[][Estimated mixing matrix $\bA$ (dashed: ground truth $\bA^*$)]{
		\centering
		\includegraphics[width=.4\linewidth]{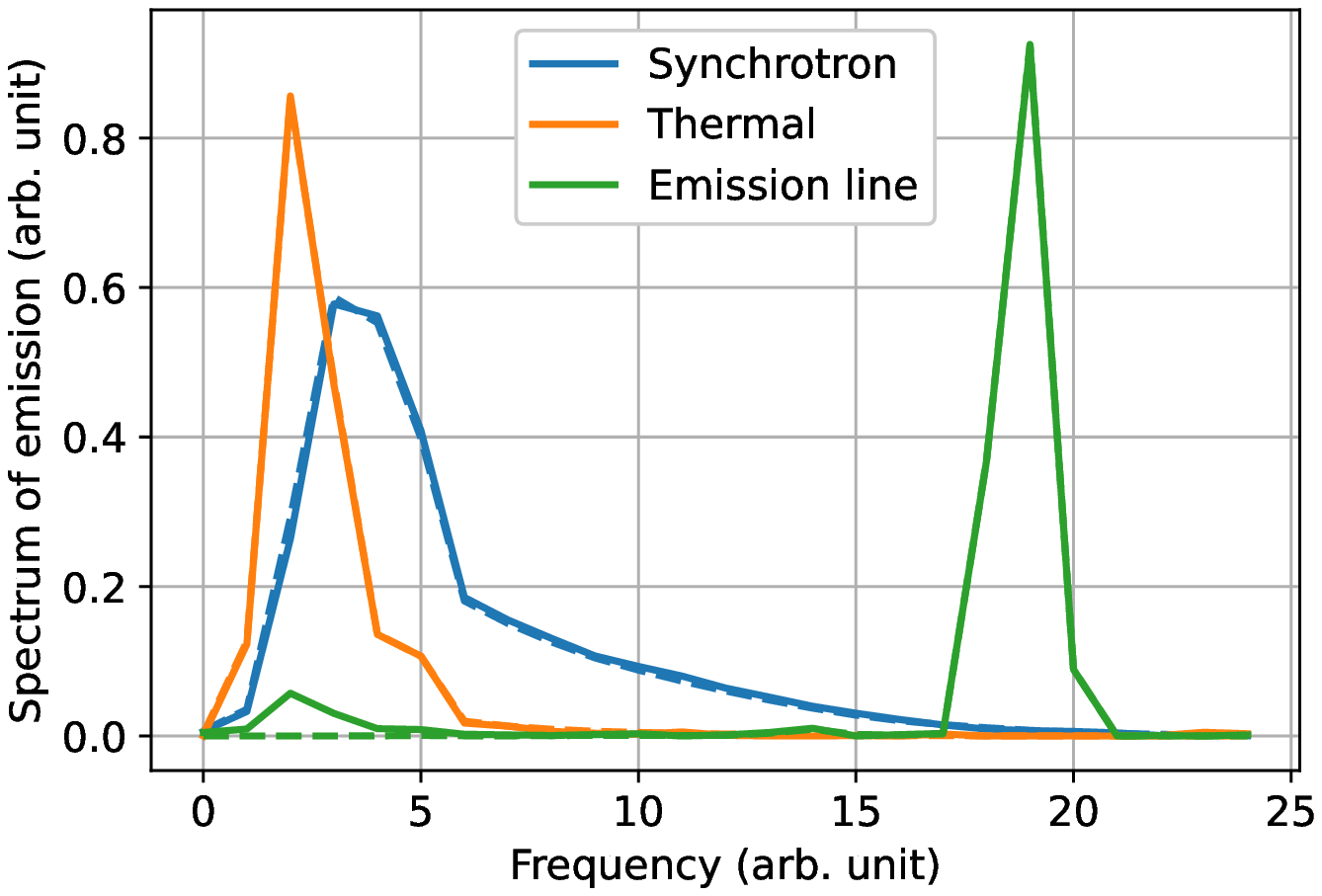}
		\label{fig:planckAe}
	}
	\caption{DBSS example with SDecGMCA (sources: arbitrary unit, logarithmic scale)}
	\label{fig:plancksdecgmca}
\end{figure}

We firstly consider that the observations are corrupted by a noise of 10 dB (see two observations on Fig.~\ref{fig:planckY2} and \ref{fig:planckY19}). Figure \ref{fig:plancksdecgmca} shows a solution given by SDecGMCA (with $K_{max}= 0.2$ to overcome the correlations between the sources). The errors are dominated by the deconvolution artifacts. Actually, the estimated sources are very close to the oracle estimation with the ground truth mixing matrix $\bA^*$ ($\text{NMSE} = 21.17$ dB \textit{vs.}~oracle $\text{NMSE} = 22.58$ dB). Moreover, putting aside a small leakage in the lower frequencies of the emission line source in the synchrotron emission, the spectra of the three sources are well reconstructed ($\text{\ca} = 18.02$ dB).
 
\begin{figure}
	\subfloat[][Oracle]{
		\centering
		\includegraphics[width=.495\linewidth]{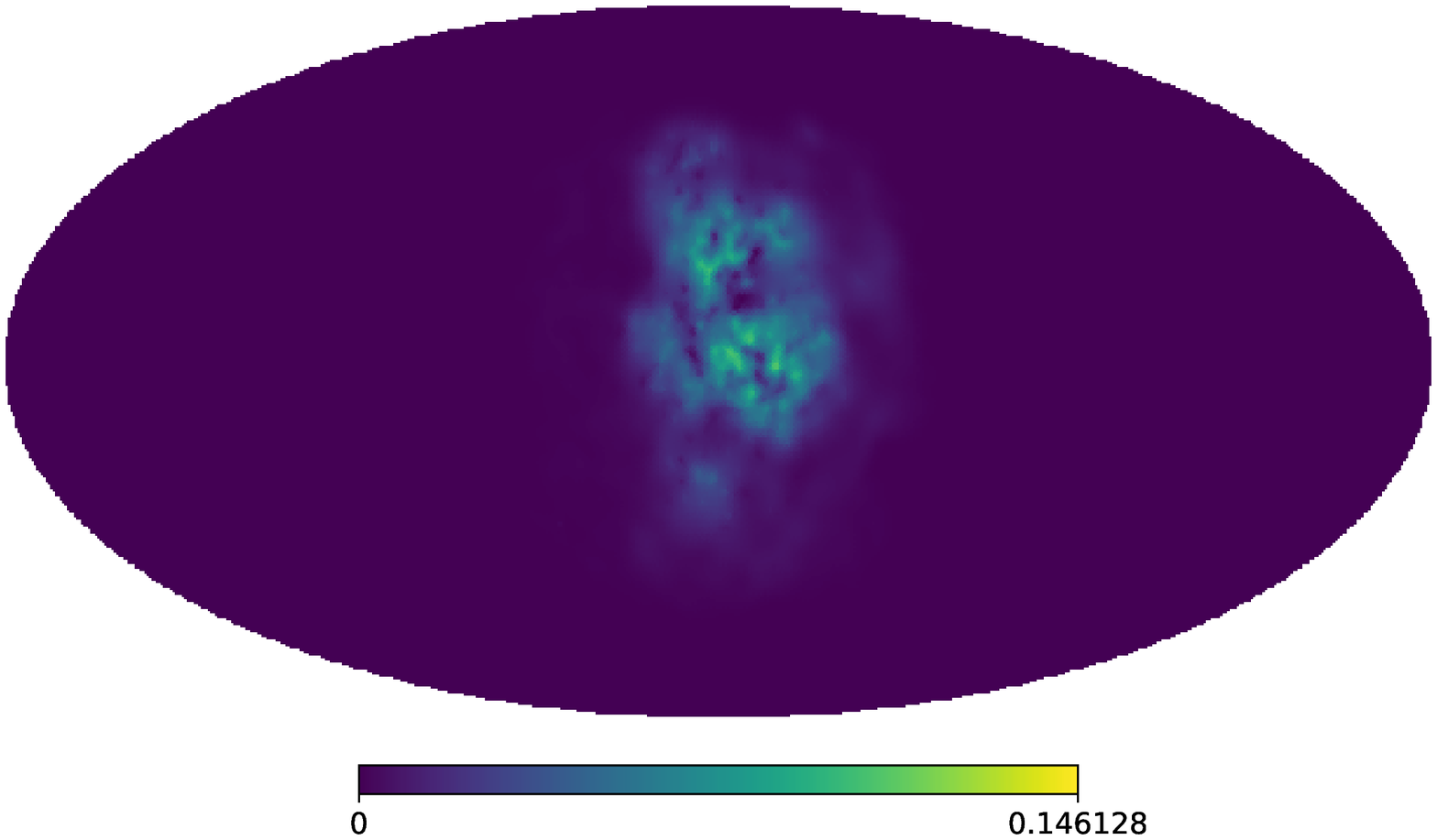}
	} 
	\hfill
	\subfloat[][SDecGMCA]{
		\centering
		\includegraphics[width=.495\linewidth]{planck_Se01.eps}
	} \\[-.3cm]
	\subfloat[][oDecGMCA]{
		\centering
		\includegraphics[width=.47\linewidth]{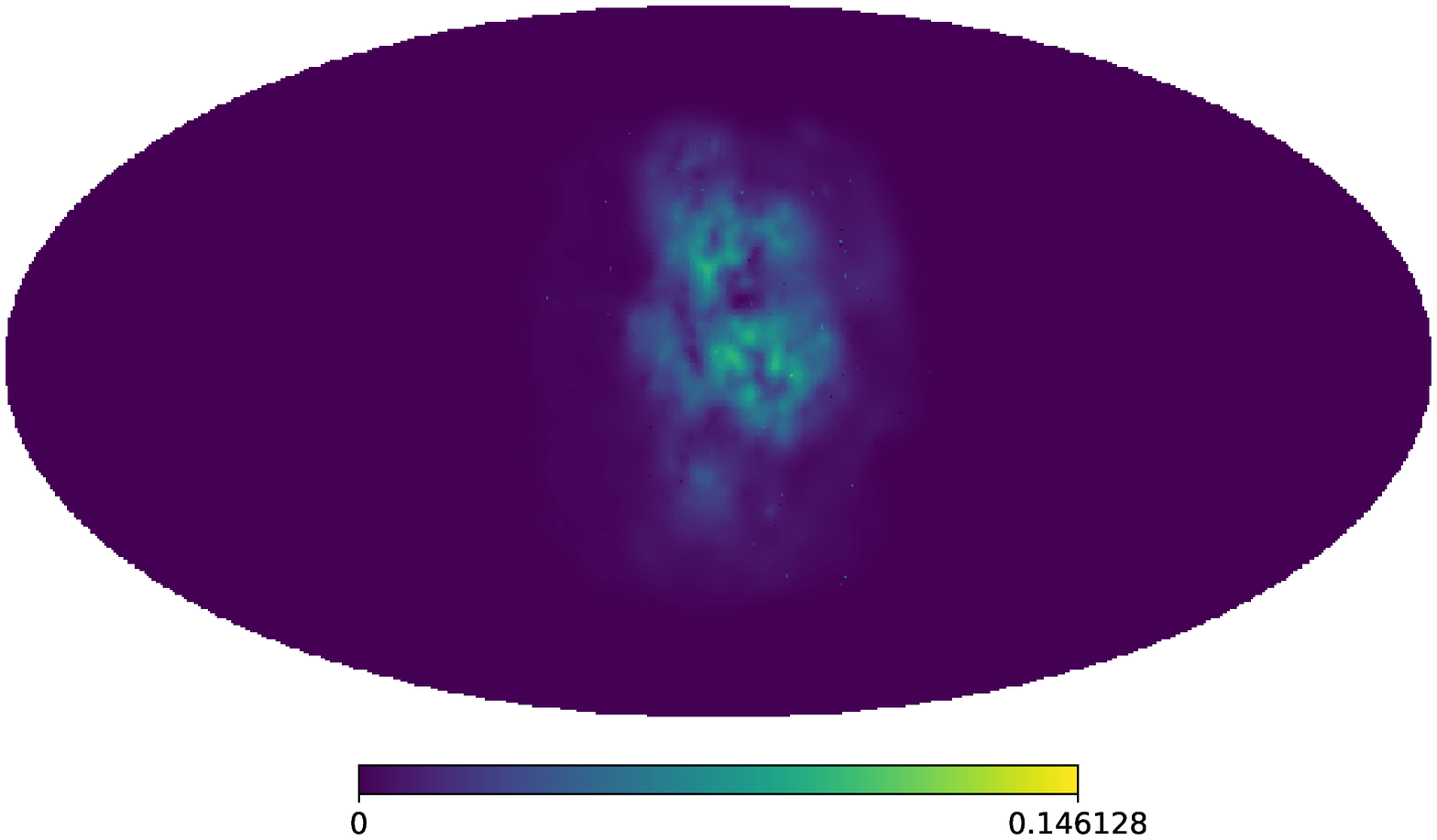}
	}
	\hfill
	\subfloat[][GMCA]{
		\centering
		\includegraphics[width=.47\linewidth]{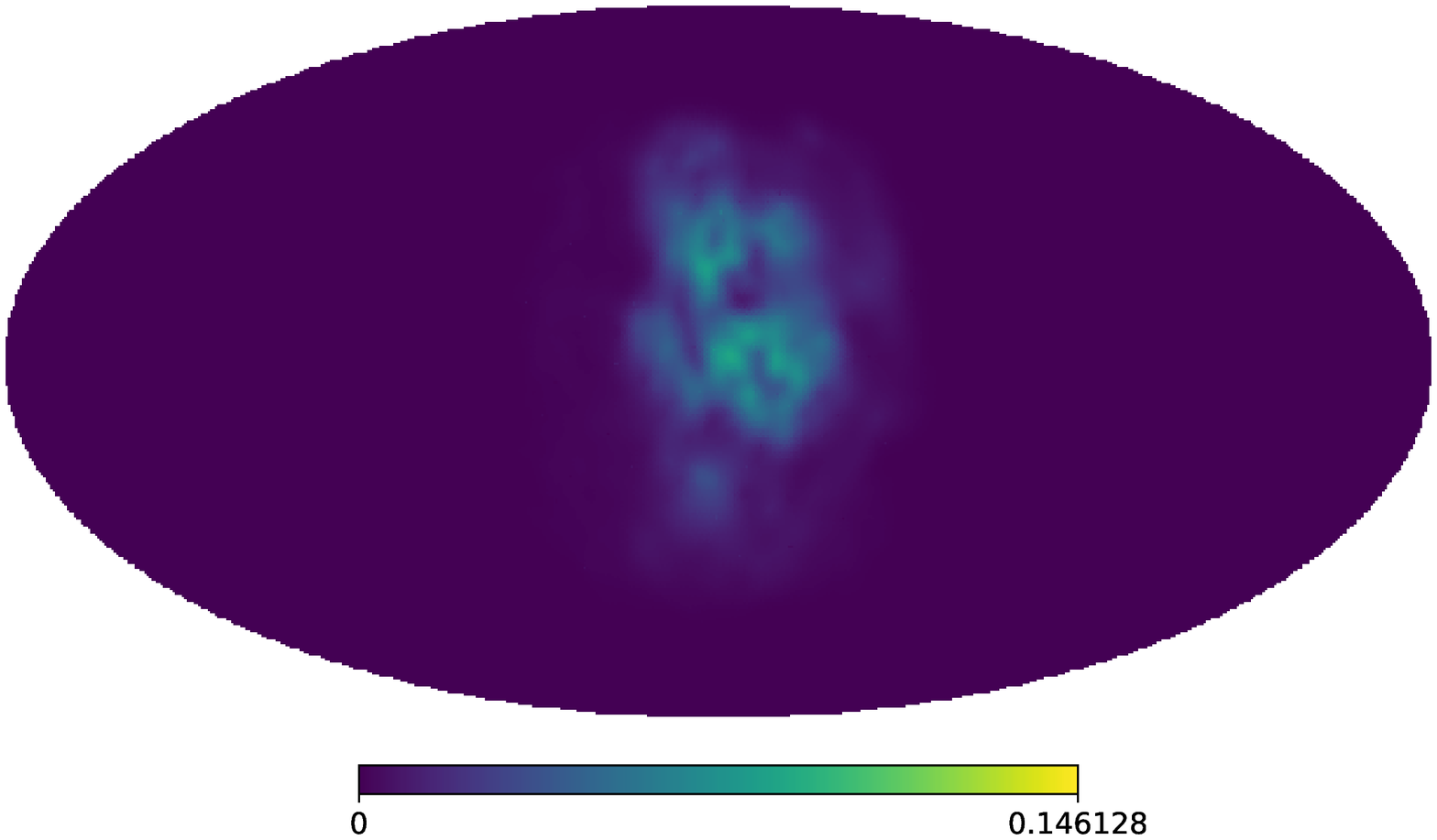}
	}\\[-.3cm]
	\subfloat[][HALS]{
		\centering
		\includegraphics[width=.47\linewidth]{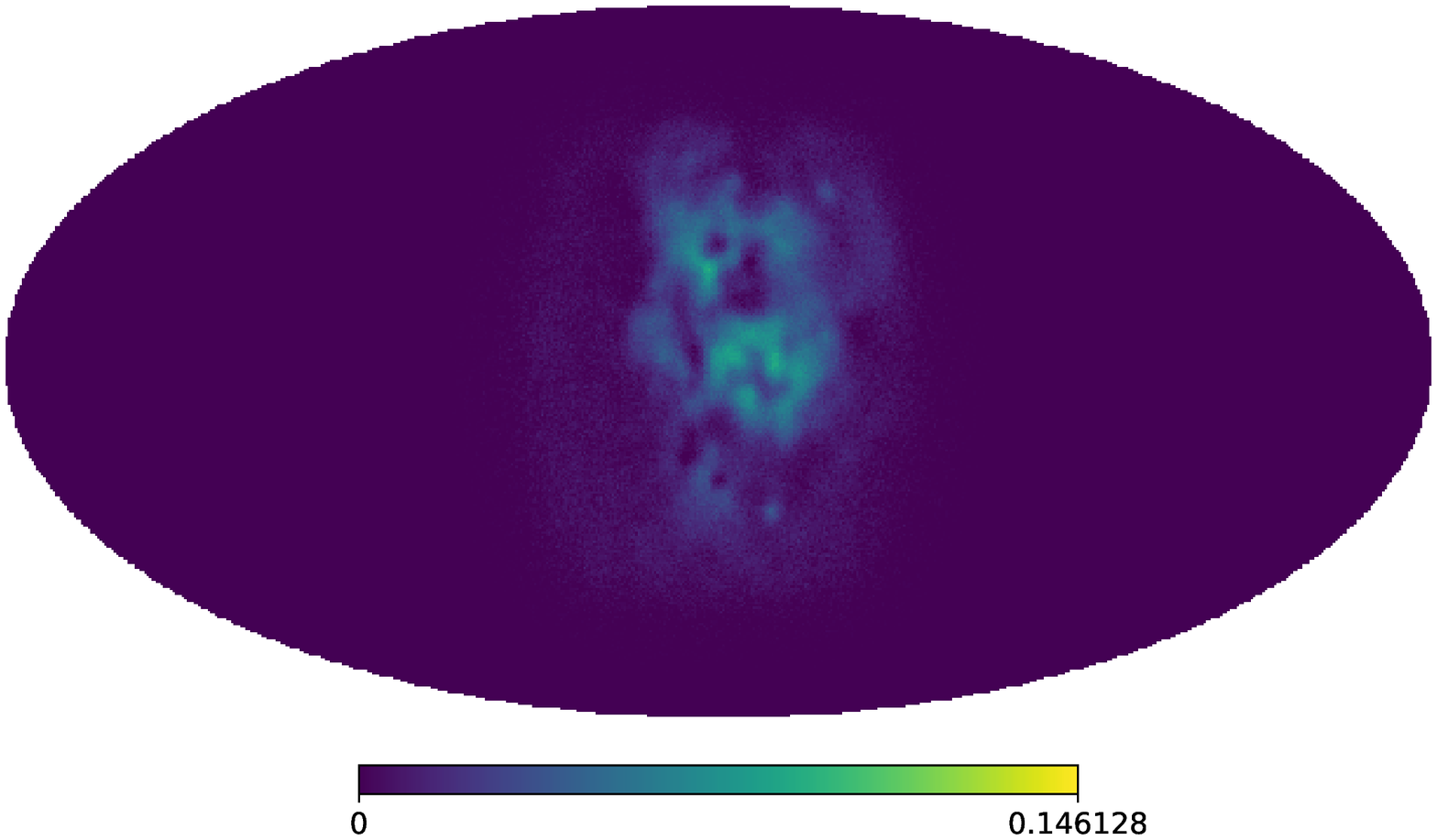}
	} 	
	\hfill
	\subfloat[][$\beta$-SNMF]{
	\centering
	\includegraphics[width=.47\linewidth]{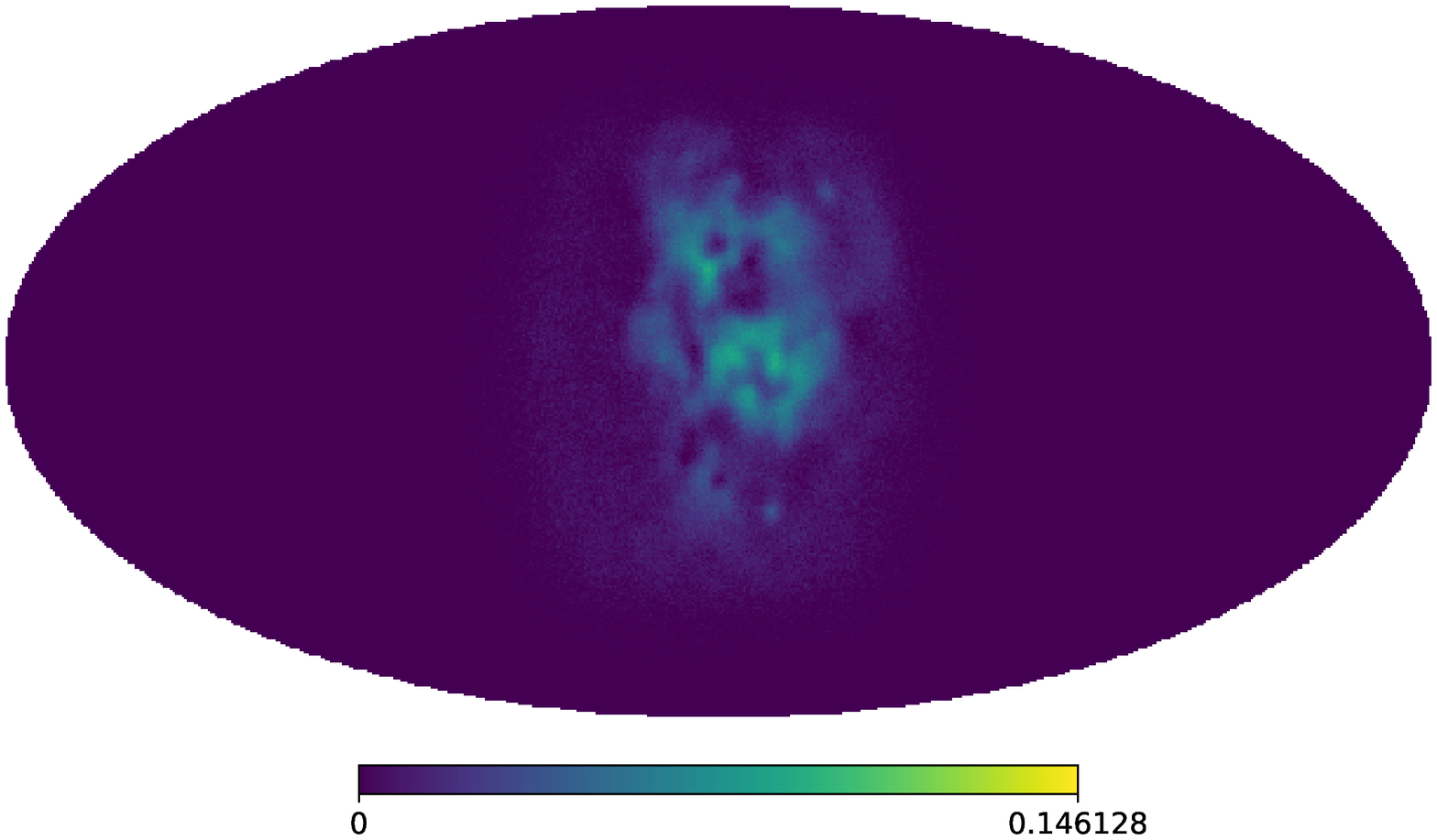}
	}
	\caption{Estimated source $\bS_2$ by different algorithms (arbitrary unit, logarithmic scale)}
	\label{fig:planckcomp}
\end{figure}

Let us compare the results returned by the previously considered DBSS and BSS algorithms. The mean performance metrics over 50 noise realizations are reported in Table \ref{tab:compReal}. Equivalently to the synthetic data, SDecGMCA markedly overcomes oDecGMCA in terms of estimation error on both $\bS$ and $\bA$.\\
\begin{table}
	\centering
	\small
	\begin{tabular}{@{}llll@{}} 
		\toprule
		& $\text{C\textsubscript{A}}$ & NMSE\textsubscript{w}& NMSE \\
		\midrule
		SDecGMCA                & \textbf{18.02}  & \textbf{26.92} & \textbf{21.17}  \\
		oDecGMCA 				& 15.78 & 25.89 & 18.16 \\
		GMCA                       & 16.60  & 25.84 & N/A \\
		HALS                        & 7.78 & 10.29  & N/A \\
		$\beta$-SNMF        & 7.93  & 10.38   &  N/A \\
		\bottomrule
	\end{tabular}
	\caption{Mean performance metrics in dB, over 50 realizations, achieved by different algorithms. The oracle mean NMSE is 22.58~dB.}
	\label{tab:compReal}
\end{table}
The estimates of the thermal source by the five algorithms are reported in Fig.~\ref{fig:planckcomp}, along with the oracle estimation. As said earlier, the SDecGMCA and oracle estimates are particularly similar. oDecGMCA reconstructs the source with a slightly lower resolution. More importantly, it is contaminated by outlier pixels, which are likely due to the regularization favoring the higher frequencies combined with a too low threshold. All the finer details are lost in the GMCA estimate, because the observations are degraded beforehand. Both HALS and $\beta$-SNMF are unable to correctly denoise the sources; this highlights the advantage of the sparsity constraint in a transformed domain. \\

\begin{figure}
	\centering
	\includegraphics[width=.495\linewidth]{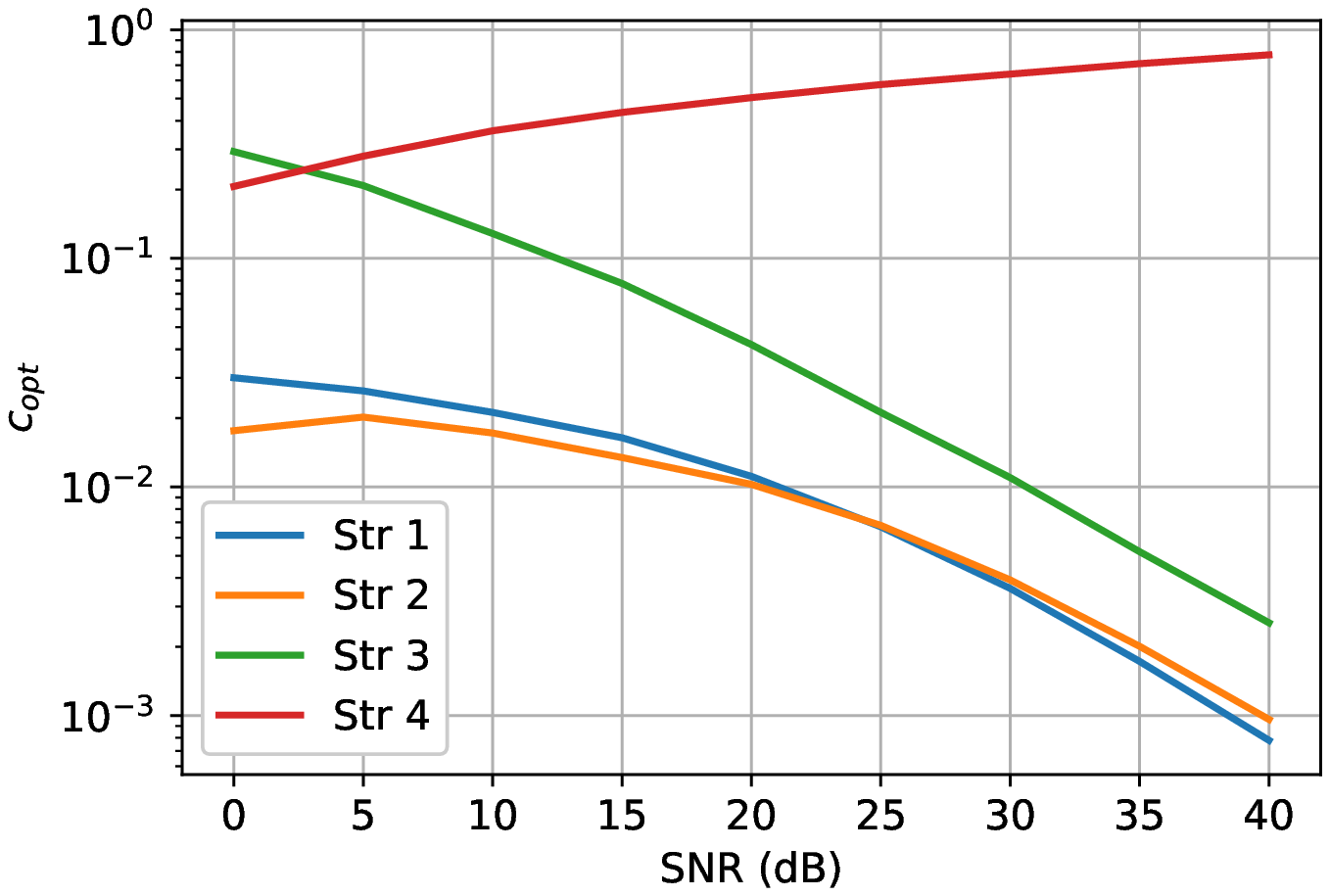}
	\caption{Mean optimal regularization hyperparameter of the non-blind problem, over 20 noise realizations, as a function of the SNR and the regularization strategy}
	\label{fig:planck_snr_copt}
\end{figure}

\begin{figure}
	\subfloat{
		\includegraphics[width=.495\linewidth]{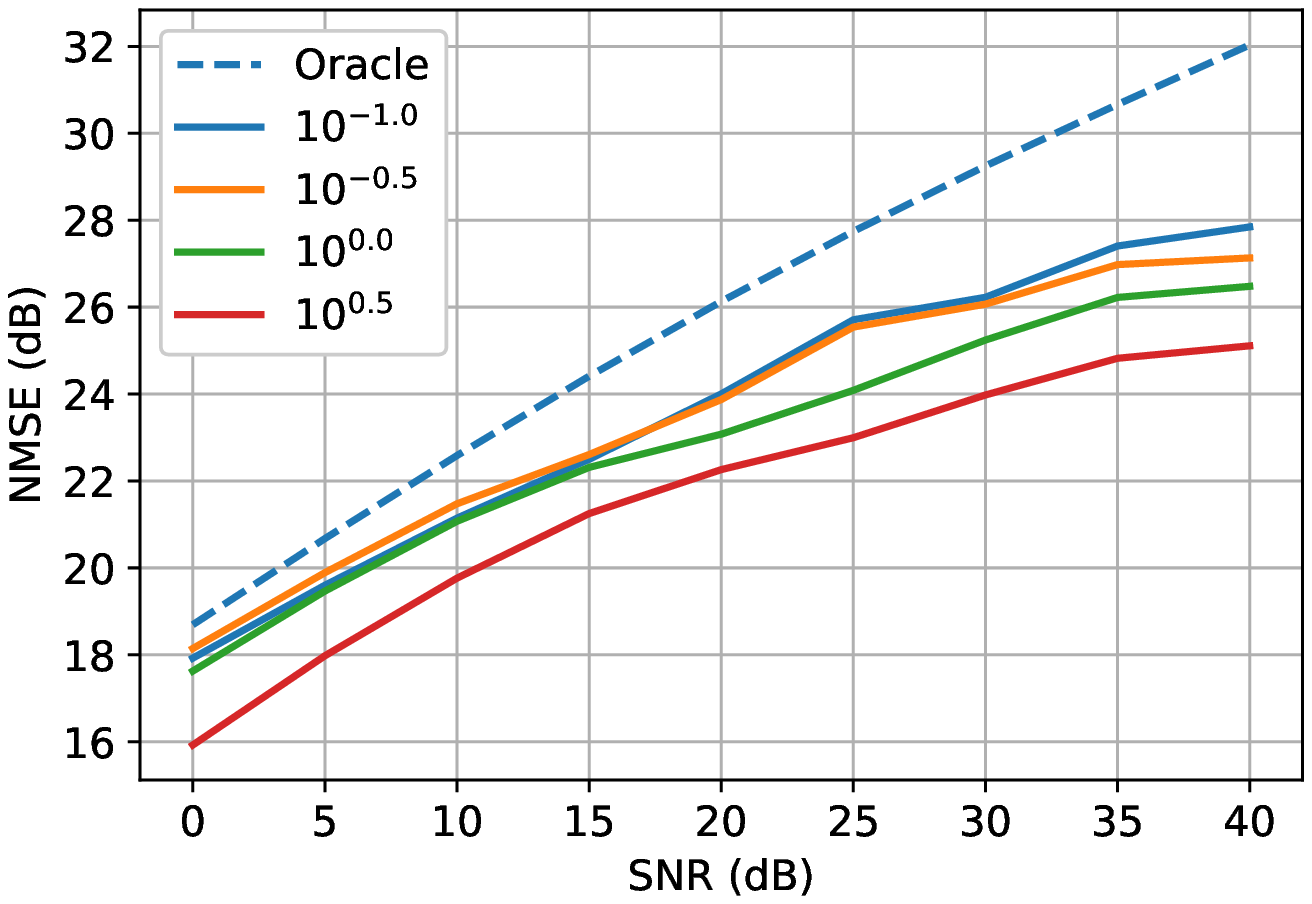}
	} 
	\hfill
	\subfloat{
		\includegraphics[width=.495\linewidth]{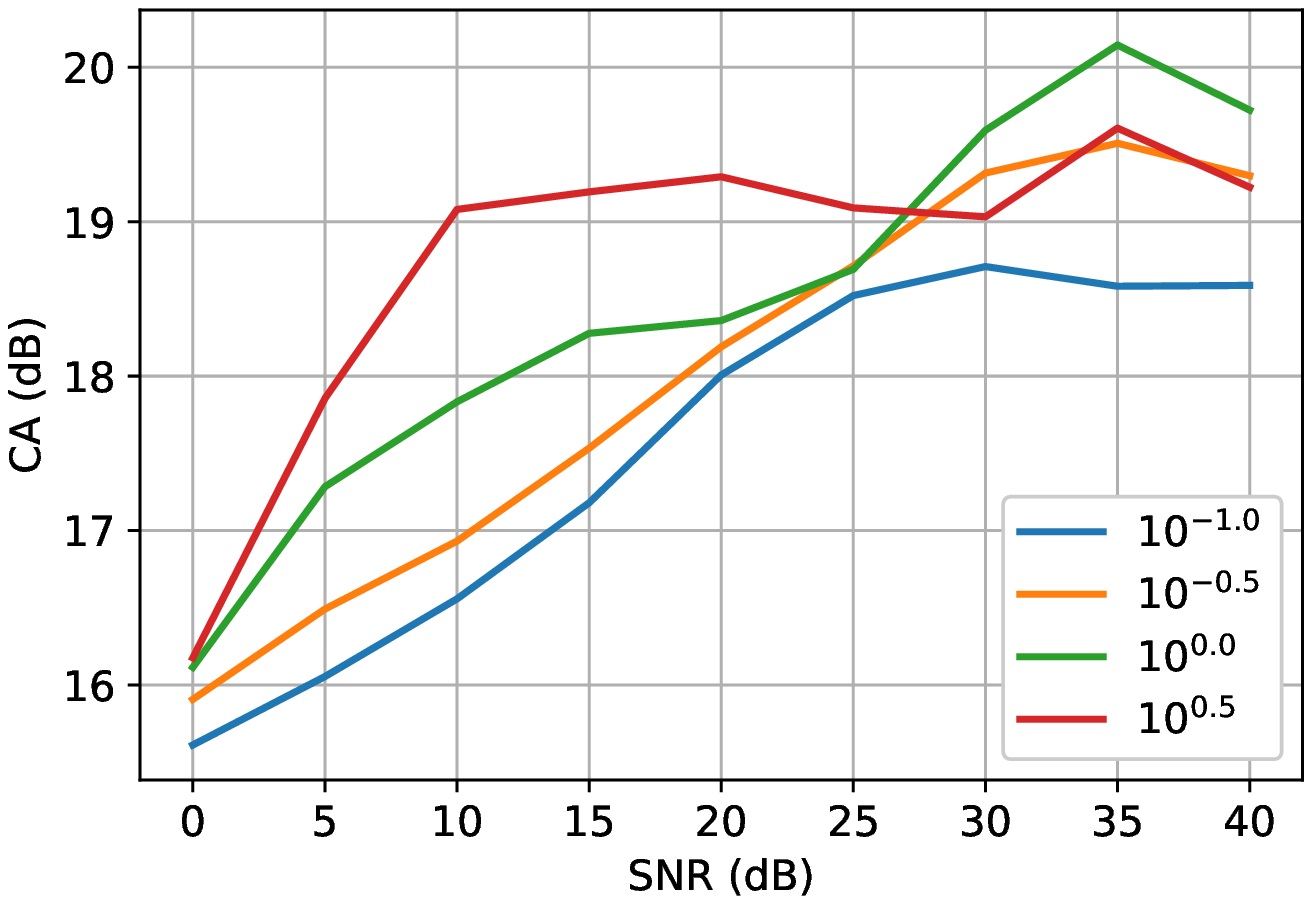}
	}
	\caption{Mean performance metrics over 20 noise realizations performed by SDecGMCA as a function of the SNR, with $c_{wu} = 1\mathrm{e}{-3}$ at warm-up and with different regularization hyperparameters $c_{ref}$ at refinement, which are indicated as multiples of ${c_{ref}}_{opt}$ in the legends.}
	\label{fig:planck_snr_perf}
\end{figure}

\begin{figure}
	\subfloat{
		\includegraphics[width=.47\linewidth]{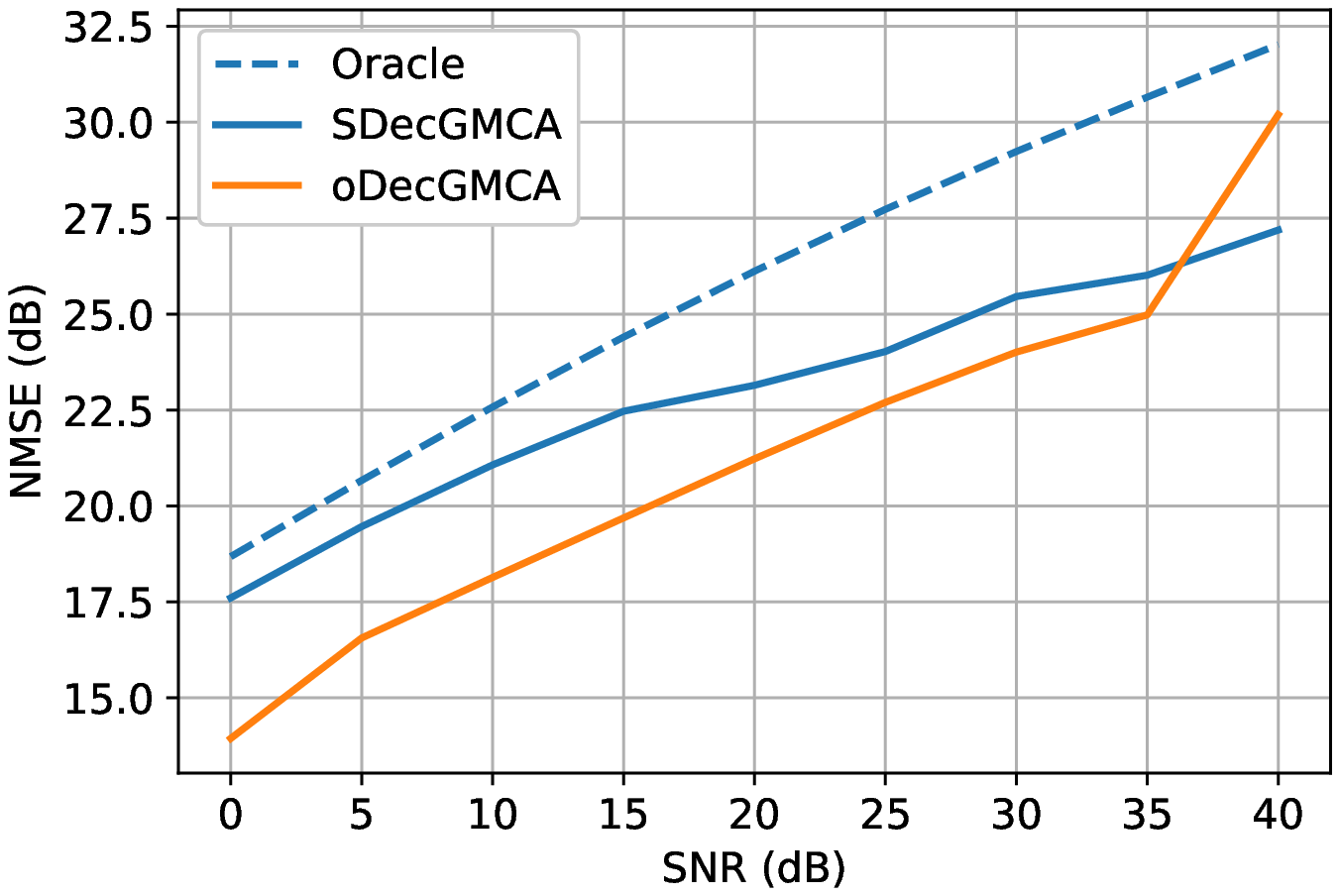}
	} 
	\hfill
	\subfloat{
		\includegraphics[width=.47\linewidth]{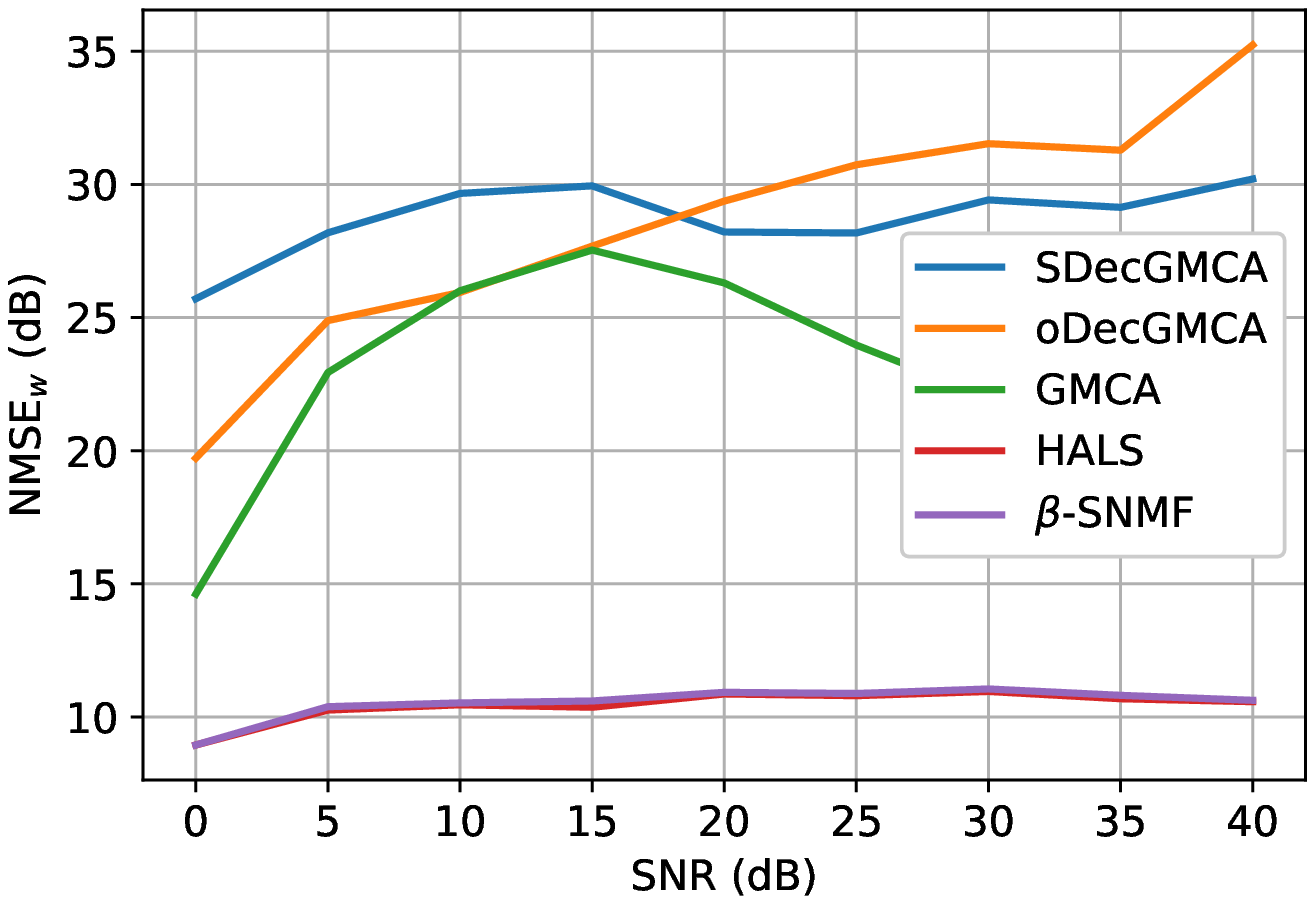}
	}\\[-.3cm]
	\subfloat{
		\includegraphics[width=.47\linewidth]{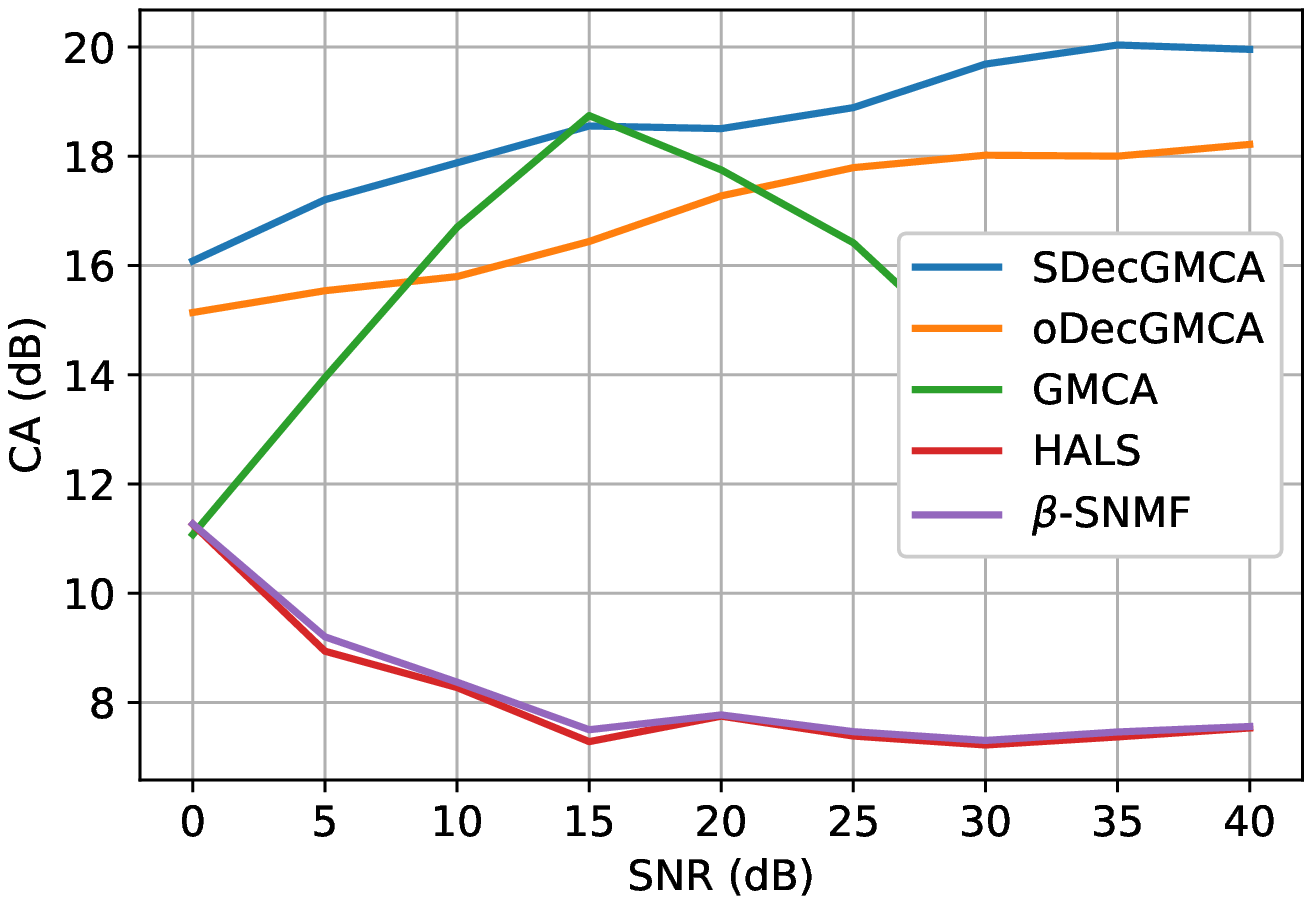}
	}
	\hfill
	\subfloat{
		\includegraphics[width=.47\linewidth]{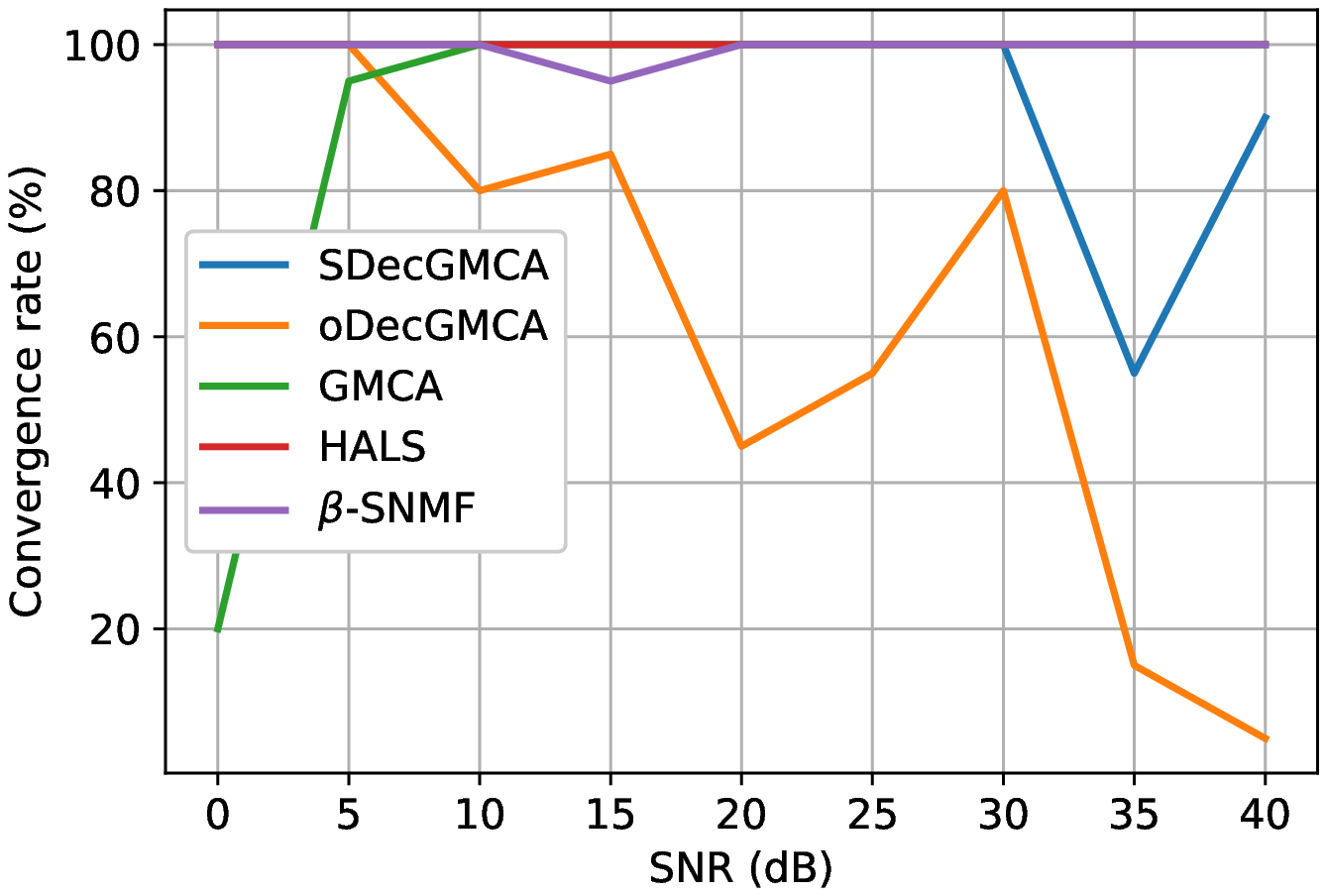}
	}
	\caption{Mean performance metrics and convergence rate over 20 noise realizations performed by different DBSS and BSS algorithms as a function of the SNR}
	\label{fig:planck_snr_perf_comp}
\end{figure}

We then focus on the behavior of SDecGMCA with respect to the SNR; the SNR is in fact the observation parameter that has the greatest impact on the results and the robustness of the algorithm. In the first place, the optimal regularization parameters are estimated (by resolving the non-blind problem); the results are plotted in Fig.~\ref{fig:planck_snr_copt}. As observed with the synthetic data, the optimal regularization hyperparamater of strategy \#4 (refinement strategy) is relatively insensitive to the SNR and is worth approximately 0.5. The optimal regularization hyperparameter at warm-up (strategy \#3) is more sensitive to the SNR. However, as with the synthetic data, it has limited influence on the result of the separation (thereafter, we set $c_{wu} = 1\mathrm{e}{-3}$, independently of the SNR).\\
The performance metrics of SDecGMCA as a function of the SNR and for different regularization hyperparameters at refinement are plotted in Fig.~\ref{fig:planck_snr_perf}. Similarly to the synthetic data, the NMSE and \ca~stabilize when there is little noise. In overall, the choice of $c_{ref}$ has little impact on the finale result (approximately $-2$ dB at most for both metrics). Overestimating $c_{ref}$ improves the estimation of $\bA$ to the detriment of the estimation of $\bS$, and \textit{vice versa}.\\
The different DBSS and BSS algorithms are compared for different noise levels; the results are reported in Fig.~\ref{fig:planck_snr_perf_comp}. Compared to the synthetic sources, the realistic sources are more correlated and less sparse in the starlet domain. This can result in a decrease in robustness. In order to compare the different methods, we calculated the mean performance metrics only over the successful realizations (that is with a $\text{C\textsubscript{A}}$ close enough to the maximum $\text{C\textsubscript{A}}$ along the realizations). SDecGMCA demonstrates a satisfactory robustness in terms of convergence to noise. As with the synthetic data, the performance metrics tend to stabilize when there is little noise. These experiments highlight that oDecGMCA is not robust at a low noise level; the few cases which converge return good metrics, hence the apparent better mean $\text{C\textsubscript{A}}$ and NMSE at high SNR than SDecGMCA. The performances of GMCA notably decrease at a low noise level. Indeed, the degree of sparsity of the sources is markedly decreased at low resolution; when combined with a low noise level (and thus a smaller threshold regularization), the separation process is deteriorated. 

\section{Conclusion}
In this article, we investigate a new joint deconvolution and sparse blind source separation algorithm to analyze multichannel spherical data. We first thoroughly investigate the impact of the regularization scheme in the least-squares minimization, which is proved to largely impact the quality of the separation process. We further introduce dedicated regularization schemes that much better adapt to the statistics of the sources to be estimated. Based on these regularization techniques, we introduce a two-step minimization algorithm coined SDecGMCA, which is showed to provide a robust and effective minimization procedure. Numerical experiments on both toy examples and realistic astronomical simulations are achieved to evaluate how the proposed algorithm performs in a wide range of mixing scenarios. Comparisons with standard BSS methods are further carried out, which show that the proposed SDecGMCA algorithm is competitive.

\section*{Acknowledgement}
This work is supported by the European Community through the grant LENA (ERC StG - contract no. 678282).

\bibliography{biblio}

\appendix

\section{Open source code} \label{app:code}

The code is open source and can be found online at \url{https://github.com/RCarloniGertosio/SDecGMCA} on version 3 of the LGPL.

\section{Implementation of SDecGMCA with a multiresolution representation} \label{app:multires}

Multiresolution representations ({\it e.g.}~wavelets, curvelets, to only name two) are versatile and yet effective analytic signal representations that are well adapted to provide sparse representations for a wide range of natural data. Consequently, we choose to use the spherical starlet representation \cite{Starck05} for the numerical experiments of the present paper. In this appendix, we describe in detail the implementation of SDecGMCA with a multiresolution representation.

\paragraph*{Dealing with the coarse resolution} As detailed in \cite{Starck_10_SparseImageand}, such signal representations decompose the data into detail scales and a coarse scale. The former bears invaluable information to disentangle the sources while the latter is a mere low frequency approximation of the signals that is useless for the separation process. Therefore, the coarse scales of the observations are removed during the separation procedure (that is the warm-up and refinement stages described in Algorithm \ref{alg:SDecGMCA}). The low frequency information included in the coarse scales of the raw data $\mathbf{X}$ is then reincorporated in the sources with the obtained mixing matrix. An iterative procedure is still necessary due to the reweighting and the regularization based on the spectra (strategy \#4). Moreover, as described earlier, the maximum source support is set to $K=1$ during this finale step in order to improve the estimation of the sources. 

\paragraph*{Non-negativity constraint on $\bS$} The non-negativity of the sources is applied by projecting the sources, expressed in the direct space, on the positive orthant. To do so, it is necessary to estimate the sources in the pixel or sample domain. Therefore, at each step of the SDecGMCA algorithm, each source is fully reconstructed, which requires an extra projection in the harmonic space per iteration. In total, two spherical harmonics transforms are performed at each iteration: one for the estimation of the sources and one for the update of the mixing matrix.

\paragraph*{Managing the regularization thresholds in a multiresolution representation} Selecting adequate thresholds plays a key role in the achievement of a separation. As pointed out earlier, the thresholds depend on the level of the noise corrupting the estimated sources. In the case where the level of the noise which contaminates the data is known, it is possible to determine analytically the noise level affecting the sources in the chosen multiresolution representation, thus ensuring a suitable thresholding.\\
We recall that the observations are contaminated by a white Gaussian noise $\mathbf{N}$ of variance $\sigma^2$. 
Quite similarly to the plane case, the harmonic transform of such a noise is Gaussian; its covariance matrix multipole by multipole is:
\begin{equation}
	\expect\left[\mathbf{\hat{N}}^{l,m}{{}\mathbf{\hat{N}}^{l,m}}^\dagger\right] = \frac{4\pi}{N_p} \sigma^2 \textbf{I}.
\end{equation}
The least-squares update of the sources colors and correlates the noise:
\begin{equation}
	\label{eq:covSourceSH}
	\expect\left[\mathbf{\hat{P}}^{l,m}{{}\mathbf{\hat{P}}^{l,m}}^\top\right] = \frac{4\pi}{N_p} \sigma^2 \left(\mathbf{M}[l] +  \diag\limits_{n\in[1,N_s]} \left(\varepsilon_{n,l}\right)\right)^{-1} \mathbf{M}[l]  \left(\mathbf{M}[l] +  \diag\limits_{n\in[1,N_s]} \left(\varepsilon_{n,l}\right)\right)^{-1}
\end{equation}
where $\mathbf{P}$ is the noise projected in the source domain by Eq.~\eqref{eq:updateS2} and $\mathbf{M}[l] = \mathbf{A}^\top \diag\left(\mathbf{\hat{H}}^l\right)^2 \mathbf{A}$. The multiresolution analysis preceding the thresholding further colors the noise; more specifically, the term $\abs*{ \mathbf{h_j}_l}^2$ is added to each detail scale $j$, where $\mathbf{h_j}$ is the filter giving the $j$\textsuperscript{th} detail scale of the multiresolution representation $\mathbf{\Phi}$. \\
The constraint term on $\mathbf{S}$ (Eq.~\refeq{eq:consS}) does not take into account the noise correlation in the source domain (principally for computational reasons, in order for the associated proximal operator to have an analytical form and be applied fast). Thus, the non-diagonal terms of the covariance matrix are neglected\footnote{In the numerical experiments, the non-diagonal terms are in fact small compared to the diagonal terms.}. This leads to the following variance in the direct space for source $n$ and scale $j$:
\begin{equation}
\expect\left[{{}\left[\mathbf{P \Phi^\top}\right]^p_{n,j}}^2\right] = \frac{\sigma^2}{N_p} \sum_{l=0}^{l_{max}} (2l+1) \left[\left(\mathbf{M}[l] +  \diag\limits_{n\in[1,N_s]} \left(\varepsilon_{n,l}\right)\right)^{-1} \mathbf{M}[l]  \left(\mathbf{M}[l] +  \diag\limits_{n\in[1,N_s]} \left(\varepsilon_{n,l}\right)\right)^{-1}\right]_n^n \abs*{ \mathbf{h_j}_l}^2
\end{equation}

\section{Numerical comparisons of the regularization strategies: detailed results} \label{app:resnonblind}
In this appendix, we present the results of the numerical experiments performed to compare the regularization strategies. To do so, we recall that we resort to the non-blind separation case, \textit{ie.}~with the ground truth mixing matrix $\mathbf{A^*}$. 

\begin{figure}
	\subfloat{
		\centering
		\includegraphics[width=.495\linewidth]{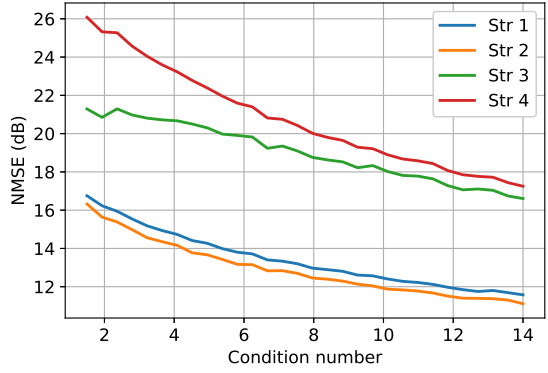}
	} \qquad
	\subfloat{
		\centering
		\includegraphics[width=.495\linewidth]{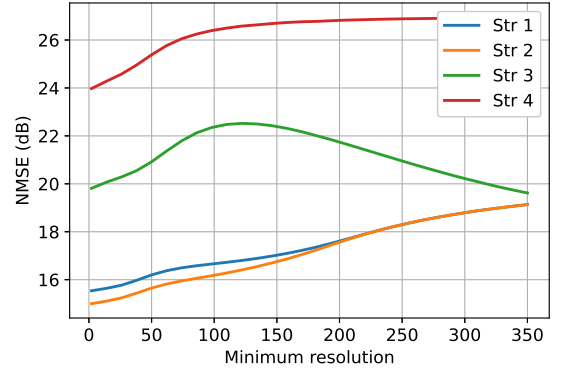}
	}\\
	\subfloat{
		\centering
		\includegraphics[width=.495\linewidth]{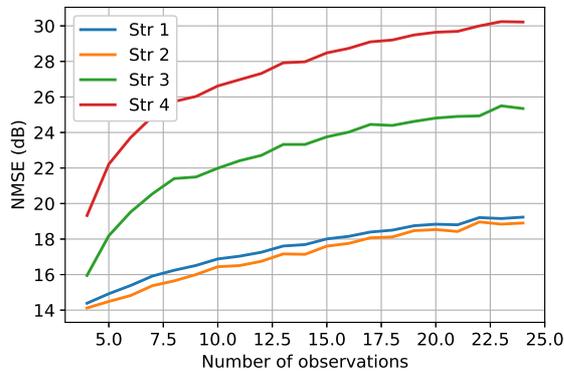}
	}\qquad
	\subfloat{
		\centering
		\includegraphics[width=.495\linewidth]{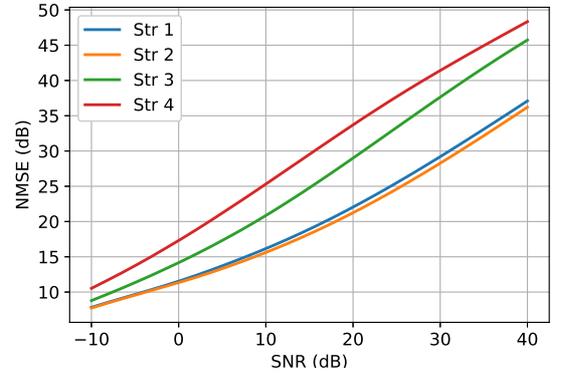}
	}
	\caption{Mean NMSE over 100 realizations as a function of observation parameters. The default values of the parameters are summarized in Section~\ref{sec:nonblindpb}. The decrease in NMSE for str.~3 is due to the decrease of $\varepsilon_{n,0}-\varepsilon_{n,l_{max}}$, which reduces the ability of the Tikhonov  regularization to discriminate the higher frequencies compared to the lower frequencies.}
	\label{fig:nonblind_rmse}
\end{figure}

\begin{figure}
	\subfloat{
		\centering
		\includegraphics[width=.495\linewidth]{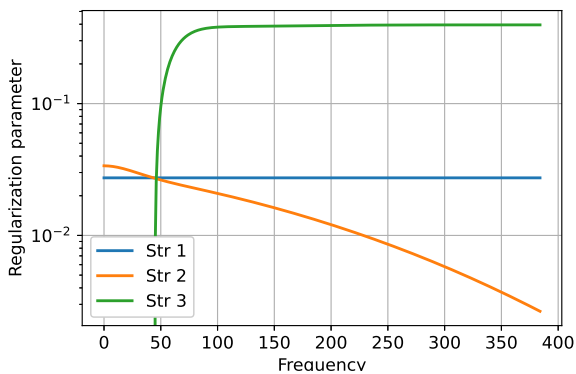}
	}\qquad
	\subfloat{
		\centering
		\includegraphics[width=.495\linewidth]{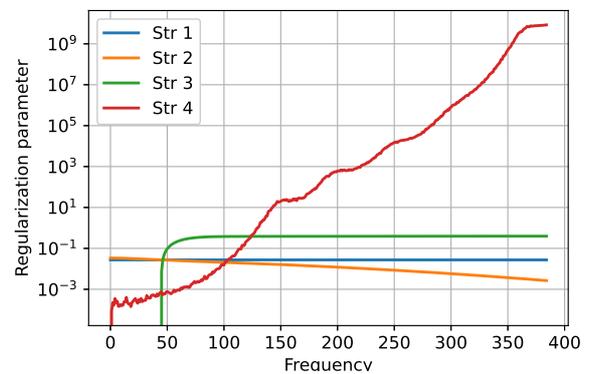}
	}
	\caption{Example of regularization parameters $\varepsilon_{n,l}$ as a function of the frequency $l$ for the default observation parameters and with the mean optimal hyperparameters.}
	\label{fig:nonblind_eps}
\end{figure}

Figure \ref{fig:nonblind_rmse} shows the mean NMSE of $100$ random trials as a function of the four aforementioned observation parameters: number of observations, mixing matrix condition number, minimum resolution and SNR. Unsurprisingly, strategy \#4 clearly provides the best reconstruction qualities. Among the other strategies, that do not assume the sources to be known, strategy \#3 achieves better results. It is mostly thanks to the non-linear maximum operator, which allows to keep the lower frequencies unbiased, where most of the sources energy is located (see example Fig. \ref{fig:nonblind_eps}, where $\varepsilon_{n,l} = 0$ for $l \leq 44$). Strategy \#2 gives poor results; indeed, it biases more significantly the lower frequencies than the higher ones.


Table \ref{table:range_copt} shows the range of variation of the mean optimal regularization hyperparameters when the observation parameters vary. Contrary to the 3 first strategies, the optimal regularization hyperparameter for strategy \#4 is rather insensitive to the observation parameters (typically $c_{opt} \sim 0.5$). The noticeable exception is regarding the SNR. When the SNR is low, the denoising by the sparsity regularization is particularly promoted, hence a marked decrease of $c_{opt}$.

\begin{table}
	\centering
	\small
	\begin{tabular}{@{}llcccc@{}}
		\toprule
		\multirow{2}{*}{Parameter} & \multirow{2}{*}{Range} & \multicolumn{4}{c}{Regularization strategy} \\
		\cmidrule{3-6}
		& & \#1        & \#2       & \#3 & \#4      \\
		\midrule
		$\cond(\mathbf{A})$  & 1.5 to 14  		& 1.04 & 1.84 & 0.31 & $\mathbf{0.10}$  \\
		$r_{min}$ 					&  2 to 350    	  & $>3.30$ & $>2.60$ & 0.75 & \textbf{0.11 }\\ 
		$N_c$ 						 & 4 to 24         	& 0.49 & 2.02 & 1.11 & \textbf{0.25} \\
		SNR (dB) 					& $-10$ to 40  & 1.18 & 1.08 & \textbf{0.41} & 0.76 \\
		\bottomrule
	\end{tabular}
	\caption{Range of variation of the mean optimal regularization hyperparameter $c_{opt}$, in terms of order of magnitude, when the observation parameters vary. In some cases, $c_{opt}$ is smaller than the lower bound of the research interval (\num{1e-5}), hence the lower bound for the range of variation.}
	\label{table:range_copt}
\end{table}

\end{document}